\documentclass[11pt,a4paper]{article}

\usepackage{jheppub}

\makeatletter
\gdef\@fpheader{} 
\makeatother

\usepackage{amsmath,amssymb,amsthm,mathtools,bm}

\usepackage{adjustbox}
\usepackage{graphicx}
\usepackage{subcaption}   
\usepackage{booktabs}     
\usepackage{tabularx}

\usepackage{microtype}
\usepackage{parskip}
\usepackage{physics}

\usepackage{dsfont}

\usepackage{enumitem}

\usepackage{xcolor}
\newcommand{\emphdark}[1]{\textcolor{black!80}{#1}}

\usepackage[colorlinks=true]{hyperref}
\usepackage[capitalize,noabbrev]{cleveref}

\newcommand{\HH}{\mathrm{HH}}

\numberwithin{equation}{section}

\begin{document}

{%

\title{``It from Bit'': The Hartle--Hawking state\\ and quantum mechanics for de Sitter observers}
\author[a]{Ying Zhao}

\affiliation[a]{MIT Center for Theoretical Physics --- a Leinweber Institute,\\
Massachusetts Institute of Technology, Cambridge, MA 02139, USA }

\emailAdd{zhaoying@mit.edu}

\abstract{
The one-state statement for closed universes has sparked considerable discussion. In this paper, we examine its physical meaning in the context of the Hartle--Hawking state and de Sitter space. We argue that the one-state property of closed universes is fully compatible with the finite-dimensional quantum mechanics experienced by observers inside de Sitter space, and that this compatibility requires neither mixing of $\alpha$-sectors nor any modification of the rules of the gravitational path integral. The apparent tension is resolved by sharply distinguishing the baby-universe Hilbert space (the space of closed universes viewed from the \emph{outside}) from the bulk Hilbert space that governs quantum mechanics for an observer \emph{inside} a single de Sitter universe. 

The baby-universe Hilbert space, together with its commutative operator algebra, is not a quantum-mechanical Hilbert space: it is merely a mathematical repackaging of classical probability theory and carries no quantum-mechanical structure at all, a direct consequence of the one-state property of closed universes. Accordingly, attempting to formulate quantum mechanics directly on the baby-universe Hilbert space conflates two logically distinct structures and leads to physically incorrect conclusions. By contrast, the quantum mechanics experienced by an observer inside de Sitter space emerges from the classical statistics encoded in the baby-universe Hilbert space, providing a concrete realization of Wheeler’s idea of ``It from Bit.'' We demonstrate these features by completely solving a topological toy model of one-dimensional de Sitter spacetime. Along the way, we clarify the physical meaning of de Sitter entropy, showing that it corresponds to the coarse-grained entropy of the underlying state.

}

\maketitle
}
	
\flushbottom

\section{Introduction}

The holographic principle states that the number of degrees of freedom needed to describe a gravitational region is upper bounded by the boundary area of that region in Planck units \cite{tHooft:1993dmi,Susskind:1994vu}. A classical example is the black hole entropy formula, which states that the entropy of a black hole is given by its horizon area in Planck units \cite{Bekenstein:1973ur,Hawking:1975vcx}. Another concrete realization of this principle is the AdS/CFT duality \cite{Maldacena:1997re,Gubser:1998bc,Witten:1998qj}: the fundamental description of a gravitational theory in an asymptotically AdS spacetime is a quantum mechanical theory living on its spatial boundary. We emphasize that the term ``degrees of freedom'' here refers to quantum mechanical degrees of freedom (``It from Qubit'').

A closed universe, by definition, does not have a spatial boundary, or equivalently, the area of its spatial boundary is zero. The holographic principle therefore appears to imply that the number of fundamental quantum mechanical degrees of freedom describing the entire closed universe is zero. This conclusion, though surprising, has received increasing support from recent work from a variety of perspectives: gravitational path integral \cite{Almheiri:2019hni,Penington:2019kki,Marolf:2020xie,Usatyuk:2024mzs}, factorization property of CFT \cite{Maldacena:2004rf,Usatyuk:2024isz}, black hole unitarity \cite{Akers:2022qdl}, constructions in AdS/CFT \cite{Antonini:2023hdh,Antonini:2024mci,Engelhardt:2025vsp,Gesteau:2025obm}, and the uniqueness of string theory \cite{McNamara:2020uza}. 
This is the one-state statement for closed universes that we explore in this paper.\footnote{Zero quantum mechanical degrees of freedom means that there is only a single state, which is effectively a number.}

It has long been known that gravity can admit superselection sectors, known as $\alpha$-sectors \cite{Coleman:1988cy,Giddings:1988cx}. Most explicit and well-controlled examples of such $\alpha$-sectors arise in low-dimensional gravity theories, where the gravitational path integral can be defined precisely \cite{Saad:2019lba,Marolf:2020xie,Chandra:2022bqq}. As superselection sectors, $\alpha$-sectors never mix, and the presence of multiple $\alpha$-sectors does not increase the number of quantum mechanical degrees of freedom in each sector. One possible scenario for closed universes, then, is the existence of many $\alpha$-sectors (many bits), while each sector contains nothing more than a set of numbers --- a realization of Wheeler's idea of ``It from Bit'' \cite{Wheeler:1989ftm, Wheeler:1990}. In this paper we explore this possibility, particularly in the context of the Hartle--Hawking state and de Sitter space. 

\emph{The main claim of this paper is that the one-state property of closed universes is fully compatible with the finite-dimensional quantum mechanics experienced by observers inside de Sitter space, and that this compatibility requires neither mixing of $\alpha$-sectors nor any modification of the rules of the gravitational path integral.} The apparent tension is resolved once one recognizes that the one-state statement is a constraint on the baby-universe Hilbert space (namely, the absence of $\alpha$-mixing) while bulk quantum mechanics does not live on the baby-universe Hilbert space at all, even though that space encodes all the information needed to reconstruct it.

There is a sharp distinction between these two Hilbert spaces. The baby-universe Hilbert space, which arises from the gravitational path integral, encodes classical statistical correlations among boundary conditions. As physicists, we are naturally inclined to associate the term ``Hilbert space'' with quantum mechanics, but in this case that intuition is misleading. The baby-universe Hilbert space, together with its commutative operator algebra, is simply a mathematical repackaging of classical probability theory. Physically, it can be considered as the space of all closed universes viewed from the \emph{outside}. It has infinite dimension, a commutative operator algebra, and carries no quantum-mechanical structure --- a direct consequence of the one-state property of closed universes.

By contrast, the quantum mechanics experienced by an observer \emph{inside} a single de Sitter universe is governed by a finite-dimensional Hilbert space. This bulk Hilbert space does not arise from a quantum theory of baby universes; instead, it emerges from the classical statistical data encoded in the baby-universe Hilbert space.

Accordingly, formulating quantum mechanics directly on the baby-universe Hilbert space conflates two logically distinct structures and is therefore conceptually misleading.\footnote{These conceptually misguided attempts include my own.}

To make this structure explicit, we study a simple one-dimensional topological model of de Sitter space. The model is specified by a single parameter $Z_L$, the analog of the sphere partition function in higher-dimensional de Sitter theories. We show explicitly that this de Sitter universe admits a quantum-mechanical description in terms of a finite-dimensional Hilbert space of dimension $Z_L$. The Hartle--Hawking no-boundary condition corresponds to an unnormalized Haar-random vector in this Hilbert space, while more general backgrounds arise via conditioning. Within this model, de Sitter entropy\footnote{By ``de Sitter entropy'' we mean the logarithm of the Euclidean partition function associated with a given classical background.} acquires a clear interpretation: it is the coarse-grained entropy of the underlying state. Introducing features such as defects or black holes in higher-dimensional semiclassical gravity corresponds to conditioning that reduces this coarse-grained entropy.

Armed with this concrete example, we examine the relationship and distinction between baby universes and bulk quantum mechanics. We show explicitly how vectors in the baby-universe Hilbert space define quantum states on the bulk Hilbert space, thereby translating the classical statistics of the gravitational path integral into the quantum mechanics experienced by a bulk observer. The lesson is conceptual rather than technical: there is no quantum mechanics of closed universes viewed from \emph{outside}. Quantum mechanics appears only for observers who are part of the universe.

Throughout this paper, our main tool is the gravitational path integral. The path integral is well behaved in low-dimensional gravity theories, including the simple one-dimensional toy model studied here. A major open issue and perhaps the weakest point of this approach is that the gravitational path integral may not be well defined in higher-dimensional theories. Our hope is that by studying low-dimensional toy models, we can extract general lessons that remain valid in more realistic settings, or at least gain insight into the general structure of quantum gravity in de Sitter space.

The rest of the paper is organized as follows. In \cref{sec:rev}, we briefly review the one-state statement for closed universes and discuss some of its consequences for AdS closed universes. In \cref{sec:dS}, we highlight the key differences between de Sitter spacetime and AdS closed universes, and develop a general framework for recovering the quantum mechanics experienced by a de Sitter observer. \cref{sec:toy_model} contains the main technical results of the paper. After introducing the toy model, we analyze quantum mechanical systems of different sizes that can exist in this simple de Sitter space. We then present a unified description of physics in terms of a quantum mechanical theory whose Hilbert-space dimension is set by the partition function $Z_L$. By studying conditionings that correspond to different classical backgrounds in higher-dimensional de Sitter theories, we identify de Sitter entropy with the coarse-grained entropy of the underlying quantum state. In the final part of \cref{sec:toy_model}, we clarify the relationship and distinction between the baby-universe Hilbert space (``Bit'') and physics inside de Sitter space (``It'').

\textbf{Note:} This paper grew out of initial collaboration and extensive discussions with Daniel Harlow, but reflects my own perspective and choices of emphasis.

\section{Review: One-state problem and its consequences for AdS closed universes}
\label{sec:rev}

\subsection{One-state property of closed universes}

There are several independent arguments supporting the one-state property for closed universes. By applying the island formula to closed-universe settings, it was argued that a closed universe cannot have nontrivial entanglement with anything outside \cite{Almheiri:2019hni}. Such a statement can only be consistent if the closed universe has a single quantum state. More general arguments based on the gravitational path integral were given in \cite{Penington:2019kki} and further developed in \cite{Marolf:2020xie,Usatyuk:2024mzs}. The one-state property is also required if the radiation from a completely evaporated black hole is to be pure \cite{Akers:2022qdl}. From the perspective of holography, the existence of a unique closed-universe state follows from factorization \cite{Maldacena:2004rf,Usatyuk:2024isz}, making the one-state property closely related to the factorization problem in AdS/CFT. Further support comes from recent concrete constructions within AdS/CFT \cite{Antonini:2023hdh,Antonini:2024mci,Engelhardt:2025vsp,Gesteau:2025obm}. Finally, from a different viewpoint, the one-state property can be understood as a consequence of the uniqueness of string theory \cite{McNamara:2020uza}.

While each fixed holographic theory (that is, each individual $\alpha$-sector) admits only a single closed-universe state, there can nevertheless exist many distinct $\alpha$-sectors. JT gravity and the topological model studied in \cite{Saad:2019lba,Marolf:2020xie} provide concrete examples of this structure. We emphasize that the existence of many $\alpha$-sectors does not contradict the one-state statement. When we say that a closed universe has one state, we mean that each $\alpha$-sector is spanned by a single state (a complex number). Because different $\alpha$-sectors do not mix, the collection of all $\alpha$-sectors may span a large Hilbert space but there is no quantum mechanics within this space. It is sometimes suggested that allowing different $\alpha$-sectors to mix would effectively introduce nontrivial quantum mechanics in a large Hilbert space and thereby recover rich physics. As we will see in \cref{sec:toy_model}, such mixing is not necessary: with sufficiently many $\alpha$-parameters, rich quantum mechanics can already be recovered for a bulk observer without introducing mixing between $\alpha$-sectors.

In what follows, we study various features of closed universes using the gravitational path integral, which is well defined in simple models of low-dimensional gravity.

\subsection{AdS closed universes}
\label{sec:AdS}

\begin{figure}[!htbp]
  \centering
  \includegraphics[width=0.86\linewidth]{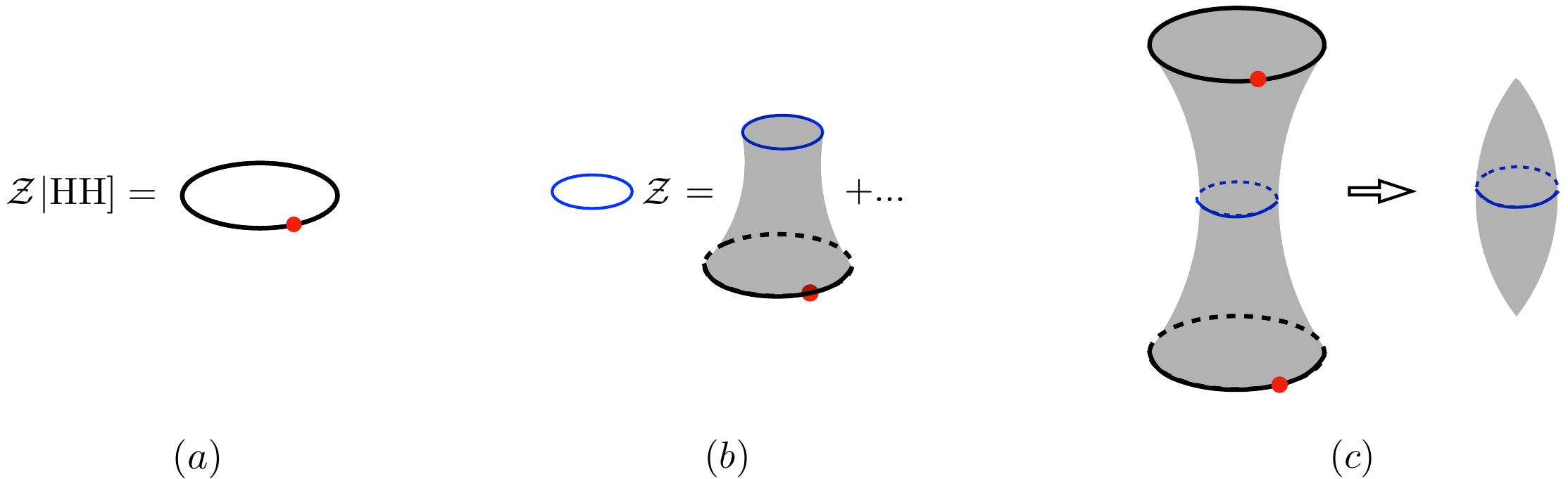}
  \caption{Preparing an AdS closed-universe state via a path integral.}
  \label{fig:AdS_state}
\end{figure}

The phase space of an AdS closed universe in JT gravity was studied in
\cite{Navarro-Salas:1992bwd,Henneaux:1985nw,Harlow:2019yfa,Usatyuk:2024mzs}. In the Euclidean path integral, one can construct the Hartle--Hawking state $|\HH]$ by imposing no-boundary conditions in the Euclidean past \cite{Hartle:1983ai}. A related but slightly different question is the following: in what sense does an asymptotic Euclidean AdS boundary condition $\mathcal{Z}$ define a closed-universe state (\cref{fig:AdS_state}(a))? The key point is that the Euclidean path integral with an asymptotic boundary prepares a wavefunction on any bulk cut: If we choose an arbitrary closed slice $\mathcal{C}$ (for example, the blue circle in \cref{fig:AdS_state}(b)) and perform the path integral with boundary conditions given by $\mathcal{Z}$ and $\mathcal{C}$, we obtain a wavefunction $\Psi_{\mathcal{Z}}[\mathcal{C}]$ on the space of closed slices, as illustrated in \cref{fig:AdS_state}(b). In the semiclassical limit, the dominant saddle contributing to this wavefunction can be a smooth geometry whose end is a closed slice $\mathcal{C}$ of definite size. Upon analytic continuation across $\mathcal{C}$, this construction prepares a Lorentzian closed universe (\cref{fig:AdS_state}(c)) \cite{Maldacena:2004rf,Usatyuk:2024mzs}.

Let us pause here to comment on notation. We use $|\HH]$ to denote the Hartle--Hawking no-boundary condition, rather than the standard Dirac ket notation used for quantum-mechanical states. The reason is that $|\HH]$ is a vector in the baby-universe Hilbert space, on which no quantum mechanics is performed. As we will argue later, the one-state statement implies that quantum mechanics on the baby-universe Hilbert space is trivial, or equivalently, that there is no mixing between different $\alpha$-sectors. At this stage, $|\HH]$ should be understood simply as a boundary condition in the gravitational path integral. An expression of the form $\mathcal{Z}|\HH]$ means that the boundary conditions in the path integral consist of an asymptotic boundary circle labeled by $\mathcal{Z}$ (see \cref{fig:AdS_state}(a)). When no confusion can arise, we will sometimes omit the explicit factor of $|\HH]$ and write only the nontrivial boundary conditions, as in \cref{fig:AdS_state}(b).

Throughout this paper, we generally denote objects associated with the baby-universe description using calligraphic symbols, such as $\mathcal{O}$, $\mathcal{Z}$, and $\mathcal{V}$. The only exception is the vector corresponding to the Hartle--Hawking no-boundary condition, for which we use $|\HH]$ in order to make its meaning explicit.

It was shown that, in this AdS closed-universe setup, various quantities such as inner products exhibit large ensemble fluctuations as a consequence of the one-state property~\cite{Harlow:2025pvj}. To facilitate comparison with the de Sitter case in later sections, we instead focus on a slightly different quantity: the expectation value of a patch operator $\mathcal{O}$.\footnote{As will become clear later, the inner products and expectation values considered here are all taken in the baby-universe Hilbert space. Nevertheless, they encode information about the quantum mechanics experienced by a bulk observer, which is associated with a different Hilbert space.} We begin by defining patch operators.

\subsection{Patch operators}

\label{sec:patch_0}

A patch operator $\mathcal{O}$ is defined in the path integral as follows. As with an asymptotic boundary circle $\mathcal{Z}$, we treat $\mathcal{O}$ as a type of boundary condition. When evaluating its value, we sum over all configurations in which $\mathcal{O}$ is inserted at an arbitrary location in the bulk. This construction parallels the definition of vertex operators in worldsheet string theory. We refer to such objects as \emph{patch operators}.

In our context, one example of a patch operator is a geodesic slice of fixed size $b$, possibly with matter operators inserted:
\begin{equation}
	\mathcal{O}=\ \adjincludegraphics[width=0.55in,valign=m]{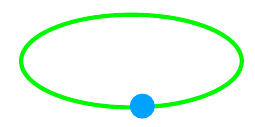}
	\label{eq:patch_b}
\end{equation}

In general, $\mathcal{O}$ does not have to be an entire closed slice. It may instead be a local operator whose position is integrated over spacetime.\footnote{Examples in JT-like theories include defect operators \cite{Mertens:2019tcm,Witten:2020wvy,Turiaci:2020fjj}.} We use the term ``patch operator'' to emphasize this feature. Further details will be discussed in \cref{sec:dS}, and explicit examples will be studied in \cref{sec:toy_model}.

\subsection{Bulk observers in AdS closed universe}
\label{sec:AdS_observer}

Having defined the patch operators $\mathcal{O}$ (\cref{eq:patch_b}), we now compute their expectation value on a background describing an AdS closed universe $\mathcal{Z}$. We will see that the one-state property has concrete physical consequences for a bulk observer. 

The boundary condition for the path integral is
\begin{equation}
\mathcal{Z}^*\mkern0.2mu\mathcal{O}\mkern0.2mu\mathcal{Z} \;=\; \adjincludegraphics[width=2.22in,valign=m]{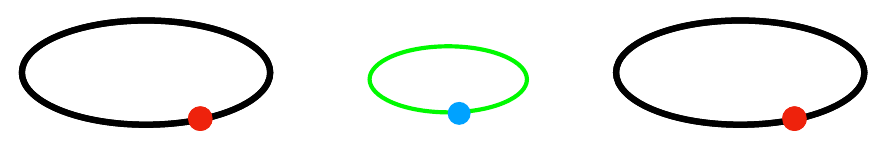}
\label{eq:exp_bd_AdS}
\end{equation}

As a notational comment, when we evaluate a path integral with boundary condition $\mathcal{B}$, the result will be denoted by $\overline{\mathcal{B}}$. Note also that on the left-hand side of \cref{eq:exp_bd_AdS}, all factors commute, since they simply represent different path-integral boundary conditions.

\begin{figure}[!htbp] 
\centering                     
      \includegraphics[width=0.99\linewidth]{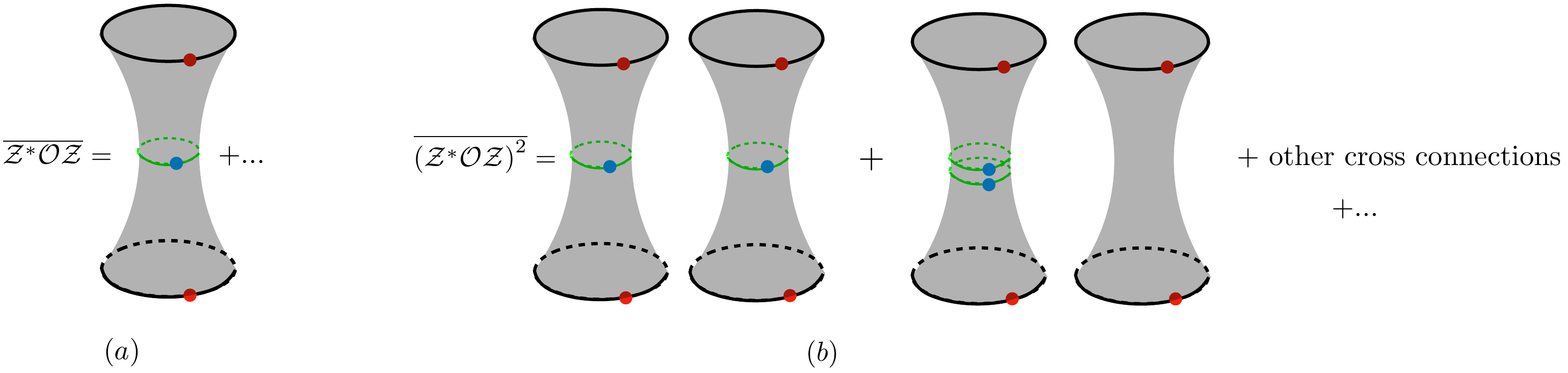}
  \caption{Expectation value (a) and expectation value squared (b) of a patch operator in an $AdS$ closed universe.}
 \label{fig:exp_AdS}
\end{figure}

\cref{fig:exp_AdS}(a) shows the computation of $\overline{\mathcal{Z}^*\mkern0.2mu\mathcal{O}\mkern0.2mu\mathcal{Z}}$, while \cref{fig:exp_AdS}(b) shows the computation of its square, $\overline{\qty(\mathcal{Z}^*\mkern0.2mu\mathcal{O}\mkern0.2mu\mathcal{Z})^2}$. We use ``$\cdots $'' to denote higher-genus contributions suppressed by $e^{-2S_0}$. To ensure that the AdS closed-universe saddle dominates, we assume that the one-boundary amplitudes vanish or are extremely small, so that the leading contribution is the cylinder. This situation can arise in low-dimensional JT-like models. In the evaluation of $\overline{\qty(\mathcal{Z}^*\mkern0.2mu\mathcal{O}\mkern0.2mu\mathcal{Z})^2}$ shown in \cref{fig:exp_AdS}(b), the first term is simply the square of \cref{fig:exp_AdS}(a). However, there are additional configurations whose contributions are not topologically suppressed and are of the same order. Only one representative example is drawn in \cref{fig:exp_AdS}(b). Following \cite{Harlow:2025pvj}, we refer to such configurations as cross-connections. Because of these cross-connections, $\overline{\qty(\mathcal{Z}^*\mkern0.2mu\mathcal{O}\mkern0.2mu\mathcal{Z})^2}\neq \overline{\mathcal{Z}^*\mkern0.2mu\mathcal{O}\mkern0.2mu\mathcal{Z}}^{\mkern1mu 2}$, with an order-one correction.

Wormhole corrections are familiar from black hole computations \cite{Penington:2019kki,Almheiri:2019qdq}, but in that context they are typically small, being suppressed by $e^{-S_{\text{BH}}}$, where $S_{\text{BH}}$ is the black hole entropy. These corrections give rise to the Page curve, demonstrating that black hole evaporation is a unitary process \cite{Penington:2019npb,Almheiri:2019psf}. In contrast, for AdS closed universes the wormhole corrections are large --- in fact, of order one. These large non-perturbative effects truncate the Hilbert space dimension to one \cite{Usatyuk:2024mzs}.

An obvious question is: what are the physical consequences of these large non-perturbative corrections for an observer living inside an AdS closed universe? It was known that spacetime wormholes induce uncertainties in bulk parameters
\cite{Coleman:1988cy,Giddings:1988cx,Polchinski:1994zs,Marolf:2020rpm,Marolf:2021ghr}. In Appendix~\ref{app:wormhole_AdS} we review various arguments and discuss their consequences when applied to the case of AdS closed universe. In short, some bulk parameters, for example, coupling constants or expectation values of low-energy operators, will exhibit large uncertainties for an observer, and the values she inferred after repeated experiments may deviate significantly from what's predicted by effective field theory.

Whether these issues should be regarded as fatal is ultimately a matter of interpretation. After all, we do not live in an AdS closed universe. Various options have been proposed to ameliorate these problems, for example by treating observers as classical (or otherwise modifying the treatment of observers in gravitational path integral) \cite{Harlow:2025pvj,Abdalla:2025gzn}. In the next section, we will see that de Sitter spacetime is different and does not suffer from these issues.


\section{De Sitter (sphere) is different}

\label{sec:dS}

In this section, we perform analogous calculations in de Sitter space and show that, although non-perturbative effects persist and the one-state statement continues to hold, the resulting physics is qualitatively different. We emphasize two key features of de Sitter space; taken together, they distinguish its physics from that of AdS closed universes.

\begin{enumerate}[label=(\Roman*)]
	\item Euclidean de Sitter space is a sphere and, in particular, has no asymptotic boundaries. In certain theories of gravity it appears as a saddle under the Hartle--Hawking no-boundary condition, which by definition corresponds to an empty boundary condition in the path integral.
	\label{feature:1}	
	
	\item The sphere partition function is typically large. In JT-like theories it carries a large topological factor $e^{2S_0}$. More generally, we expect it to scale as $e^{S_{\text{dS}}}$, where $S_{\text{dS}}$ is of order the de Sitter entropy. As a consequence, contributions to the path integral from multiple spheres are much larger than those from a single sphere (see \cref{fig:sphere_0}).
	\label{feature:2}

	\begin{figure}[!htbp] 
	\centering                     
	      \includegraphics[height=0.8in]{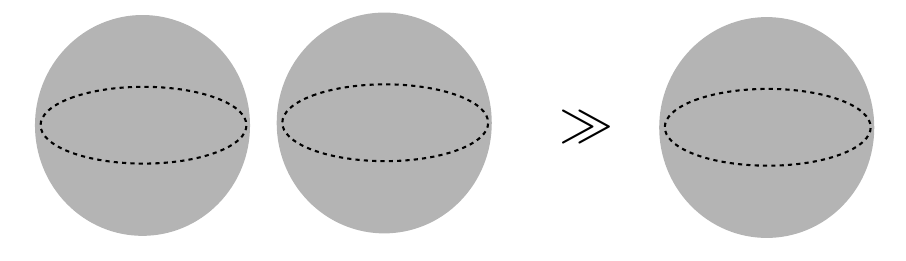}
	  \caption{Because the sphere partition function is large, the contribution of two spheres to the path integral is much larger than that of a single sphere.}
	 \label{fig:sphere_0}
	\end{figure}
\end{enumerate}

\subsection{Path integral calculation}
\label{sec:GPI}

We again consider the expectation value of the patch operator $\mathcal{O}$, defined in \cref{eq:patch_b}, now evaluated under the Hartle--Hawking no-boundary condition. In this case, the boundary condition for the path integral consists solely of a single patch operator insertion, as in \cref{eq:patch_b}, with no additional boundaries. Evaluating \eqref{eq:patch_b} in the path integral, we obtain\footnote{Such contributions were absent in the AdS closed-universe case, since there is no negatively curved sphere saddle and we assumed $\overline{\mathcal{O}}=0$ in that context.}
\begin{align}
	\overline{\mathcal{O}} &= \adjincludegraphics[height=0.8in,valign=m]{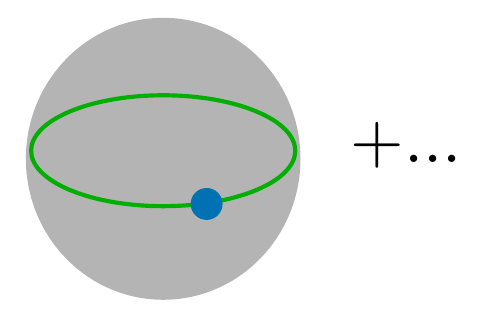} \, ,
	\label{eq:exp_HH_1}\\
	\overline{\mathcal{O}^2} &= \adjincludegraphics[height=0.8in,valign=m]{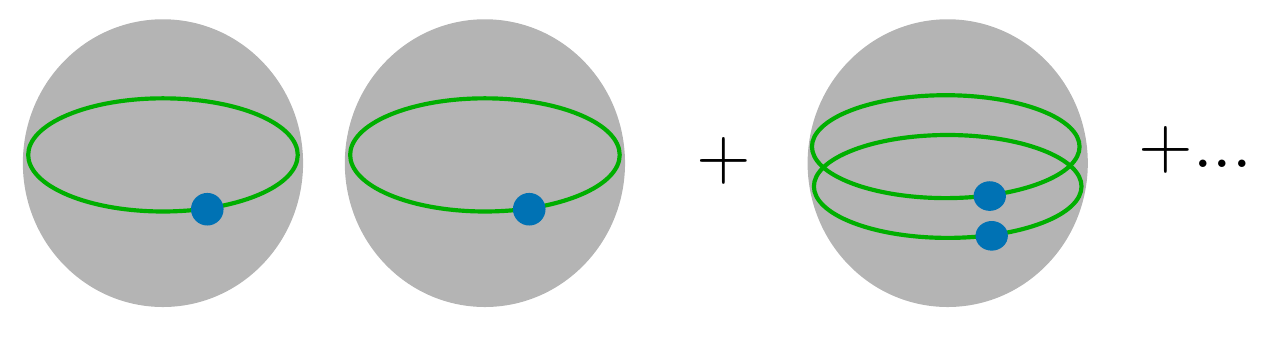} \, .
	\label{eq:exp_HH_2}
\end{align}

Notice that the second term in \cref{eq:exp_HH_2} is suppressed by $e^{-S_{\text{dS}}}$ relative to the first. Consequently, \cref{eq:exp_HH_2} agrees with the square of \cref{eq:exp_HH_1} up to corrections of order $e^{-S_{\text{dS}}}$.\footnote{In some theories there can also be higher-topology contributions, but they are similarly suppressed.} In this case, wormhole corrections are therefore highly suppressed. This observation was pointed out to us by Juan Maldacena.

\begin{figure}[!htbp] 
\centering                  
      \includegraphics[height=0.9in]{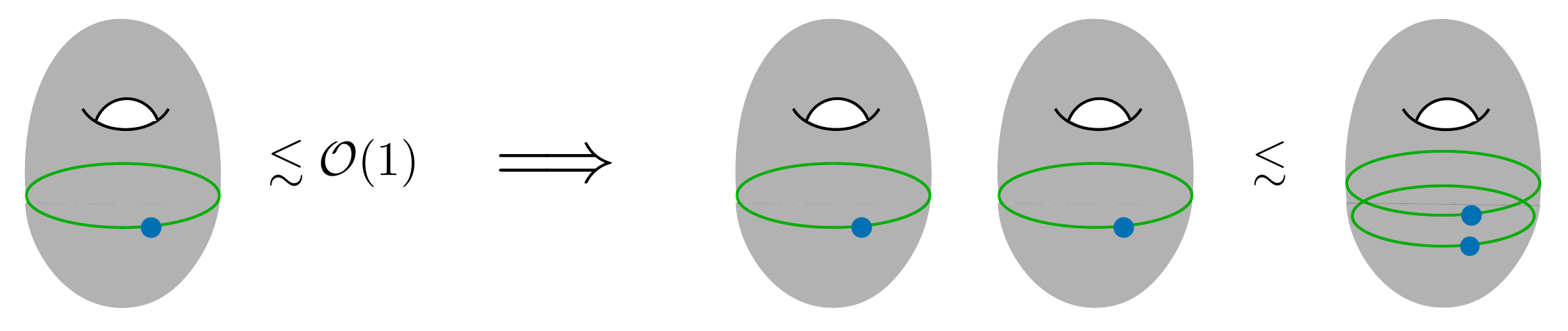}
  \caption{If the partition function is not large, wormhole contributions can produce large ensemble fluctuations.}
 \label{fig:fluctuation_torus}
\end{figure}

Let us pause to ask why this happens. The one-state conclusion for closed universes still holds.\footnote{The path-integral argument in \cite{Penington:2019kki} used de Sitter as an example.} As a result, patch operators with generic boundary conditions still exhibit large fluctuations. To suppress wormhole corrections, however, we need both conditions stated at the beginning of this section. Without Condition~\ref{feature:1}, patch operators can connect to different asymptotic boundaries, and the cross-connections discussed in \cref{sec:rev} can appear. As for Condition~\ref{feature:2}, if the dominant configuration has a small partition function (i.e.\ not parametrically large), then wormhole corrections are again at least order one (see \cref{fig:fluctuation_torus}).

This illustrates that the \emph{Hartle--Hawking no-boundary condition}, when dominated by \emph{sphere topology (Euclidean de Sitter)}, is special. Perhaps it is not an accident that we live in a universe with a positive cosmological constant. Note also that we are using ``de Sitter'' in a broad sense: it need not be eternal de Sitter space. For example, slow-roll inflationary models and Coleman--De~Luccia bubbles also have sphere topology and satisfy our conditions. All that is required is the absence of Euclidean asymptotic boundaries together with a parametrically large partition function.

\subsection{Emergence of quantum mechanics for a de Sitter bulk observer}
\label{sec:qm}

Our ultimate interest is the experience of a bulk observer living inside de Sitter space. In particular, the theory should reproduce standard quantum mechanics for such an observer to very high accuracy.

Starting from the Hartle--Hawking no-boundary condition, we consider a universe containing an observer and a spin. The question is whether the observer experiences ordinary quantum mechanics. To address this, the first step is to condition on a universe that contains both the observer and the spin. We introduce the following patch operator:
\begin{align}
	\mathcal{O}_{P,X} = \adjincludegraphics[width=0.8in,valign=m]{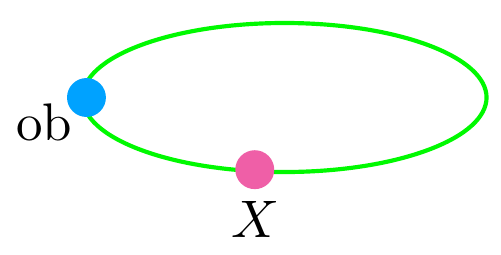}
	\label{eq:patch_b_2}
\end{align}
In \cref{eq:patch_b_2}, the green circle represents a spatial slice, the blue dot represents the observer, and the pink dot represents the spin. The operator $P$ conditions on the existence of the observer--spin system, while $X$ acts on that system. Although $\mathcal{O}_{P,X}$ depends only on the combination $PXP$, we keep $P$ and $X$ separate in the subscript to emphasize that the operator \(X\) is defined relative to the conditioning $P$.

For a fixed conditioning $P$, there can be many choices of $X$, forming an operator algebra $\mathcal{A}_P$. In later sections, when no confusion can arise, we will sometimes write $\mathcal{O}_{P,X}$ simply as $\mathcal{O}_{PXP}$.

The patch operator defined in \cref{eq:patch_b_2}, like that in \cref{eq:patch_b}, specifies the configuration on an entire closed spatial slice. As mentioned in \cref{sec:patch_0}, we will also consider patch operators localized to a finite spacetime region. Physically, no realistic observer can probe the entire universe; instead, she can access only a limited set of observables within the portion of the universe visible to her. To incorporate this, we modify the definition of the patch operator in \cref{eq:patch_b_2} as follows:
\begin{align}
	\adjincludegraphics[width=1.86in,valign=m]{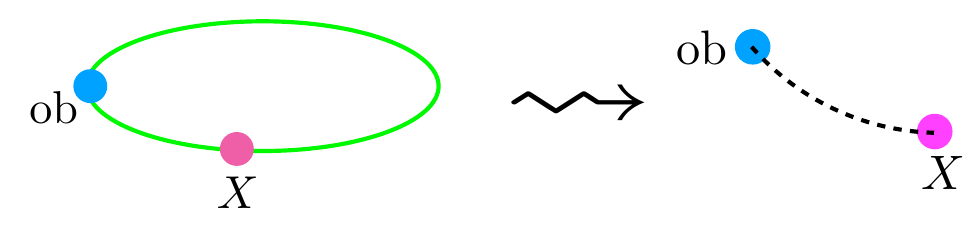} \;\equiv\; \mathcal{O}_{P,X}.
	\label{eq:patch_b_3}
\end{align}

In the revised definition \cref{eq:patch_b_3}, we condition only on the existence of the observer and the spin (and their connectivity). The operator $X$ acts on this observer--spin system, while the rest of the universe is left unspecified.

Mathematically, $PXP$ is an operator in the effective (bulk) description. When evaluating $\mathcal{O}_{P,X}$ in the path integral, we sum over all configurations with an insertion of the operator $PXP$:
\begin{align}
	\overline{\mathcal{O}_{P,X}} \equiv \overline{\mathcal{O}_{PXP}}
	\coloneq
	\sum_{\mathcal{M}}\int \mathcal{D}g\,\mathcal{D}\phi\;
	\qty(\int_{\mathcal{M}} \!dx\,\bigl(PXP\bigr)(x))\,
	e^{-S_E[g,\phi;\,\mathcal{M}]}\, .
	\label{eq:patch_def_path_integral}
\end{align}
In \cref{eq:patch_def_path_integral}, the integral and sum run over all metric and field configurations, including different topologies, and also over all possible locations at which $PXP$ is inserted. This is analogous to the treatment of vertex operators in worldsheet string theory.

Recall that such one-patch operators, when the path integral is dominated by sphere topology, generically exhibit small fluctuations and are self-averaging. We now consider the quantity
\begin{align}
	\expval{X}_{P} \coloneq \frac{\mathcal{O}_{P,X}}{\mathcal{O}_{P,I}}
	= \adjincludegraphics[width=0.9in,valign=m]{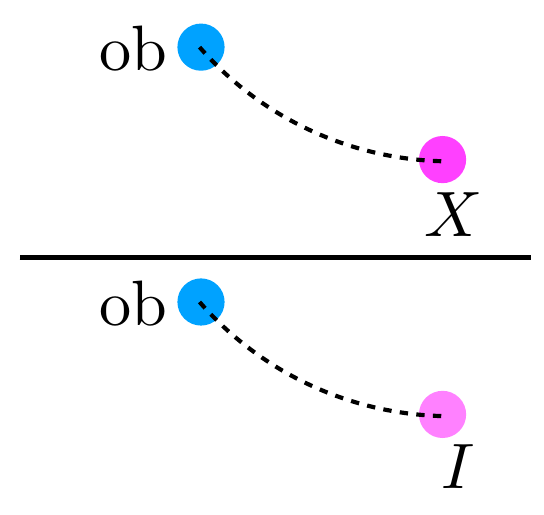}\, ,
	\label{eq:exp_X}
\end{align}
where the subscript $P$ on the left-hand side of \cref{eq:exp_X} emphasizes that $\expval{X}_P$ is defined with respect to the conditioning $P$. At this stage, $\expval{X}_P$ is simply a path-integral quantity defined by the right-hand side of \cref{eq:exp_X}. We will explain its physical interpretation and the choice of notation shortly.

A full evaluation of \cref{eq:exp_X} in the path integral requires the use of the replica trick. In \cref{sec:toy_model}, we will study the statistics of such quantities in detail using a toy model. Here we restrict ourselves to a qualitative discussion. Since both the numerator and denominator of \cref{eq:exp_X} are self-averaging, we can evaluate them separately before taking the ratio and obtain a good approximation:
\begin{align}
	\overline{\expval{X}_{P}} = \adjincludegraphics[width=2in,valign=m]{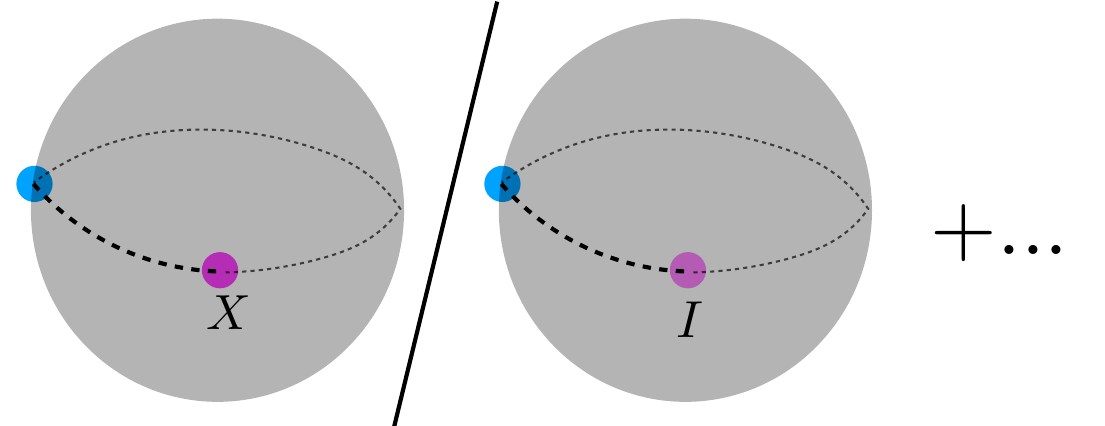}
	\label{eq:exp_X_evaluation}
\end{align}
The first term on the right-hand side of \cref{eq:exp_X_evaluation} gives the expectation value of the operator $X$ acting on the observer--spin system in a de Sitter background. This is the intended meaning of the notation $\expval{\cdot}_P$. The ellipsis ``$\ldots$'' denotes non-perturbative corrections suppressed by $e^{-S_{\text{dS}}}$. 

Finally, note that here we use standard angle brackets rather than the square brackets $|\HH]$ used for the vector corresponding to the Hartle--Hawking no-boundary condition. The reason is that $X$ is an operator acting on a quantum-mechanical system, and $\expval{\cdot}_P$ denotes an ordinary quantum operator expectation value.

As a sanity check, we examine the quantum fluctuations of the operator $X$. The relevant patch operators are
\begin{equation}
\begin{aligned}
\expval{X^2}_{P}
 & = \frac{\mathcal{O}_{P,X^2}}{\mathcal{O}_{P,I}}
   = \adjincludegraphics[width=0.9in,valign=m]{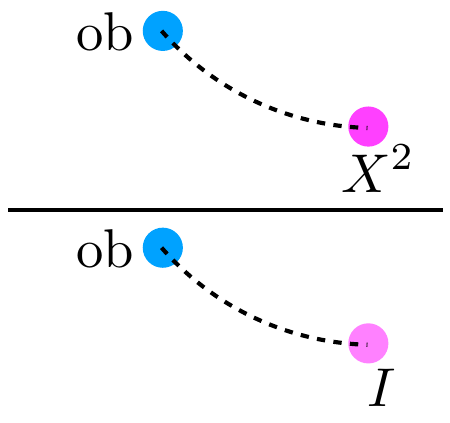},
\\
\expval{X}_{P}^{\,2}
  &= \qty(\frac{\mathcal{O}_{P,X}}{\mathcal{O}_{P,I}})^2
   = \adjincludegraphics[width=0.9in,valign=m]{exp_X_2.pdf}\,
     \adjincludegraphics[width=0.9in,valign=m]{exp_X_2.pdf}.
\label{eq:exp_X_square}
\end{aligned}
\end{equation}

As before, we evaluate the numerators and denominators separately, obtaining
\begin{equation}
\begin{aligned}
\overline{\expval{X^2}_{P}} &= \adjincludegraphics[width=2in,valign=m]{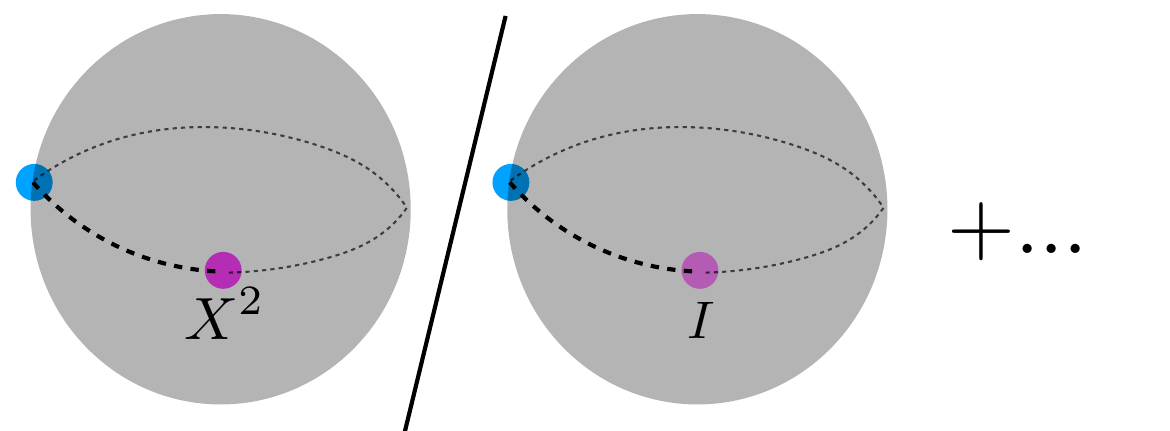},\\
\overline{\expval{X}_{P}^{\,2}} &= \adjincludegraphics[width=2in,valign=m]{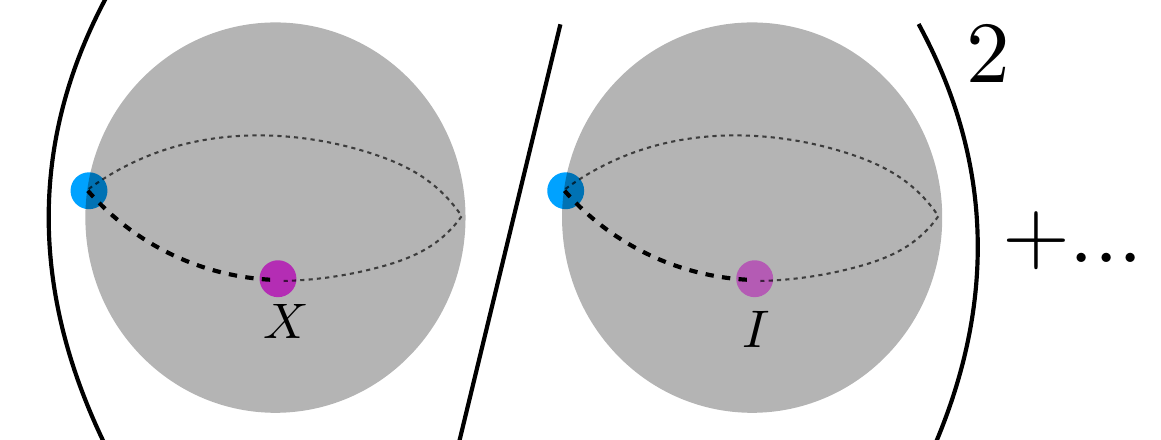}.
\end{aligned}
\label{eq:expX2_and_expX_sq}
\end{equation}

From \cref{eq:expX2_and_expX_sq}, we see that the quantum fluctuations of the operator $X$ computed in the path integral agree with the effective field theory prediction up to corrections of order $e^{-S_{\text{dS}}}$.

A key feature of our construction is that, when computing expectation values, we only consider one-point functions (equivalently, one-patch operators, i.e.\ operators supported on a single patch). In the definition of $\expval{X}_P$ in \cref{eq:exp_X}, the operator $X$ may be arbitrarily complicated as an operator on the observer--spin system, but it is still supported on a single patch. For example, $X = X_1 \cdots  X_n$ may be a product of any number of unitaries and/or projections corresponding to multiple experiments performed by the observer. These unitaries and projections are packaged into a single patch operator,
\begin{equation}
\mathcal{O}_{PXP} = \mathcal{O}_{P X_1 \cdots  X_n P}.
\label{eq:one_patch_only}
\end{equation}

Another important feature is the following. We use the term ``patch operator'' to emphasize that the operator need not be supported on an entire spatial slice, and that $PXP$ also need not be localized in either space or time. For the particular patch operators shown in \cref{eq:patch_b_2,eq:patch_b_3}, the observer and the spin lie on a single spatial slice, but more general configurations are allowed. For example, to describe an observer performing two measurements on the spin separated by a time interval $\Delta t$, one may consider two copies of~\eqref{eq:patch_b_3} separated by a timelike geodesic of length $\Delta t$. As discussed in \cref{sec:qm_ob}, the operators $X$ may be arbitrary elements of an operator algebra, including products of an arbitrary number of operators, and need not be local in general.

Once we have all correlation functions $\expval{\cdot}_P$ on the observer--spin system, a Hilbert space can be constructed via the GNS construction. In this sense, quantum mechanics emerges. As an example, consider the transition amplitude of the spin from the state $\ket{+\hat z}$ (the $+1$ eigenstate of the Pauli operator $\sigma_z$) to the state $\ket{+\hat x}$. It is given by
\begin{align}
 	\frac{\expval{\Pi_{+\hat x}\,\Pi_{+\hat z}}_{P}}{\expval{\ket{+\hat x}\mkern-5mu\bra{+\hat z}}_{P}}
 	\stackrel{?}{=}\bra{+\hat x}\ket{+\hat z}\, ,
 	\label{eq:transition_amplitude}
\end{align}
where $\Pi_{+\hat z}$ and $\Pi_{+\hat x}$ are the projection operators onto the spin states $\ket{+\hat z}$ and $\ket{+\hat x}$, respectively.\footnote{There is an implicit identity operator acting on the observer, which we do not write explicitly.}

Note that in \cref{eq:transition_amplitude}, the left-hand side is a path-integral quantity encoding the bulk observer's physics, while the right-hand side is the standard quantum-mechanical answer. In what sense does the left-hand side reproduce the right-hand side, and with what accuracy? We have already seen that the path-integral evaluation of the left-hand side is sharply peaked around the value of the right-hand side, with deviations suppressed exponentially as $e^{-S_{\text{dS}}}$. We will return to this question in \cref{sec:toy_model}, where we will see that the equality holds exactly within each $\alpha$-sector.

\subsection{``It from Bit'': Part I --- Patch operators and quantum mechanics}
\label{sec:baby_1}

It is sometimes suggested that, since quantum mechanics in a one-dimensional Hilbert space is trivial (with no noncommuting operators), there can be no nontrivial physics unless one mixes different $\alpha$-sectors. How does our construction circumvent this issue? The key point is to distinguish two conceptually different aspects:
\begin{enumerate}[label=(\Roman*)]
	\item ``Bit'': no quantum mechanicsi. for closed universes as viewed from the \emph{outside}.
	\label{aspect:1}
	\item ``It'': quantum mechanics \emph{inside} de Sitter space, i.e.\ the quantum mechanics experienced by a bulk observer.
	\label{aspect:2}
\end{enumerate}

The one-state property of closed universes is fundamentally a statement about Aspect~\ref{aspect:1}. What is relevant for a bulk observer's experience, however, is Aspect~\ref{aspect:2}. Throughout this paper, we use Dirac notation $\ket{\cdot}$ for standard quantum mechanics associated with Aspect~\ref{aspect:2}, while square-bracket notation $\lvert \cdot ]$ is reserved for Aspect~\ref{aspect:1}, to emphasize that no genuine quantum mechanics is involved there. We will comment on general features of both aspects. Some of the discussion in this section may appear abstract, but its meaning should become clearer when we return to it in \cref{sec:baby_more} after analyzing a concrete example.

Mathematically, patch operators provide the bridge between these two aspects. In the gravitational path integral, a patch operator defined as in \cref{eq:patch_b_3,eq:patch_def_path_integral} is simply a complex-valued random variable. Despite the terminology, it does not act as an operator on a quantum-mechanical Hilbert space; rather, it takes values in $\mathbb{C}$.\footnote{We will later see that it acts on the baby-universe Hilbert space, which has a commutative operator algebra and hence no nontrivial quantum mechanics.} This feature reflects the one-state property of closed universes (Aspect~\ref{aspect:1}). At the same time, patch operators encode the physics relevant to Aspect~\ref{aspect:2}.

\subsubsection{Patch operators and bulk observer's quantum mechanics}
\label{sec:qm_ob}

Recall that $X$, as an element of the operator algebra $\mathcal{A}_P$, is a physical operator acting on the observer--spin system. In the definition of the patch operator $\mathcal{O}_{P,X}$ in \cref{eq:patch_b_3,eq:patch_def_path_integral}, the operator $X$ appears as a label of the random variable $\mathcal{O}_{P,X}$.

A physically useful way to phrase this is as follows. Let $\Omega$ denote the space of $\alpha$-parameters. We can collect all patch operators with the same conditioning $P$ into a map from the operator algebra $\mathcal{A}_P$ to the space of complex-valued random variables $M(\Omega,\mathbb{C})$:
\begin{equation}
\begin{aligned}
	\mathcal{M}_{P}:\ \mathcal{A}_{P} &\longrightarrow M(\Omega,\mathbb{C}) ,\\
	X &\longmapsto \mathcal{O}_{P,X} .
	\label{eq:map}
\end{aligned}
\end{equation}
Restricting to a fixed $\alpha$, the map in \cref{eq:map} reduces to
\begin{equation}
\begin{aligned}
	(\mathcal{M}_{P})_{\alpha}:\ \mathcal{A}_{P} &\longrightarrow \mathbb{C} ,\\
	X &\longmapsto (\mathcal{O}_{P,X})_{\alpha} .
	\label{eq:map_alpha}
\end{aligned}
\end{equation}

As we will see shortly, these maps are linear in $X$. As linear functionals on the operator algebra $\mathcal{A}_P$ satisfying appropriate properties, $(\mathcal{M}_P)_\alpha$ in \cref{eq:map_alpha} specifies a quantum state, while \cref{eq:map} describes a classical ensemble of such states with a probability distribution.

In a simple example studied in \cref{sec:toy_model}, we will see explicitly that the collection of maps in \cref{eq:map,eq:map_alpha}, for all possible conditionings $P$, completely characterizes the physics of de Sitter space. In fact, when the operator $PXP$ is localized on a single spatial slice, the corresponding patch operator $\mathcal{O}_{P,X}$ is closely related to the no-boundary density matrix introduced in \cite{Ivo:2024ill}. In the remainder of this section, we discuss some general properties of the family of maps $\{\mathcal{M}_P\}$.

\begin{itemize}
	\item \textbf{Patch operators are linear maps}

Let $\overline{\mathcal{O}_{P,X_1}\ldots\mathcal{O}_{P,X_k}}$ denote the result of the path-integral evaluation of the patch operators $\mathcal{O}_{P,X_1},\ldots,\mathcal{O}_{P,X_k}$ under the Hartle--Hawking no-boundary condition.\footnote{By this we mean that there are no other boundary conditions other than these $k$ patch operators.} Since these patch operators serve as boundary conditions for the path integral, their ordering is irrelevant.

From the definition of patch operators in \cref{eq:patch_def_path_integral}, evaluating $\overline{\mathcal{O}_{P,X_1}\ldots\mathcal{O}_{P,X_k}}$ involves summing over all configurations with insertions of the operators $P X_i P$ at arbitrary locations. The contribution of each configuration is linear in each $X_i$. As a result, the path-integral value of $\overline{\mathcal{O}_{P,X_1}\ldots\mathcal{O}_{P,X_k}}$ is linear in $X_i$ for every $i$.

In particular, replacing the first patch operator $\mathcal{O}_{P,X_1}$ by $\mathcal{O}_{P,aY_1+bY_2}$, we find
\begin{align}
	\overline{\mathcal{O}_{P,aY_1+bY_2}\ldots\mathcal{O}_{P,X_k}}
	= a\,\overline{\mathcal{O}_{P,Y_1}\ldots\mathcal{O}_{P,X_k}}
	+ b\,\overline{\mathcal{O}_{P,Y_2}\ldots\mathcal{O}_{P,X_k}} \, .
	\label{eq:linear_1}
\end{align}

On the other hand, by the definition of $\alpha$-sectors, a patch operator $\mathcal{O}_{P,X}$ takes a fixed value $(\mathcal{O}_{P,X})_{\alpha}$ in each $\alpha$-sector. The gravitational path-integral evaluation of $\mathcal{O}_{P,X_1}\cdots  \mathcal{O}_{P,X_k}$ then computes the ensemble average of the product $(\mathcal{O}_{P,X_1})_{\alpha}\cdots  (\mathcal{O}_{P,X_k})_{\alpha}$:
\begin{align}
	\overline{\mathcal{O}_{P,X_1}\ldots\mathcal{O}_{P,X_k}}
	= \int d\alpha\, p_{\alpha}\,
	(\mathcal{O}_{P,X_1})_{\alpha}\cdots (\mathcal{O}_{P,X_k})_{\alpha} .
	\label{eq:eq_1}
\end{align}
As a result,
\begin{equation}
	a\,\overline{\mathcal{O}_{P,Y_1}\ldots\mathcal{O}_{P,X_k}}
	+ b\,\overline{\mathcal{O}_{P,Y_2}\ldots\mathcal{O}_{P,X_k}}
	= \int d\alpha\, p_{\alpha}\,
	\bigl[a(\mathcal{O}_{P,Y_1})_{\alpha}
	+ b(\mathcal{O}_{P,Y_2})_{\alpha}\bigr]
	\cdots (\mathcal{O}_{P,X_k})_{\alpha} .
	\label{eq:linear_2}
\end{equation}

Comparing \cref{eq:linear_1,eq:linear_2}, we conclude that
\begin{align}
	(\mathcal{O}_{P,aY_1+bY_2})_{\alpha}
	= a\,(\mathcal{O}_{P,Y_1})_{\alpha}
	+ b\,(\mathcal{O}_{P,Y_2})_{\alpha} ,
\end{align}
that is, viewed as a random variable, $\mathcal{O}_{P,X}$ is linear in $X$. We will make essential use of this linearity later.

\item \textbf{Patch operators encode algebraic properties of the observer's operator algebra}

The next question is: beyond linearity, how are the nontrivial algebraic properties of $\mathcal{A}_P$ encoded in the family of patch operators $\{\mathcal{O}_{P,X}\mid X\in\mathcal{A}_P\}$?

The operator algebra $\mathcal{A}_P$ can of course contain noncommuting operators, say $X$ and $Y$. By contrast, there is nothing noncommutative about the corresponding complex numbers $\mathcal{O}_{P,X}$ and $\mathcal{O}_{P,Y}$. Instead, noncommutativity is reflected in correlation functions. More precisely, there exist operators $A,B\in\mathcal{A}_P$ and an $\alpha$-sector such that
\begin{align}
	(\mathcal{O}_{P,AXYB})_\alpha \neq (\mathcal{O}_{P,AYXB})_\alpha .
\end{align}

More generally, the algebraic structure of $\mathcal{A}_P$ is captured by the maps $\mathcal{M}_P$. For example, if $\mathcal{A}_P$ satisfies an algebraic relation $[X,Y]=Z$, then this is encoded as the equality of random variables
\begin{align}
	\mathcal{O}_{P,A[X,Y]B} = \mathcal{O}_{P,AZB}
\end{align}
for all $A,B\in\mathcal{A}_P$.

In fact, up to the $X$-independent normalization factor $\mathcal{O}_{P,I}$, the map $\mathcal{M}_P$ in \cref{eq:map,eq:map_alpha} computes the expectation values of all operators $X\in\mathcal{A}_P$.

\end{itemize}

\subsubsection{No quantum mechanics on the baby-universe Hilbert space}

The quantum mechanics discussed above is the quantum mechanics experienced by a bulk observer \emph{inside} de Sitter space. It is not quantum mechanics on the space of all closed universes, or the baby-universe Hilbert space.\footnote{Sometimes it's also called the third quantized Hilbert space. This is not an appropriate name, as there is no quantum mechanics on this Hilbert space.}

Let us first ask: do we have, or even need, a baby-universe quantum mechanics? To frame the issue, consider an extreme version of the Boltzmann brain problem. Specifically, consider the following ``transition amplitude'' between two different closed-universe spatial slices.

\begin{figure}[!htbp] 
\centering                     
      \includegraphics[width=0.5\linewidth]{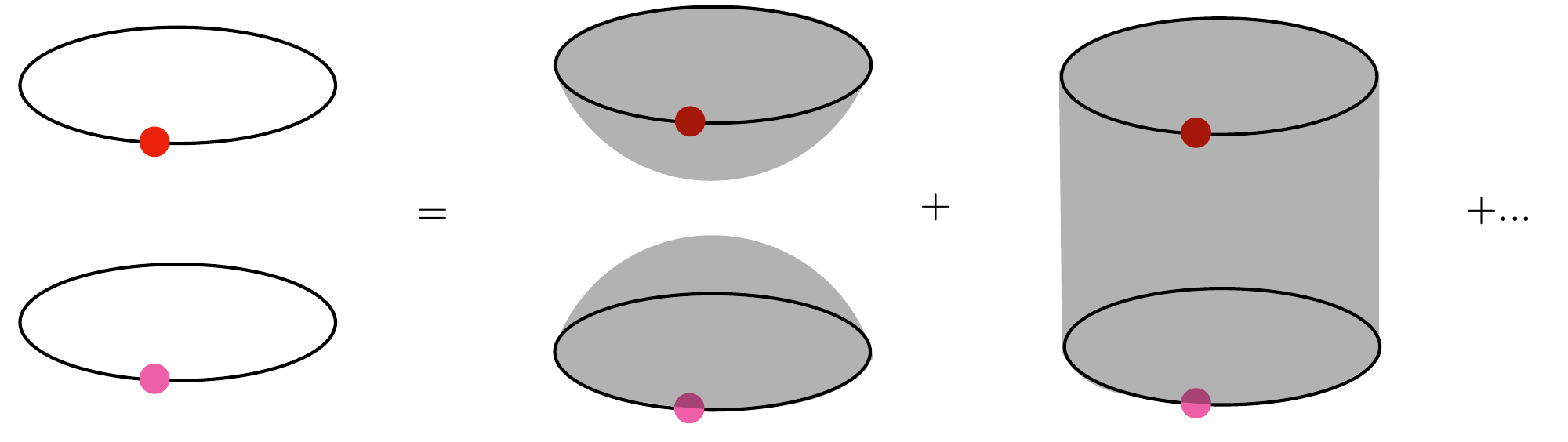}
  \caption{The ``transition amplitude'' between two entire closed-universe slices appears to be dominated by a process in which the universe disappears and then reappears.}
 \label{fig:transition_Boltzmann}
\end{figure}

Figure~\ref{fig:transition_Boltzmann} shows the path-integral evaluation of such a quantity. When the one-boundary amplitudes are large, the dominant contribution comes from the first term on the right-hand side, which naively suggests that the entire universe can disappear and then reappear as a fluctuation. Mathematically, if we normalize the ``states'' in \cref{fig:transition_Boltzmann}, any such ``transition amplitude'' has modulus $1$, which is a direct consequence of having only one state in each $\alpha$-sector.
Alternatively, one can perform a different computation in which cross-connections are forbidden when evaluating the square of such a ``transition amplitude''; the resulting modulus can then differ from $1$, but it remains exponentially close to $1$ when the one-boundary amplitudes are exponentially large.

Our point is that, while these different prescriptions compute different quantities characterizing the statistical properties of baby universes, none of them is directly relevant to the experience of de Sitter bulk observers. They are quantities associated with Aspect~\ref{aspect:1}, whereas a de Sitter bulk observer's experience concerns Aspect~\ref{aspect:2}. In particular, these are \emph{not} transition amplitudes measured or observed by anyone living \emph{inside} the universe (hence the quotation marks around ``transition amplitude'' above).

Let us make a few philosophical remarks. We usually speak of transition amplitudes only when we have nontrivial quantum mechanics, and nontrivial quantum mechanics is operationally tied to the presence of an observer \cite{Bohr1949,LandauLifshitzQM}. From this perspective, there is no reason for an entire closed universe to constitute a quantum-mechanical system unless there exists a \emph{super-observer}. By a super-observer we mean an agent that remains outside any closed universe and can, in principle, watch different closed universes disappear and reappear. In other words, this extreme version of the Boltzmann brain problem is a problem only for such super-observers attempting to do nontrivial quantum mechanics with different closed universes, and it has nothing to do with a bulk observer's experience. Indeed, one way to phrase the one-state property is that the entire closed universe, viewed as a quantum-mechanical system from the outside, is trivial --- equivalently, there is no super-observer \cite{Harlow:2025pvj}.

It is useful to contrast this with worldsheet string theory. There, too, one sums over all topologies, but one never encounters an analogous one-state property. One obvious reason is that the rules of computation differ. In worldsheet string theory, when we compute any quantity such as a string scattering amplitude, we sum over all worldsheet geometries connecting the vertex operators. When computing the square of that amplitude, however, we simply take the square of the resulting number. Unlike in the gravitational path integral, there is no additional ``wormhole correction.'' This is the mathematical difference. 

The physical difference is that in worldsheet string theory there \emph{are} super-observers -- observers living in the target space, and the goal is precisely to describe their quantum mechanics. In the closed-universe context, without analogs of target space or target-space observers, there is no quantum mechanics of different closed universes as seen from the outside. As we will see in \cref{sec:baby_more}, what is physically important instead are the classical statistical properties of various quantities associated with baby universes, which can be formalized using the language of the baby-universe Hilbert space.

\subsubsection{Patch operators as diagonal operators on the baby-universe Hilbert space}

We now return to the role played by patch operators in connection with Aspect~\ref{aspect:1}. One may consider the space of all closed universes that can be prepared by gravitational path integral. As shown in \cite{Marolf:2020xie} and discussed in more detail in \cref{sec:baby_more}, this space admits a Hilbert-space structure and is, in an appropriate sense, spanned by the $\alpha$-sectors.

Patch operators can be regarded as operators acting on this baby-universe Hilbert space, but only in a highly restricted manner: they act by multiplication by a c-number within each $\alpha$-sector. As a result, all patch operators commute with one another. The underlying physical reason is again the one-state property of closed universes and the consequent absence of $\alpha$-mixing.

From this perspective, the map in \cref{eq:map} can be viewed as a map from $\mathcal{A}_P$ to $\mathcal{A}_{\mathrm{BU}}$, the commutative operator algebra of the baby-universe Hilbert space. Although patch operators do not generate nontrivial quantum mechanics on the baby-universe Hilbert space, this map nevertheless encodes the nontrivial quantum mechanics experienced by a bulk observer living inside a closed universe.

 \subsection{Explicit realization via quantum codes}
\label{sec:code}

In \cite{Akers:2022qdl} it was shown that quantum codes can reproduce gravitational path integral results to high accuracy. From this perspective, many of the features discussed in the previous sections admit an explicit realization in terms of quantum codes. In particular, one convenient way to represent the classical statistics we have been discussing is via a quantum code that computes patch operators and the corresponding expectation values of quantum operators.

\begin{equation}
	\mathcal{O}_{P,X}
	= \adjincludegraphics[width=0.66in,valign=m]{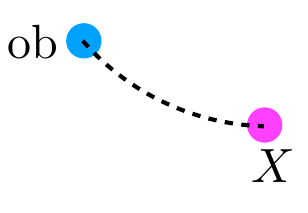}
	= \adjincludegraphics[height=2.2in,valign=m]{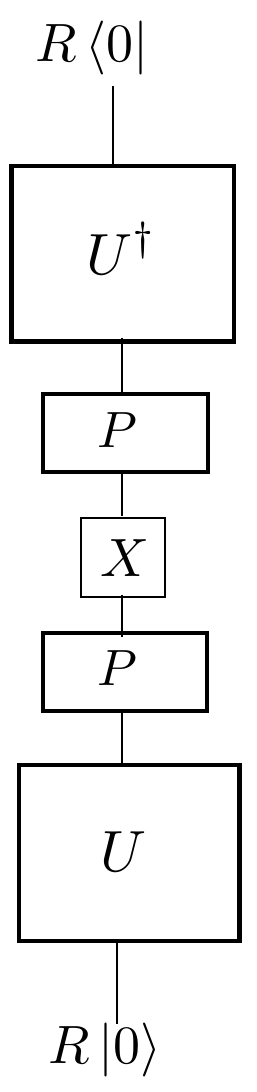}
	\qquad
	\expval{X}_P
	= \adjincludegraphics[width=0.76in,valign=m]{exp_X_2.pdf}
	= \adjincludegraphics[height=2.2in,valign=m]{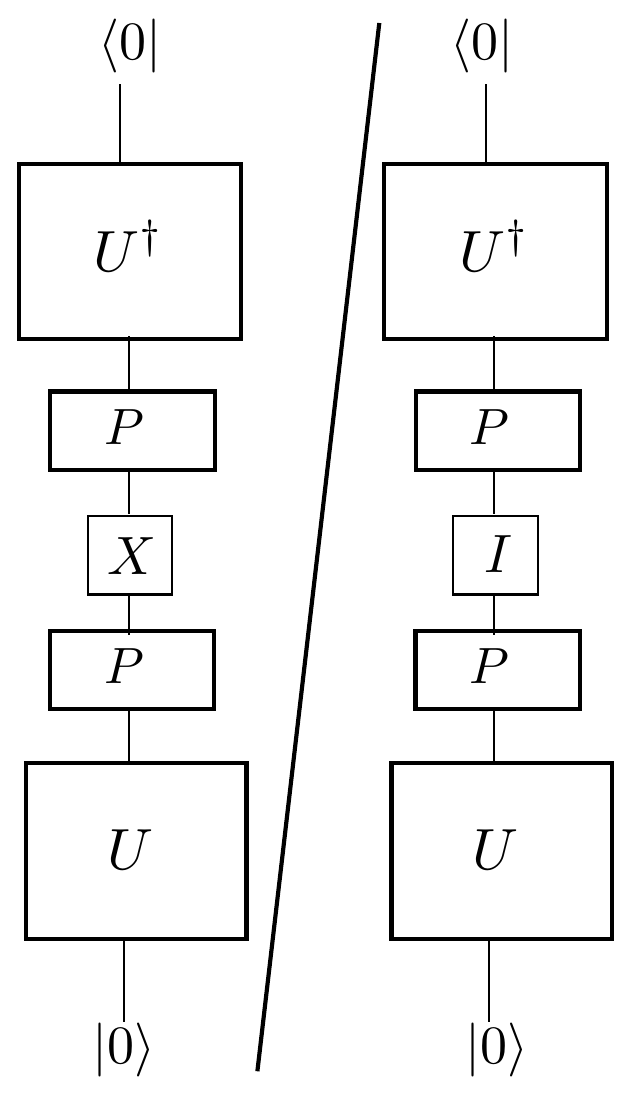}
	\label{eq:code}
\end{equation}

In \cref{eq:code}, $U \sim \text{Haar}\big(SU(d)\big)$ with $d = e^{S_{\mathrm{dS}}}$. The state $\ket{0}$ is a normalized vector in a Hilbert space of dimension $d$. The variable $R$ is a positive random scalar with distribution $R \sim \frac{1}{\sqrt{2}}\chi_{2d}$, equivalently $R^2 \sim \text{Gamma}(d,1)$. More explicitly,
\begin{align}
	p_{\scriptscriptstyle R}(r) = \frac{2}{\Gamma(d)}\, r^{2d-1} e^{-r^2}.
\end{align}

From the code in \cref{eq:code}, it is straightforward to see that fluctuations of $\mathcal{O}_{P,X}$ are suppressed by $1/\operatorname{rank} P$. This suppression factor corresponds to the de Sitter partition function with conditioning $P$, as we will show in \cref{sec:conditioning}.

As discussed in \cref{sec:qm_ob}, patch operators $\mathcal{O}_{P,X}$ are random variables. In this code, the randomness arises from the choice of both $U$ and $R$. The space of $\alpha$-parameters $\Omega$ is the space of unnormalized states $U R \ket{0}$:
\begin{align}
	\Omega
	= \Big\{ R, \{U_{a0}\}
	\;\Big|\;
	R>0,\;
	\sum_a |U_{a0}|^2 = 1,\;
	U_{a0} \sim e^{i\theta} U_{a0}
	\Big\}.
\end{align}

The detailed structure of the quantum code is model dependent. The construction in \cref{eq:code} is the appropriate code for the toy model studied in \cref{sec:toy_model}. Note that we use Haar-random matrices from the unitary group rather than the orthogonal group. The reason is that introducing an observer carrying a clock explicitly breaks $\mathcal{CRT}$ \cite{Harlow:2023hjb,Susskind:2026fjc}. In \cref{sec:code_real}, we will rewrite the code in terms of $\mathcal{CRT}$-invariant observables using orthogonal matrices $O$, for physical observables $X$ that are Hermitian.

\section{A toy model of de Sitter spacetime}
\label{sec:toy_model}

In this section, we study a one-dimensional topological model of de Sitter space in detail, using the framework outlined in \cref{sec:dS}. This model exhibits several instructive features. In particular, a parameter $Z_L$ associated with the loop partition function in this simple model (the analog of the sphere partition function $e^{S_{\mathrm{dS}}}$ in higher-dimensional de Sitter theories) determines, when it takes integer values, the dimension of the largest quantum-mechanical system that can exist in this toy de Sitter space. With more general conditionings, the de Sitter entropy is identified with the coarse-grained entropy of the underlying state.

This section is somewhat lengthy and is organized as follows. In \cref{sec:one_bit,sec:set_up} we motivate and define the setup of the model. In \cref{sec:subsystems} we study the properties of various systems that can exist in this de Sitter space. We will see that, depending on the value of $Z_L$, some systems are allowed while others are not. The underlying reason becomes clear in \cref{sec:QM_U}, where we present a unified description of the physics inside this de Sitter space. In \cref{sec:CRT} we give a manifestly $\mathcal{CRT}$-invariant formulation. Finally, in \cref{sec:baby_more}, armed with a concrete example, we return to the relationship and distinction between the physics inside de Sitter space and the baby-universe Hilbert space, briefly discussed in \cref{sec:baby_1}.

Throughout \cref{sec:one_bit,sec:set_up,sec:subsystems,sec:QM_U,sec:CRT}, we work under Hartle--Hawking no-boundary condition. More general boundary conditions are discussed only in \cref{sec:baby_more}.

\subsection{Warm-up: A classical two-state system}
\label{sec:one_bit}

In one spacetime dimension, Euclidean de Sitter space is a circle. Since our model is topological, the length of the circle plays no role. On the other hand, the number of circles is dynamical, and wormhole fluctuations correspond to the splitting and joining of circles. Each circle contributes a factor of $Z_L$ to the path integral.

As a warm-up, we first consider a classical two-state system in this simple de Sitter space. This classical system is not physical. Studying this unphysical system serves two purposes. Mathematically, it provides a warm-up for more involved calculations. Conceptually, it is useful for making contact with later discussions of $\mathcal{CRT}$-invariant fomulation.

The two states of this classical system are labeled by $\ket{-\hat y}_0$ and $\ket{+\hat y}_0$, which represent the direction of the ambient coordinate time.\footnote{The choice of notation may look unusual, but it is designed to match the $\mathcal{CRT}$-invariant description in \cref{sec:CRT}. It is merely a label and carries no physical meaning. All quantities associated with this unphysical degree of freedom carry a subscript $0$ to distinguish them from the physical spins discussed later in \cref{sec:spin}.} It should be clear why this classical system cannot be physical: the direction of de Sitter coordinate time has no invariant meaning. As emphasized in \cite{Harlow:2023hjb}, $\mathcal{CRT}$ is a gauge symmetry in quantum gravity. We introduce this classical system here to facilitate the discussion in \cref{sec:CRT}, where we first enlarge the Hilbert space before imposing gauge invariance.

The commutative operator algebra $\mathcal{A}_0$ of a two-state classical system is spanned by $\mathbb{Z}_2 = \{(I_2)_0, (\sigma_y)_0\}$, where $\sigma_y$ is the Pauli matrix. Two important operators are
\begin{align}
(P_+)_0 = \frac{(I_2)_0 + (\sigma_y)_0}{2} = \ket{+\hat y}_{\!\scriptscriptstyle 0}\mkern-6mu\bra{+\hat y}
\quad\text{and}\quad
(P_-)_0 = \frac{(I_2)_0 - (\sigma_y)_0}{2} = \ket{-\hat y}_{\!\scriptscriptstyle 0}\mkern-6mu\bra{-\hat y} ,
\end{align}
where $(P_+)_0$ projects onto one coordinate-time direction and $(P_-)_0$ projects onto the other.

To analyze this system mathematically, we consider the patch operators $\mathcal{O}_{\scriptscriptstyle(I_2)_{\scriptscriptstyle 0}}$ and $\mathcal{O}_{\scriptscriptstyle(\sigma_y)_{\scriptscriptstyle 0}}$ and study their statistics.\footnote{Since there is no conditioning in this example, there is no $P$ label.} For example, consider evaluating $\overline{\mathcal{O}_{\scriptscriptstyle(I_2)_{\scriptscriptstyle 0}}^m\,\mathcal{O}_{\scriptscriptstyle(\sigma_y)_{\scriptscriptstyle 0}}^{\,n}}$ in the path integral. The boundary condition consists of $m$ insertions of $\mathcal{O}_{\scriptscriptstyle(I_2)_{\scriptscriptstyle 0}}$ (shown as green dots in \cref{eq:rule_PI_0}) and $n$ insertions of $\mathcal{O}_{\scriptscriptstyle(\sigma_y)_{\scriptscriptstyle 0}}$ (red dots in \cref{eq:rule_PI_0}). As boundary conditions of the path integral, all dots are treated as distinct.

In evaluating $\overline{\mathcal{O}_{\scriptscriptstyle(I_2)_{\scriptscriptstyle 0}}^m\,\mathcal{O}_{\scriptscriptstyle(\sigma_y)_{\scriptscriptstyle 0}}^{\,n}}$, we distribute these $m+n$ patch operators among an arbitrary number of circles. A given circle contributes nontrivially only if it contains an even number of $\mathcal{O}_{(\sigma_y)_{\scriptscriptstyle 0}}$ insertions. Circles without any insertions do not contribute, as they correspond to vacuum bubbles associated with the normalization of the underlying probability distribution on the space of $\alpha$-parameters.\footnote{Specifically, the sum of all vacuum bubbles computes $\int d\alpha\, \tilde p_{\alpha}$ where $\tilde p_{\alpha}$ is some unnormalized probability distribution. Once we normalize it and assume $\int d\alpha \,p_{\alpha} = 1$, we no longer need to include vacuum bubbles.}

\begin{equation}
	\adjincludegraphics[width=1.4in,valign=m]{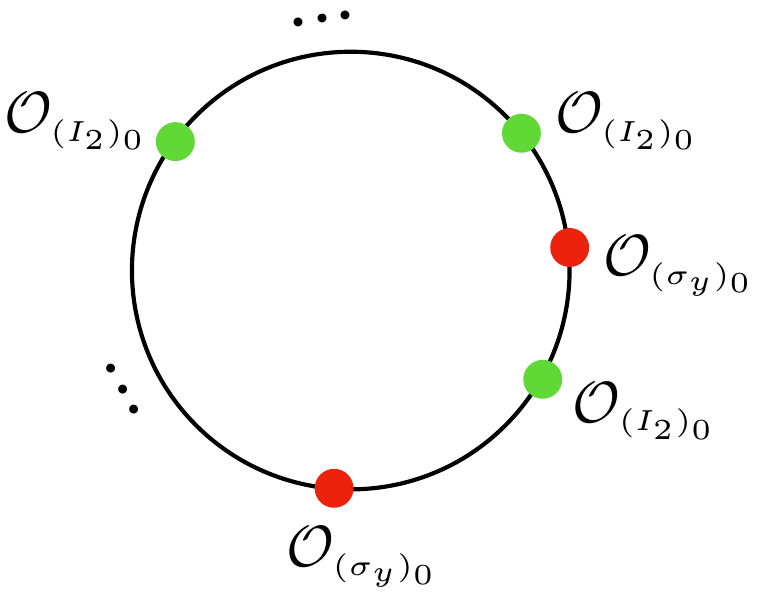} = \begin{cases}
		Z_L & \text{Even number of $\mathcal{O}_{\scriptscriptstyle (\sigma_y)_0}$ insertions}\\
		0   &  \text{Otherwise}
	\end{cases}
	\label{eq:rule_PI_0}
\end{equation}

In fact, a more convenient basis for solving this two-state classical system is $(P_{+})_0$ and $(P_{-})_0$, since one can easily verify that the corresponding patch operators $\mathcal{O}_{\scriptscriptstyle(P_{+})_0}$ and $\mathcal{O}_{\scriptscriptstyle(P_{-})_0}$ are independent random variables. Nevertheless, to facilitate comparison with later calculations, we will work in the $(I_2)_{ 0}, (\sigma_y)_0$ basis in this section and determine the statistics of $\mathcal{O}_{\scriptscriptstyle(I_2)_{\scriptscriptstyle 0}}$ and $\mathcal{O}_{\scriptscriptstyle(\sigma_y)_{\scriptscriptstyle 0}}$ directly by brute-force combinatorics. Since these properties are unphysical, we will not present the results here; the details are given in Appendix~\ref{app:classical}.

As emphasized at the beginning of this section, $\mathcal{CRT}$ is a gauge symmetry in quantum gravity, and the ambient time direction has no physical meaning \cite{Harlow:2023hjb}. In the remainder of the discussion, we fix this gauge by choosing a preferred ambient time direction labeled by $\ket{-\hat y}_0$ \cite{Susskind:2026fjc}. Once this choice is made, it also defines the forward time direction for any quantum-mechanical system (including observers) inside this de Sitter space.

\subsection{More general setup of the model}
\label{sec:set_up}

We now study the quantum mechanics \emph{inside} this de Sitter space. As a consequence of the $\mathcal{CRT}$ gauge fixing discussed earlier, the Euclidean circle acquires an orientation. Although we will not always refer to observers explicitly in this section, it should be understood that the quantum mechanics discussed here is the quantum mechanics experienced by a de Sitter bulk observer who is part of the system under consideration.

\begin{equation}
	\adjincludegraphics[width=1.2in,valign=m]{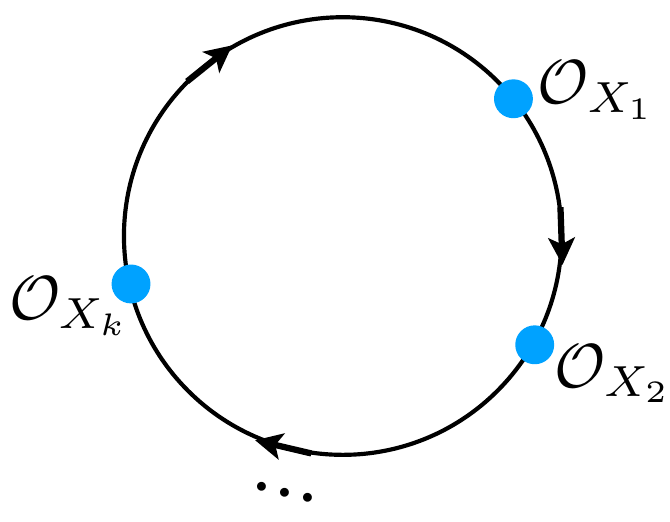}
	= \tr(X_1 X_2 \cdots  X_k)\, Z_L
	\label{eq:rule_PI}
\end{equation}

\cref{eq:rule_PI} illustrates a contribution to the path integral with several patch operators inserted. The blue dots denote patch operators $\mathcal{O}_{X_i}$, where each $X_i$ is an operator in the effective description of the theory.\footnote{$X_i$ plays the role analogous to $PXP$ in \cref{sec:qm}.} Here $\tr$ denotes the normalized trace, defined so that $\tr(I_d)=1$ for a system of any dimension $d$. The theory is specified by a single parameter $Z_L>0$. On each circle, the patch operators enjoy cyclic symmetry but not order-reversal symmetry, reflecting the orientation of the circle. These properties are captured by the trace appearing on the right-hand side of \cref{eq:rule_PI}.

As an example, let us evaluate the product of three patch operators $\mathcal{O}_{X_1}\mathcal{O}_{X_2}\mathcal{O}_{X_3}$ in the path integral. With three patch operators, the boundary condition consists of three distinguishable insertions, which can be distributed among one, two, or three circles:
\begin{equation}
	\begin{aligned}
	 &\ \overline{\mathcal{O}_{X_1}\mathcal{O}_{X_2}\mathcal{O}_{X_3}}\\
	=\ & \adjincludegraphics[height=0.72in,valign=m]{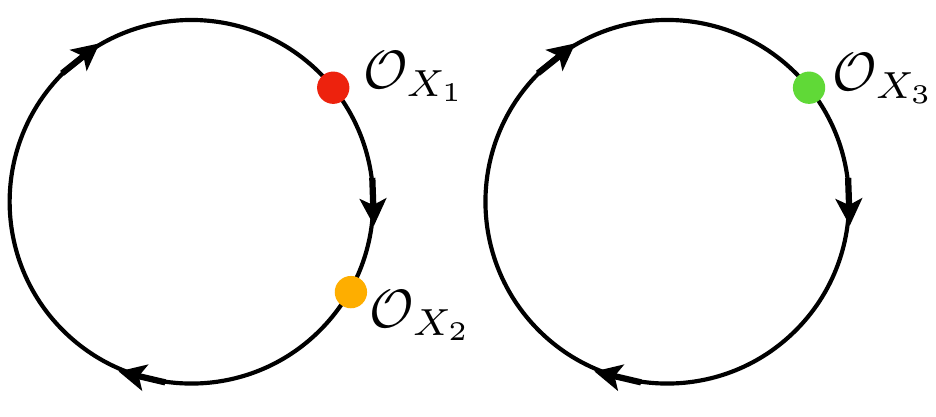}\ +\ \adjincludegraphics[height=0.72in,valign=m]{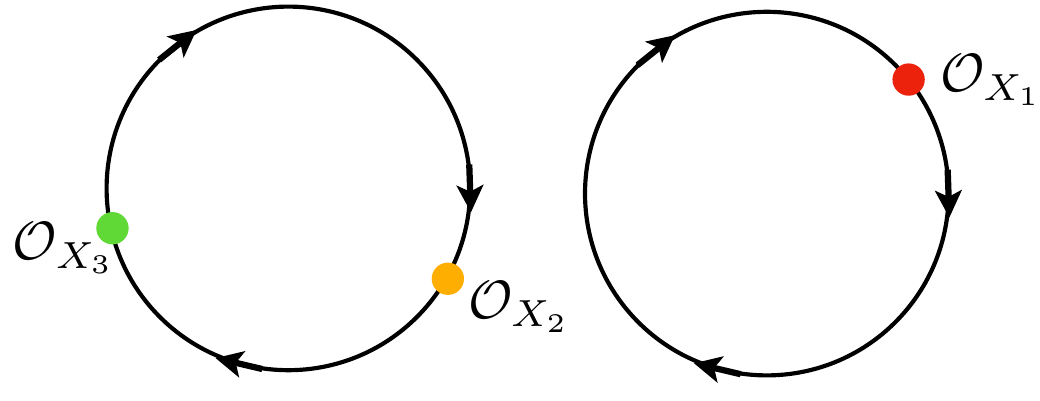}\ +\  \adjincludegraphics[height=0.72in,valign=m]{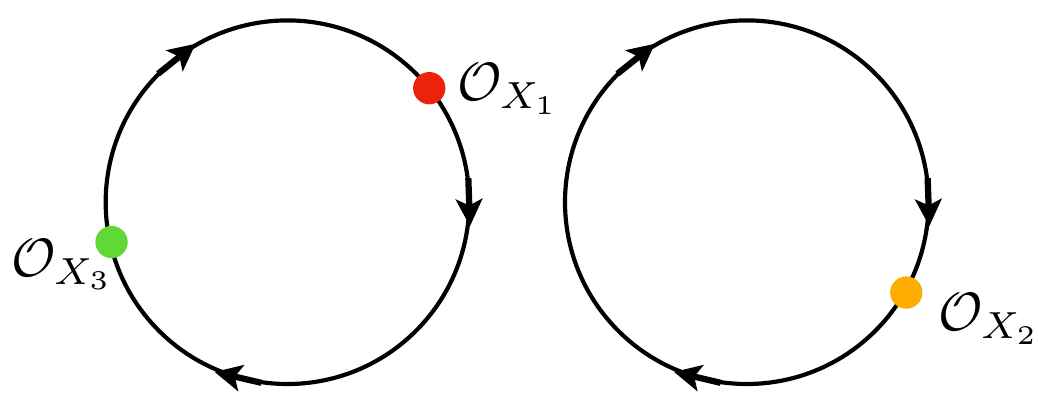}\\
	&+\adjincludegraphics[height=0.72in,valign=m]{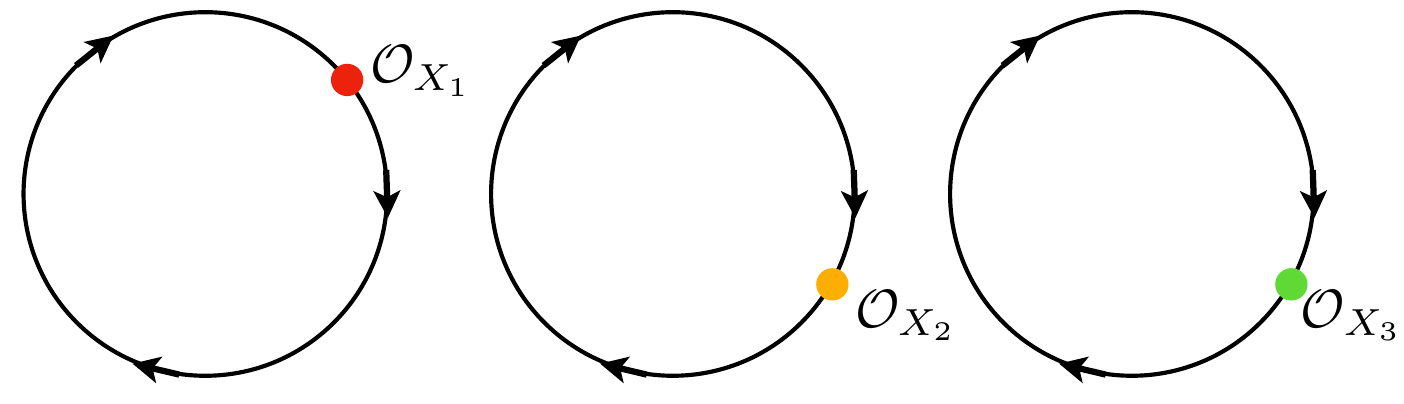}\ +\ \adjincludegraphics[height=0.72in,valign=m]{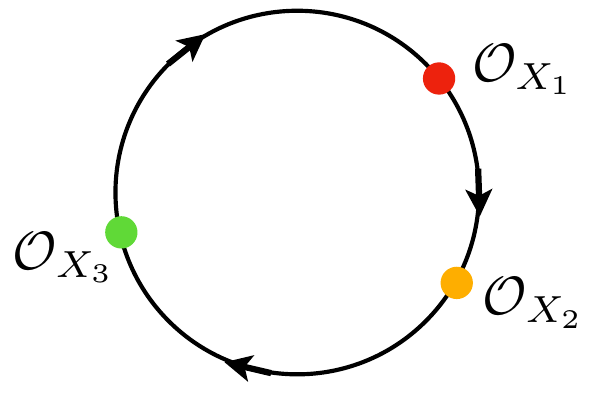}\ +\ \adjincludegraphics[height=0.72in,valign=m]{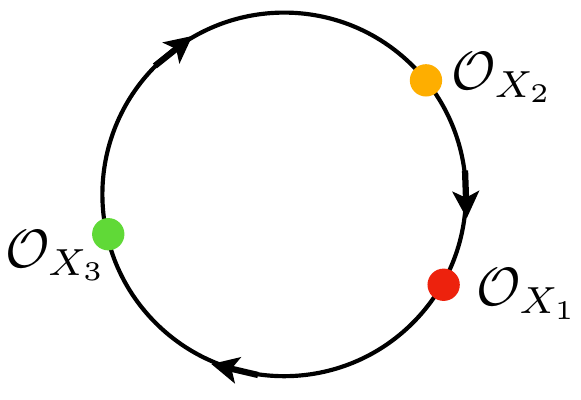}\\
	=\ &\tr(X_1X_2)\tr(X_3)Z_L^2\ +\ \tr(X_2X_3)\tr(X_1)Z_L^2\ +\ \tr(X_1X_3)\tr(X_2)Z_L^2\\
	&+\tr(X_1)\tr(X_2)\tr(X_3)Z_L^3\ +\ \tr(X_1X_2X_3)Z_L\ +\ \tr(X_2X_1X_3)Z_L \, .
	\end{aligned}
	\label{eq:example}
\end{equation}
In \cref{eq:example}, there are three configurations in which two insertions lie on one circle and the remaining insertion lies on a second circle, and one configuration in which each insertion occupies its own circle. When all three insertions lie on a single circle, there are two inequivalent cyclic orderings, reflecting cyclic symmetry.

Before ending this section, we comment on the path-integral rule in \cref{eq:rule_PI}. At first sight, \cref{eq:rule_PI} may look like an arbitrary choice. However, as we will see later, in such a simple model its form is essentially fixed. The only free parameter is a positive number $Z_L$.

\begin{itemize}
    \item \textbf{Why take a trace?}

    This is required by rotational symmetry of the Euclidean circle (equivalently, the boost symmetry of Lorentzian de Sitter), which is a gauge symmetry.\footnote{One may argue that one could take the trace of each operator $X_i$ separately, i.e., $C\tr(X_1)...\tr(X_n)$ for some constant $C$, but such an expression will have more symmetry than that of the configuration in \eqref{eq:rule_PI}. The configuration in \eqref{eq:rule_PI} has cyclic symmetry among the patch operators but in general is not invariant under the exchange of two patch operators. Also note such a choice is not completely specific to this simple $1+0$ dimensional toy model. More realistic spacetime does not have rotational symmetry, but we can take a trace along an observer's worldline \cite{workinprogress1}. However see \cite{Chen:2026boh} for potential obstructions.}

	\item \textbf{Why use the normalized trace?}

    One motivation is that, as shown in \cite{Chandrasekaran:2022cip}, the de Sitter algebra is of type $\mathrm{II}_1$. Setting that aside, since this is a finite, topological model there is no mathematical obstruction to considering the canonical trace that grows with the dimension of the system. One reason to prefer the normalized trace is that we do not want the effective loop partition function $Z_L$ to increase as we study larger and larger subsystems. More importantly, as we will see in \cref{sec:consistency}, the normalized trace is required for the model to have consistent physics, and it leads to important consequences such as a maximal dimension for systems that can fit into this toy de Sitter space.

    \item \textbf{Choice of $Z_L$}

    As we will see in \cref{sec:QM_U}, when $Z_L$ is an integer it equals the dimension of the largest quantum-mechanical system that can exist in this toy de Sitter space, and it is also related to the dimension of the $\alpha$-parameter space in the model.
\end{itemize}

\subsection{Different systems in the universe}
\label{sec:subsystems}

So far we have not fully specified the bulk theory. In particular, in \cref{eq:rule_PI} we did not specify the space on which the operators $X_i$ act or over which the trace is taken. As discussed in \cref{sec:qm}, specifying the theory requires choosing an operator algebra, denoted $\mathcal{A}_P$ in \cref{sec:qm}.

In this section, we study different possible choices of this operator algebra. Physically, we assume that the universe contains systems of various dimensions and investigate the quantum mechanics associated with those systems. In \cref{sec:spin}, we begin with a single spin as a warm-up example.\footnote{Here and throughout this section, by a ``spin'' we simply mean a two-state quantum-mechanical system, the simplest system with a noncommuting operator algebra. The term ``spin'' here has nothing to do with representations of the rotation group.} In \cref{sec:general}, we then consider systems of arbitrary dimension $d$.

As we proceed, we will find that not all system sizes are mathematically allowed. The physical origin of this restriction will become clear when we show how these different systems fit together as subsystems of a single larger system in \cref{sec:unify}.

\subsubsection{Quantum mechanics of one spin}
\label{sec:spin}

We first assume that the universe contains a single spin and study the quantum mechanics of this spin. To fully characterize a two-state quantum-mechanical system, we require the operator algebra $\mathcal{A}_2 = M_2(\mathbb{C})$, the algebra of all complex $2\times 2$ matrices, generated by the Pauli group $\mathcal{P}=\{\sigma_0\equiv I_2,\sigma_1,\sigma_2,\sigma_3\}$.\footnote{The algebra $\mathcal{A}_0$ of the two-state classical system studied in \cref{sec:one_bit} is identical to the abelian subalgebra spanned by $\sigma_0$ and $\sigma_2$. With this identification, it is straightforward to see that \cref{eq:rule_PI_0} is a special case of \cref{eq:rule_PI}.} The corresponding set of patch operators for this spin system is $\{\mathcal{O}_{X_2}\mid X_2\in M_2(\mathbb{C})\}$.

By the linearity discussed in \cref{sec:qm_ob}, it suffices to study the joint statistics of the four patch operators $(\mathcal{O}_{\sigma_0},\mathcal{O}_{\sigma_1},\mathcal{O}_{\sigma_2},\mathcal{O}_{\sigma_3})$.\footnote{In this model, linearity is manifest in \cref{eq:rule_PI}.} Because of the nontrivial commutation relations among the Pauli matrices $\sigma_i$, brute-force combinatorics is no longer convenient. Instead, we exploit the group structure of the contributions that appear in the path-integral evaluation of the moment-generating function. The details of the calculation are presented in Appendix~\ref{app:one_spin}.

The joint probability distribution of $(\mathcal{O}_{\sigma_0},\mathcal{O}_{\sigma_1},\mathcal{O}_{\sigma_2},\mathcal{O}_{\sigma_3})$ is
\begin{align}
	&p_{\scriptscriptstyle \mathcal{O}_{\sigma_0},\mathcal{O}_{\sigma_1},\mathcal{O}_{\sigma_2},\mathcal{O}_{\sigma_3}}(x_0,x_1,x_2,x_3)\nonumber\\
	=\ &\frac{\Gamma\!\left(\frac{Z_L}{2}+\frac{1}{2}\right)}{\pi^{\frac{3}{2}}\,\Gamma\!\left(\frac{Z_L}{2}-1\right)\Gamma(Z_L)}\,
	e^{-x_0}\,
	\bigl(x_0^2-x_1^2-x_2^2-x_3^2\bigr)^{\frac{Z_L}{2}-2}\,
	\Theta(x_0)\,\Theta(x_0^2-x_1^2-x_2^2-x_3^2) .
	\label{eq:prob_spin}
\end{align}
As a consistency check, the distribution in \cref{eq:prob_spin} reduces to \cref{eq:prob_join_0} after integrating out $x_1$ and $x_3$.

Another physically interesting quantity is the expectation value of operators acting on this spin. Recall the definition in \cref{eq:exp_X}:
\begin{align}
	(\expval{I_2},\expval{\sigma_1},\expval{\sigma_2},\expval{\sigma_3})
	\equiv
	\left(
	\frac{\mathcal{O}_{\sigma_0}}{\mathcal{O}_{\sigma_0}},
	\frac{\mathcal{O}_{\sigma_1}}{\mathcal{O}_{\sigma_0}},
	\frac{\mathcal{O}_{\sigma_2}}{\mathcal{O}_{\sigma_0}},
	\frac{\mathcal{O}_{\sigma_3}}{\mathcal{O}_{\sigma_0}}
	\right) .
	\label{eq:exp_spin_def}
\end{align}
From \cref{eq:exp_spin_def}, the probability distribution of the expectation values of the Pauli operators is
\begin{align}
	&p_{\scriptscriptstyle \expval{I_2},\expval{\sigma_1},\expval{\sigma_2},\expval{\sigma_3}}(y_0,y_1,y_2,y_3)\nonumber\\
	=\ &\frac{\Gamma\!\left(\frac{Z_L}{2}+\frac{1}{2}\right)}{\pi^{\frac{3}{2}}\Gamma\!\left(\frac{Z_L}{2}-1\right)}\,
	\bigl(1-y_1^2-y_2^2-y_3^2\bigr)^{\frac{Z_L}{2}-2}\,
	\Theta(1-|\vec y|^2)\,\delta(y_0-1) .
	\label{eq:prob_spin_2}
\end{align}

As argued in \cref{sec:qm}, the physics of the spin in this toy de Sitter space is completely encoded in the properties of the patch operators in Equations~\eqref{eq:prob_spin} or \eqref{eq:prob_spin_2}. We now examine these results in more detail.
\begin{itemize}
	\item \textbf{Range of the parameter}

	First, note that for the probability density \eqref{eq:prob_spin} to be nonnegative and normalizable, we require $Z_L\geq2$, which matches the complex dimension of a single spin.

	\item \textbf{Initial state of the spin}

	By the \emph{initial state} we mean the state of the spin under the Hartle--Hawking no-boundary condition, without any additional conditioning. It is completely characterized by \cref{eq:prob_spin_2}. Here are its properties.
	
	\begin{itemize}
\item \emphdark{\textbf{Density matrix of the spin}}

The vector $\vec y=(y_1,y_2,y_3)$, interpreted as the expectation value of the Pauli operators $\vec\sigma$ on this spin, always lies inside the Bloch ball. For each $\vec y$, the corresponding density matrix of the spin is
\begin{align}
	(\rho_{\scriptscriptstyle 2})_{\vec y}
	= \frac{I_2+\vec y\cdot\vec\sigma}{2} .
	\label{eq:spin_density}
\end{align}
With \eqref{eq:spin_density}, the distribution in \cref{eq:prob_spin_2} can be interpreted as an ensemble of spin density matrices. The ensemble average of these density matrices is maximally mixed:
\begin{align}
	\overline{\rho_{\scriptscriptstyle 2}}=\frac{I_2}{2}.
	\label{eq:spin_average}
\end{align}

\item \emphdark{\textbf{Angular distribution}}

This ensemble is completely isotropic, i.e.\ there is no preferred direction in spin space. However, this does not by itself imply that the spin is unpolarized in individual realizations.

\item \emphdark{\textbf{Radial distribution}}

To assess whether the spin is polarized, we examine the distribution of its purity,
$\Tr\!\left((\rho_{\scriptscriptstyle 2}^{\,\scriptscriptstyle 2})_{\vec y}\right)=(1+|\vec y|^2)/2$,\footnote{$\Tr$ denotes the canonical trace, with $\Tr(I_2)=2$.}
which is determined by the distribution of $|\vec y|^2$:
\begin{align}
	|\vec y|^2 &\sim \text{Beta}\!\left(\frac{3}{2},\frac{Z_L}{2}-1\right),\\
	p_{|\vec y|^2}(s)
	&= \frac{2\,\Gamma\!\left(\frac{Z_L}{2}+\frac{1}{2}\right)}{\sqrt{\pi}\,\Gamma\!\left(\frac{Z_L}{2}-1\right)}
	s^{\frac{1}{2}}(1-s)^{\frac{Z_L}{2}-2}\,\mathds{1}_{[0,1]}(s).
\end{align}
The mean and variance of $|\vec y|^2$ are
\begin{align}
	\overline{|\vec y|^2}=\frac{3}{Z_L+1},\qquad
	\overline{|\vec y|^4}-\overline{|\vec y|^2}^{\,2}
	=\frac{6(Z_L-2)}{(Z_L+1)^2(Z_L+3)} .
\end{align}

For large $Z_L$, $|\vec y|^2$ is concentrated near zero, and each member of the ensemble is close to maximally mixed. Since the typical size of $y_i$ is of order $\sqrt{1/Z_L}$, the averaged density matrix in \cref{eq:spin_average} provides an excellent approximation. Fluctuations are suppressed by the loop partition function, as discussed in \cref{sec:dS}. In this regime, the spin is almost certainly unpolarized.

By contrast, as $Z_L\to 2^+$, $|\vec y|^2$ becomes concentrated near $1$, and the ensemble consists of spins uniformly distributed on the Bloch sphere. Each member of the ensemble is then a pure state polarized in some direction. In this regime there are order-one ensemble fluctuations in the spin direction, and in particular the averaged density matrix in \cref{eq:spin_average} is not a good approximation for individual realizations. One might worry that a de Sitter observer would encounter large fluctuations analogous to those in the AdS closed-universe case discussed in \cref{sec:AdS_observer}. We will return to this issue in \cref{sec:conditioning}. For now, we simply note that the case of a small partition function $Z_L=2$ is highly unphysical.
\end{itemize}

\item \textbf{Operators on the spin}

\begin{itemize}

\item \emphdark{\textbf{Identity operator}}

We begin with $\mathcal{O}_{I_2}\equiv \mathcal{O}_{\sigma_0}$. Despite corresponding to the identity operator, $\mathcal{O}_{I_2}$ has nontrivial statistics. It follows a Gamma distribution with shape parameter $Z_L$ and scale $1$. Its mean is $Z_L$, and the distribution becomes sharply peaked as $Z_L$ increases:
\begin{equation}
\begin{aligned}
	\mathcal{O}_{I_2} &\sim \text{Gamma}(Z_L,1),\\
	p_{\scriptscriptstyle \mathcal{O}_{I_2}}(x_0)
	&= \frac{1}{\Gamma(Z_L)}\,x_0^{Z_L-1}e^{-x_0}\Theta(x_0).
	\label{eq:prob_spin_identity}
\end{aligned}
\end{equation}

From \cref{eq:patch_def_path_integral,eq:rule_PI}, $\mathcal{O}_{I_2}$ is the loop partition function of this de Sitter space containing a spin. Its saddle-point value is
\begin{align}
	\overline{\mathcal{O}_{I_2}} = Z_L = \exp\!\qty(\log Z_L).
	\label{eq:partition_spin}
\end{align}
Here $\log Z_L$ plays a role analogous to the cosmic-horizon area or the de Sitter entropy in higher-dimensional de Sitter space.

\item \emphdark{\textbf{More general operators}}

Now consider a general operator on the spin,
$X_2=\sum_{\mu}X_{\mu}\sigma_{\mu}\in\mathcal{A}_2=M_2(\mathbb{C})$.
By linearity of patch operators, the patch operator corresponding to $X_2$ is
\begin{align}
	\mathcal{O}_{X_2}=\sum_{\mu}X_{\mu}\mathcal{O}_{\sigma_{\mu}}.
	\label{eq:spin_patch_general}
\end{align}

Define the random $2\times2$ Hermitian matrix
\begin{align}
	\Sigma_2 \equiv \frac{1}{2}\sum_{\mu}\mathcal{O}_{\sigma_{\mu}}\sigma_{\mu}.
	\label{eq:def_Sigma_spin}
\end{align}
Then \cref{eq:spin_patch_general} can be rewritten as
\begin{align}
	\mathcal{O}_{X_2} = \Tr(\Sigma_2 X_2),
	\label{eq:map_spin}
\end{align}
where $\Tr$ denotes the canonical trace on the two-dimensional Hilbert space.

The interpretation of $\Sigma_2$ is now clear: it is the unnormalized density matrix of the spin. This also clarifies the motivation for the definition in \cref{eq:exp_X}: After normalizing by $\mathcal{O}_{I_2}$, we obtain the expectation value of the operator $X_2$ on the spin,
\begin{align}
	\expval{X_2}=\frac{\Tr(\Sigma_2 X_2)}{\Tr(\Sigma_2)}.
\end{align}
Since $\Sigma_2$ is Hermitian in each $\alpha$-sector, quantum mechanics for this spin is exact within each $\alpha$-sector. In particular, relations such as \cref{eq:transition_amplitude} hold exactly in every $\alpha$-sector.

\end{itemize}

\item \textbf{Connection to the no-boundary density matrix}

The no-boundary density matrix was introduced in \cite{Ivo:2024ill} as a generalization of the Hartle--Hawking wave function. Instead of specifying data on an entire closed spatial slice, one specifies data on a subregion of the universe and evaluates the path integral with no-boundary conditions.

We have already seen that $\mathcal{O}_{I_2}$ is the loop partition function of this de Sitter space containing a spin. What about more general patch operators? Let $\ket{\hat n}$ denote the pure spin state polarized along the $+\hat n$ direction. From the definition of patch operators in \cref{eq:patch_b_3,eq:patch_def_path_integral}, the quantity $\mathcal{O}_{\ket{\hat n_2}\bra{\hat n_1}}$ is precisely the $\hat n_1\hat n_2$ matrix element of the no-boundary density matrix evaluated by the path integral. Using \cref{eq:map_spin}, we find
\begin{align}
	\mathcal{O}_{\ket{\hat n_2}\bra{\hat n_1}}
	= \Tr(\Sigma_2\,\ket{\hat n_2}\bra{\hat n_1})
	= \bra{\hat n_1}\Sigma_2\ket{\hat n_2}.
\end{align}
We therefore conclude that $\Sigma_2$ is exactly the no-boundary density matrix of the universe, restricted to the spin.

Its saddle-point value is
\begin{align}
	\bra{\hat n_1}\,\overline{\Sigma_2}\,\ket{\hat n_2}
	= \frac{Z_L}{2}\,\bra{\hat n_1}\ket{\hat n_2}.
	\label{eq:no_boundary_spin_saddle}
\end{align}
As a check on numerical factors, \cref{eq:no_boundary_spin_saddle} can be obtained either from the definition in \cref{eq:def_Sigma_spin} or by directly evaluating the one-patch operator $\mathcal{O}_{\ket{\hat n_2}\bra{\hat n_1}}$ using \cref{eq:rule_PI}.

Finally, note that the no-boundary density matrix in \cref{eq:def_Sigma_spin}, being defined directly through the path integral, is unnormalized. In particular, it contains large factors of $Z_L$, as seen in \cref{eq:no_boundary_spin_saddle}. To obtain a normalized density matrix, we divide $\Sigma_2$ by $\mathcal{O}_{I_2}$, recovering the standard spin density matrix given in \cref{eq:spin_density,eq:prob_spin_2}.

\item \textbf{The space of $\alpha$-parameters}

From the probability distribution in \cref{eq:prob_spin}, the patch operators $\mathcal{O}_{\sigma_\mu}$ take continuous values, implying that there are uncountably many $\alpha$-parameters. In fact, the space of $\alpha$-parameters is
\begin{align}
	\Omega_2
	= \bigl\{(x_0,x_1,x_2,x_3)\ \big|\ x_0\ge 0,\ x_0^2\ge x_1^2+x_2^2+x_3^2,\ x_i\in\mathbb{R}\bigr\},
	\label{eq:alpha_space_spin}
\end{align}
where the subscript $2$ emphasizes that these $\alpha$-parameters are those required to characterize a single spin in the universe. As a manifold, $\Omega_2$ has real dimension $4$, equal to the number of independent real parameters of a $2\times2$ Hermitian matrix.

In fact, the $\alpha$-parameter space $\Omega_2$ is precisely the space of unnormalized density matrices for a spin. Let
\begin{align}
	\mathrm{Herm}_+(2)
	= \bigl\{H\in M_2(\mathbb{C}) \mid H=H^\dagger,\ H\succeq 0\bigr\}
\end{align}
denote the space of positive semidefinite $2\times2$ Hermitian matrices. The correspondence is
\begin{align}
	\Omega_2 &\longleftrightarrow \mathrm{Herm}_+(2), \nonumber\\
	\alpha=(x_0,\vec x)
	&\longleftrightarrow \frac{1}{2}\bigl(x_0 I_2+\vec x\cdot\vec\sigma\bigr)
	= (\Sigma_2)_\alpha .
	\label{eq:correspondence_spin}
\end{align}
It is straightforward to see that the correspondence in \cref{eq:correspondence_spin} is simply the restriction of \cref{eq:def_Sigma_spin} to each $\alpha$-sector.

One might wonder how nontrivial bulk quantum mechanics can be encoded in a fundamental description that contains only a single state. The answer is already visible here: the fundamental description contains many $\alpha$-sectors. In this simple example, the bulk physics has a Hilbert space of dimension two, yet it requires infinitely many $\alpha$-parameters. This is precisely what one expects when quantum mechanics is encoded in a classical statistical description.

\end{itemize}

\subsubsection{Quantum mechanics of a general system in the universe}
\label{sec:general}

In this section, we study a quantum-mechanical system of dimension $d$ in this de Sitter space. The most general operator algebra acting on such a system is
$\mathcal{A}_d = M_d(\mathbb{C})$, the algebra of all complex $d\times d$ matrices. We choose a set of Hermitian generators
$\{T_0\equiv I_d, T_1, T_2,\ldots,T_{d^2-1}\}$ satisfying
$\Tr(T_\mu T_\nu)=d\,\delta_{\mu\nu}$, and study the joint statistics of the corresponding patch operators
$\{\mathcal{O}_{T_\mu}\mid \mu=0,1,\ldots,d^2-1\}$.\footnote{Here $\Tr$ denotes the canonical trace with $\Tr(I_d)=d$. One convenient choice for $T_\mu$ is the standard generalized Gell--Mann matrices multiplied by $\sqrt{d/2}$.}

For general $d$, we can no longer be as explicit as in the single-spin case based on Pauli matrices. Instead, we organize the $d^2$ patch operators into a single Hermitian random matrix,
\begin{align}
	\Sigma_d \equiv \frac{1}{d}\sum_{\mu}\mathcal{O}_{T_\mu}\,T_\mu .
	\label{eq:def_Sigma_general}
\end{align}
We then study the statistics of $\Sigma_d$. The details are in Appendix~\ref{app:general_d}. From the moment-generating function
\begin{align}
	\overline{\exp\!\big[\Tr(S_d\Sigma_d)\big]}
	= \det(I_d-S_d)^{-\frac{Z_L}{d}},
	\label{eq:MGF_general}
\end{align}
we see that $\Sigma_d$ follows a complex Wishart distribution,
\begin{align}
	\Sigma_d \sim \mathcal{W}_d^{\mathbb{C}}\!\left(\frac{Z_L}{d}, I_d\right).
	\label{eq:distribution_general}
\end{align}

We now examine the physical implications of the distributions in \cref{eq:MGF_general,eq:distribution_general}.
\begin{itemize}
	\item \textbf{Range of the parameter}

	The shape parameter $Z_L/d$ of the distribution admits two qualitatively distinct regimes:
	\begin{itemize}
		\item \emphdark{$\mathbf{Z_L/d>d-1}$}.

		In this regime, the probability density of $\Sigma_d$ is
		\begin{align}
			p_{\scriptscriptstyle \Sigma_d}(W)
			= \frac{1}{\Gamma_d^{\mathbb{C}}\!\left(Z_L/d\right)}
			\det(W)^{Z_L/d-d}\,
			\exp\!\big(-\Tr W\big)\,
			\mathds{1}_{\scriptscriptstyle W\succeq 0},
		\end{align}
		where
		$\Gamma_d^{\mathbb{C}}(x)
		= \pi^{d(d-1)/2}\prod_{j=1}^d \Gamma(x-j+1)$.
		In this case, the matrix $\Sigma_d$ is almost surely of full rank.

		\item \emphdark{$\mathbf{Z_L/d=m\in\{1,2,\ldots,d-1\}}$}.

		In this regime, $\Sigma_d$ has rank $m<d$.\footnote{Mathematically $Z_L=0$ is also allowed, but in that case the model is completely trivial so we don't consider it.}
	\end{itemize}

	\item \textbf{Physical meaning of $\Sigma_d$}

From the definition in \cref{eq:def_Sigma_general} and the properties of the generators $T_\mu$, it follows that for any operator $X_d\in\mathcal{A}_d$,
\begin{align}
	\mathcal{O}_{X_d} = \Tr(\Sigma_d X_d).
\end{align}
We thus see once again that $\Sigma_d$ plays the role of the no-boundary density matrix of the universe, restricted to this $d$-dimensional system. In particular, $\mathcal{O}_{I_d}\equiv \mathcal{O}_{T_0}$ is the loop partition function of the de Sitter space containing this system.

Setting $S_d=s_0 T_0$, the moment-generating function \eqref{eq:MGF_general} reduces to
\begin{align}
	\overline{\exp(s_0\,\mathcal{O}_{I_d})} = (1-s_0)^{-Z_L},
\end{align}
which implies
\begin{align}
	\mathcal{O}_{I_d} \sim \text{Gamma}(Z_L,1).
	\label{eq:prob_general_identity}
\end{align}
The random variable $\mathcal{O}_{I_d}$ is sharply peaked around $Z_L$ when $Z_L$ is large.

\item \textbf{Initial state of the system}

The initial state of this $d$-dimensional system is encoded in the normalized density matrix
$\rho_{\scriptscriptstyle d}\coloneq \Sigma_d/\mathcal{O}_{I_d}$.
From the distributions in \cref{eq:MGF_general,eq:distribution_general}, the normalized density matrix satisfies \cite{Zyczkowski:2001fsr}
\begin{align}
	\overline{\rho_{\scriptscriptstyle d}} &= \frac{1}{d}I_d,
	\label{eq:average_general}\\
	\overline{\Tr(\rho_{\scriptscriptstyle d}^{\,2})} &= \frac{Z_L/d+d}{Z_L+1},
	\label{eq:purity_general}\\
	\overline{\rho_{\scriptscriptstyle d}\otimes\rho_{\scriptscriptstyle d}}
	&= \frac{(Z_L/d)I_{d^2}+\hat S_{d^2}}{d(Z_L+1)},
	\label{eq:tensor}
\end{align}
where $I_{d^2}$ denotes the identity operator on the doubled system and $\hat S_{d^2}$ is the swap operator. \cref{eq:tensor} follows from the unitary invariance of the distribution \eqref{eq:MGF_general} together with \cref{eq:purity_general}.

Now consider an operator $X_d\in\mathcal{A}_d$. Using \cref{eq:average_general,eq:tensor}, we obtain
\begin{align}
	\overline{\expval{X_d}}
	&= \overline{\Tr(\rho_{\scriptscriptstyle d}X_d)}
	= \tr(X_d),
	\label{eq:mean_general}\\
	\mathrm{Var}(\expval{X_d})
	&= \overline{\expval{X_d}^{\,2}}-\overline{\expval{X_d}}^{\,2}
	= \frac{\tr(X_d^{\,2})-(\tr X_d)^2}{Z_L+1},
	\label{eq:ensemble_general}
\end{align}
where $\tr$ denotes the normalized trace.

\begin{itemize}
\item \emphdark{$\mathbf{Z_L/d>d-1}$}

In this regime, the density matrix $\rho_{\scriptscriptstyle d}$ is distributed according to
\begin{align}
	p_{\scriptscriptstyle \rho_{\scriptscriptstyle d}}(\rho)
	\propto
	\det(\rho)^{Z_L/d-d}\,
	\delta(\Tr\rho-1)\,
	\mathds{1}_{\scriptscriptstyle \rho\succeq 0}.
	\label{eq:prob_normalized_general}
\end{align}
When $Z_L/d\gg d$, the system is nearly maximally mixed, and the entanglement entropy grows with the system size as $\log d$. In this regime, for any operator $X_d$ the mean value in \cref{eq:mean_general} provides an excellent approximation.

\item \emphdark{$\mathbf{Z_L/d=m\le d-1}$, $\mathbf{m\in\mathbb{Z}_{>0}}$}

In this regime, the system is no longer maximally mixed. Its entanglement entropy is approximately $\log m$ and decreases as the system dimension $d$ increases.

From \cref{eq:mean_general,eq:ensemble_general}, when $X_d$ is a typical (full-rank) operator in $M_d(\mathbb{C})$, its ensemble fluctuation is still suppressed by $\sqrt{1/Z_L}$. By contrast, if $X_d$ is a low-rank projection, as in the computation of the no-boundary density matrix, then
$\overline{\expval{X_d}}\sim 1/d$ and
$\mathrm{Var}(\expval{X_d})\sim 1/(d\,Z_L)$.
In this case the fluctuation is only suppressed by $\sqrt{d/Z_L}$.
As $d$ approaches $Z_L$, the fluctuation becomes order one. This behavior parallels the $Z_L\to 2^+$ limit discussed in \cref{sec:spin}. In \cref{sec:conditioning}, we will argue that such regimes are unphysical.
\end{itemize}

\item \textbf{The space of $\alpha$-parameters}

In \cref{eq:def_Sigma_general}, we organized the patch operators
$\{\mathcal{O}_{T_\mu}\mid \mu=0,1,\ldots,d^2-1\}$ into a single matrix $\Sigma_d$.
From the distribution \eqref{eq:distribution_general}, we see that $\Sigma_d$ takes continuous values. We now examine the structure of the $\alpha$-parameter space $\Omega_d$ in the two regimes of $Z_L$.

\begin{itemize}
\item \emphdark{$\mathbf{Z_L/d>d-1}$}

In this regime, $\Sigma_d$ is almost surely of full rank. The $\alpha$-parameter space is therefore
\begin{align}
	\Omega_d\Big|_{\scriptstyle Z_L/d>d-1}
	= \mathrm{Herm}_+(d)
	= \{H\in M_d(\mathbb{C})\mid H=H^\dagger,\ H\succeq 0\}.
	\label{eq:alpha_space_general_1}
\end{align}
This space has real dimension $d^2$.

\item \emphdark{$\mathbf{Z_L/d=m\le d-1}$, $\mathbf{m\in\mathbb{Z}_{>0}}$}

In this regime, $\Sigma_d$ has rank $m$. The $\alpha$-parameter space is
\begin{align}
	\Omega_d\Big|_{\scriptstyle Z_L/d=m\le d-1}
	=
	\Bigl\{
	U\,\mathrm{diag}(\lambda_1,\ldots,\lambda_m,\underbrace{0,\ldots,0}_{d-m})\,U^\dagger
	\ \Big|\ 
	\lambda_i\ge 0,\ U\in SU(d)
	\Bigr\},
	\label{eq:alpha_space_general_m}
\end{align}
which has real dimension $d^2-(d-m)^2=2dm-m^2$.
\end{itemize}

\end{itemize}

\subsection{Quantum mechanics inside de Sitter space}
\label{sec:QM_U}

So far, we have studied the quantum-mechanical descriptions of systems of various dimensions in this de Sitter space. For the physics of the universe to be consistent, these descriptions must all fit together. Since a bulk observer can choose to study subsystems of different sizes at will, all such systems should arise as subsystems of a single, larger quantum-mechanical system. In this section, we show that this requirement is indeed satisfied in the present toy model.

\subsubsection{Consistency across system dimensions}
\label{sec:consistency}

Consider a system $S$ in this de Sitter space, which factorizes as a tensor product of two subsystems,
$S=A\otimes B$. Their dimensions satisfy $d_S=d_A d_B$. One can either study subsystem $A$ directly, or study the larger system $S$ and then deduce the properties of $A$ by tracing out $B$. Consistency requires that these two procedures agree.

From \cref{eq:MGF_general}, if we study a system of dimension $d_A$ directly, we conclude that its no-boundary density matrix $\Sigma_A$ has moment-generating function
\begin{align}
	\overline{\exp\!\big[\Tr_A(S_A\Sigma_A)\big]}
	= \det(I_A-S_A)^{-\frac{Z_L}{d_A}} .
	\label{eq:MGF_A}
\end{align}

Now consider the larger system $S$. Choose a Hermitian basis
$\{T^A_\mu\mid \mu=0,\ldots,d_A^2-1\}$ for $\mathcal{A}_A=M_{d_A}(\mathbb{C})$ satisfying
$\Tr_A(T^A_\mu T^A_\nu)=d_A\delta_{\mu\nu}$, and a Hermitian basis
$\{T^B_a\mid a=0,\ldots,d_B^2-1\}$ for $\mathcal{A}_B=M_{d_B}(\mathbb{C})$ satisfying
$\Tr_B(T^B_a T^B_b)=d_B\delta_{ab}$. Here $T^A_0=I_A$ and $T^B_0=I_B$ are the identity operators.
Then
\begin{align}
T^S_{\mu a}\equiv T^A_\mu\otimes T^B_a,
\qquad
\mu=0,\ldots,d_A^2-1,\ \ a=0,\ldots,d_B^2-1,
\end{align}
form a Hermitian basis of $\mathcal{A}_S=M_{d_S}(\mathbb{C})$ satisfying
$\Tr_S(T^S_{\mu a}T^S_{\nu b})=d_S\,\delta_{\mu\nu}\delta_{ab}$.
Throughout, $\Tr$ denotes the canonical trace, and the subscript indicates the Hilbert space over which the trace is taken.

With this basis, we construct the no-boundary density matrix $\Sigma_S$ on $S$ from the patch operators $\{\mathcal{O}^S_{T^S_{\mu a}}\}$:
\begin{align}
	\Sigma_S=\frac{1}{d_S}\sum_{\mu,a}\mathcal{O}^S_{T^S_{\mu a}}\,T^S_{\mu a}.
\end{align}
Since $A$ is a subsystem of $S$, its reduced density matrix $\widetilde\Sigma_A$ is
\begin{align}
	\widetilde\Sigma_A \coloneq \Tr_B(\Sigma_S)
	= \frac{1}{d_A}\sum_{\mu}\mathcal{O}^S_{T^S_{\mu 0}}\,T^A_\mu .
	\label{eq:Sigma_A_reduced}
\end{align}

From \cref{eq:MGF_general}, $\Sigma_S$ has moment-generating function
\begin{align}
	\overline{\exp\!\big[\Tr_S(S_S\Sigma_S)\big]}
	= \det(I_S-S_S)^{-\frac{Z_L}{d_S}}
	\label{eq:MGF_S}
\end{align}
for any $S_S\in M_{d_S}(\mathbb{C})$. Now set $S_S=S_A\otimes I_B$ with $S_A\in M_{d_A}(\mathbb{C})$. Then
\begin{align}
	\Tr_S\!\big[(S_A\otimes I_B)\Sigma_S\big]
	&= \Tr_A(S_A\widetilde\Sigma_A),
	\label{eq:left}\\
	\det(I_S-S_A\otimes I_B)
	&= \det(I_A-S_A)^{d_B}.
	\label{eq:right}
\end{align}
Using \cref{eq:left,eq:right}, \cref{eq:MGF_S} reduces to
\begin{align}
	\overline{\exp\!\big[\Tr_A(S_A\widetilde\Sigma_A)\big]}
	= \det(I_A-S_A)^{-\frac{Z_L}{d_A}} .
	\label{eq:MGF_A_reduced}
\end{align}

Comparing \cref{eq:MGF_A,eq:MGF_A_reduced}, we see that $\Sigma_A$ and $\widetilde\Sigma_A$ have identical statistics and therefore encode the same physics. This shows that the density matrices obtained in \cref{sec:general} for different system dimensions are mutually compatible.

Finally, note that this consistency relies crucially on using the normalized trace in \cref{eq:rule_PI}. If we instead used the canonical (unnormalized) trace, the exponents on the right-hand sides of \cref{eq:MGF_A,eq:MGF_A_reduced} would not agree.

\subsubsection{A unified quantum-mechanical description}
\label{sec:unify}

We now consider the largest system $\mathrm{U}$ that can fit into this de Sitter space.
From \cref{sec:general}, for non-integer $Z_L$ the maximal dimension $d_{\mathrm{U}}$ must satisfy
$d_{\mathrm{U}}(d_{\mathrm{U}}-1)<Z_L$, while for integer $Z_L$ one may take $d_{\mathrm{U}}=Z_L$.
From now on we assume $Z_L\in\mathbb{Z}_{>0}$ and set $d_{\mathrm{U}}=Z_L$.

Let $\mathcal{H}_{\mathrm{U}}\cong\mathbb{C}^{Z_L}$ be the Hilbert space of $\mathrm{U}$, and let
$\Sigma_{\mathrm{U}}$ denote its no-boundary density matrix constructed from patch operators.
Its moment-generating function is
\begin{align}
	\overline{\exp[\Tr(S\Sigma_{\mathrm{U}})]}
	= \det(I_{Z_L}-S)^{-1},
	\label{eq:MGF_U}
\end{align}
which implies that $\Sigma_{\mathrm{U}}$ follows a rank-one singular complex Wishart distribution,
\begin{align}
	\Sigma_{\mathrm{U}}\sim\mathcal{W}_{Z_L}^{\mathbb{C}}(1,I_{Z_L}).
\end{align}

Equivalently, $\Sigma_{\mathrm{U}}$ can be written as a rank-one unnormalized density matrix constructed from an unnormalized Haar-random state. Let $\ket{\psi_{\HH}}\in\mathcal{H}_{\mathrm{U}}$ be an unnormalized Haar-random state, i.e.\ a complex Gaussian vector $\psi_{\HH}\in\mathbb{C}^{Z_L}$ with i.i.d.\ entries $(\psi_{\HH})_i\sim\mathcal{CN}(0,1)$ for $i=1,\ldots,Z_L$. Then
\begin{align}
	\Sigma_{\mathrm{U}} &= \ket{\psi_{\HH}}\mkern-4mu\bra{\psi_{\HH}},
	\label{eq:Sigma_U}\\
	\mathcal{O}_X &= \Tr(X\Sigma_{\mathrm{U}})
	= \bra{\psi_{\HH}}X\ket{\psi_{\HH}},
	\qquad X\in\mathcal{A}_{\mathrm{U}} .\label{eq:HH_exp_0}
\end{align}

Consider now a subsystem $S$ of dimension $d_S$. $\mathrm{U}=S\otimes E$ with $d_S d_E =Z_L$.
The reduced no-boundary density matrix on $S$ is then
\begin{align}
	\Sigma_S = \Tr_E(\ket{\psi_{\HH}}\mkern-4mu\bra{\psi_{\HH}}).
	\label{eq:density_S}
\end{align}
By the consistency analysis in \cref{sec:consistency}, $\Sigma_S$ has exactly the same statistics
as those derived directly for a $d_S$-dimensional system in \cref{sec:general}.

As discussed in \cref{sec:subsystems}, the random matrices $\Sigma$ play the role of no-boundary density matrices. In our approach, we do not begin by postulating a Hilbert space of physical states. Instead, we start from a single state corresponding to no-boundary conditions, compute expectation values of all operators in this state, and then construct the Hilbert space via the GNS construction. The state we start from is given by $\Sigma_{\mathrm{U}}$. From this perspective, the no-boundary density matrices, viewed as particular operators in $\mathcal{A}_{\mathrm{U}}$, are more primitive than wave functions. The state $\ket{\psi_{\HH}}$ arises in the rank-one (pure-state) limit of these density matrices, as described in \cref{eq:Sigma_U}.

Moreover, since the statistics of patch operators in \cref{eq:HH_exp_0} encode all physical information in the universe, the Gaussian vector $\psi_{\HH}$ can be used to construct the wave function of the universe under the no-boundary conditions. A complication specific to this $1{+}0$-dimensional toy model is that a spatial slice consists of two disconnected points. We will return to this issue in \cref{sec:Hartle_Hawking}.

Another natural question concerns the inner product between two physical states $\ket{\phi_1}, \ket{\phi_2} \in \mathcal{H}_{\mathrm{U}}$. We emphasize that this inner product is \emph{not} computed by the gravitational path integral. Instead, the gravitational path integral provides classical statistical data in the baby-universe Hilbert space. Inner products in the baby-universe Hilbert space therefore compute classical correlations between boundary conditions. By themselves, they do not define the inner product on the space of physical states.

More precisely, in our model physical states in $\mathcal{H}_{\mathrm{U}}$ obey the canonical inner product by construction:
\begin{align}
	\braket{\phi_1}{\phi_2}=\phi_1^{\dagger}\phi_2,
	\label{eq:inner_product_canonical}
\end{align}
which is encoded in the baby-universe Hilbert space as identities among patch operators:
\begin{align}
	\mathcal{O}_{\bigl(\ket{\psi_1}\mkern-1.6mu\bra{\phi_1}\bigr)\bigl(\ket{\phi_2}\mkern-1.6mu\bra{\psi_2}\bigr)}
	=
	\bigl(\phi_1^{\dagger}\phi_2\bigr)\,
	\mathcal{O}_{\ket{\psi_1}\mkern-1.6mu\bra{\psi_2}},
	\qquad
	\forall\,\ket{\psi_1},\ket{\psi_2}\in\mathcal{H}_{\mathrm{U}}.
\end{align}

We therefore obtain a unified quantum-mechanical description of this de Sitter space.
The total Hilbert space $\mathcal{H}_{\mathrm{U}}$ has dimension $Z_L$, the analog of $e^{S_{\mathrm{dS}}}$
in higher-dimensional theories, and operator algebra
$\mathcal{A}_{\mathrm{U}}=M_{Z_L}(\mathbb{C})$.
The Haar-random state $\ket{\psi_{\HH}}\in\mathcal{H}_{\mathrm{U}}$ fully encodes
the physics.
Equivalently, the theory is specified by the pair
\begin{align}
	\{\psi_{\HH},\,\mathcal{A}_{\mathrm{U}}=M_{Z_L}(\mathbb{C})\}.
\end{align}

As expected for a quantity defined by a gravitational path integral, $\ket{\psi_{\HH}}$ is
unnormalized.
Its norm squared satisfies
\begin{align}
	\|\psi_{\HH}\|^2
	= \sum_{i=1}^{Z_L}|(\psi_{\HH})_i|^2
	\sim \text{Gamma}(Z_L,1),
	\label{eq:norm_square}
\end{align}
which is precisely the loop partition function of this de Sitter space.
Equivalently, it is the patch operator $\mathcal{O}_I$ associated with the identity operator.

Finally, we describe the space of $\alpha$-parameters.
An unnormalized Haar-random state in $\mathbb{C}^{Z_L}$ is specified by $Z_L$ complex Gaussian
variables, modulo an overall phase.
Thus the $\alpha$-parameter space is
\begin{align}
	\Omega_{\mathrm{U}}
	= \{\psi_{\HH}\in\mathbb{C}^{Z_L}\mid \psi_{\HH}\sim e^{i\theta}\psi_{\HH}\},
	\label{eq:alpha_parameter_U}
\end{align}
which has real dimension $2Z_L-1$, in agreement with the general result
\eqref{eq:alpha_space_general_m} for $m=1$.

The state $\ket{\psi_{\HH}}$ admits an explicit representation via a quantum code,
\begin{equation}
	\ket{\psi_{\HH}}
	= \adjincludegraphics[width=0.48in,valign=m]{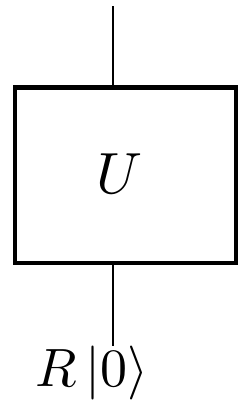},
	\label{eq:code_state}
\end{equation}
with $U\sim\text{Haar}(SU(Z_L))$ and
$R\sim \frac{1}{\sqrt{2}}\chi_{2Z_L}$.
This is precisely the code used in \cref{eq:code}.

\subsubsection{Different backgrounds from conditioning}
\label{sec:conditioning}

In \cref{sec:spin,sec:general} we studied the \emph{unconditioned} initial states of subsystems of various dimensions in the universe. By this we meant the states induced from the no-boundary condition without any additional conditioning. For a subsystem of dimension $d$ satisfying $d\le Z_L/d$, the state is almost maximally mixed. This is a direct consequence of Page’s theorem \cite{Page:1993df}, since the entire system is in a Haar-random state $\ket{\psi_{\HH}}$.

To obtain nontrivial physics, we must move away from this featureless equilibrium by introducing \emph{conditioning}. In higher-dimensional de Sitter gravity, the analog of such conditioning is to consider backgrounds with additional structure, such as conical defects (in dimension three) or Schwarzschild--de Sitter black holes (in $D>3$).

Recall from the general framework of \cref{sec:qm} that patch operators carry an additional label $P$, representing a conditioning. In the discussions so far, $P$ was implicitly taken to be the identity. More generally, $P\in M_{Z_L}(\mathbb{C})$ is a projection operator. After conditioning, the unnormalized state of the universe becomes $P\ket{\psi_{\HH}}$, while the operator algebra remains
\begin{align}
	\mathcal{A}_P = M_{Z_L}(\mathbb{C}).
\end{align}
In particular, the normalized trace appearing in \cref{eq:rule_PI} is always taken in $M_{Z_L}(\mathbb{C})$, independent of $P$. Consequently,
\begin{align}
	\tr(P) &= \frac{r_{\scriptscriptstyle P}}{Z_L},
	\label{eq:trace_conditioning}\\
	\mathcal{O}_{P,X} &= \mathcal{O}_{PXP}
	= \bra{\psi_{\HH}}PXP\ket{\psi_{\HH}} .
\end{align}

As emphasized in \cref{sec:qm_ob}, all physics inside the universe is encoded in the maps
$\mathcal{M}_P$ defined in \cref{eq:map,eq:map_alpha}. In the present toy model,
they becomes
%
\begin{align}
	\mathcal{M}_{P}:\ M_{Z_L}(\mathbb{C})
	&\longrightarrow M(\Omega_{\mathrm{U}},\mathbb{C}),
	&
	X &\longmapsto
	\bra{\psi_{\HH}}PXP\ket{\psi_{\HH}},
	\label{eq:map_toy}\\
	(\mathcal{M}_{P})_\alpha:\ M_{Z_L}(\mathbb{C})
	&\longrightarrow \mathbb{C},
	&
	X &\longmapsto
	\bra{(\psi_{\HH})_\alpha}PXP\ket{(\psi_{\HH})_\alpha}.
	\label{eq:map_alpha_toy}
\end{align}

$(\mathcal{M}_P)_\alpha$ specifies a quantum state
$P\ket{(\psi_{\HH})_\alpha}$, while $\mathcal{M}_P$ defines a classical ensemble of such states,
with the distribution inherited from the Haar measure on $\psi_{\HH}$.\footnote{These states are
unnormalized and defined only up to an overall phase.}

Expectation values are therefore given by
\begin{align}
	\expval{X}_P
	\equiv \frac{\mathcal{O}_{P,X}}{\mathcal{O}_{P,I}}
	= \frac{\bra{\psi_{\HH}}PXP\ket{\psi_{\HH}}}
	{\bra{\psi_{\HH}}P\ket{\psi_{\HH}}},
	\qquad X\in M_{Z_L}(\mathbb{C}).
\end{align}

We now analyze the physical effect of a nontrivial conditioning $P$. By a change of basis, $P$
can always be written as
\begin{align}
	P = \mathrm{diag}(0,\ldots,0,\overbrace{1,\ldots,1}^{r_{\scriptscriptstyle P}}).
	\label{eq:condition_basis}
\end{align}
In this basis, $P\ket{\psi_{\HH}}$ is manifestly an unnormalized Haar-random state in a
$r_{\scriptscriptstyle P}$-dimensional subspace. One might therefore expect conditioning to simply
replace $Z_L$ by $r_{\scriptscriptstyle P}$. However, the crucial point is that the operator
algebra remains fixed: $\mathcal{A}_P=\mathcal{A}_{\mathrm{U}}=M_{Z_L}(\mathbb{C})$. 

In fact, the moment-generating function of the unnormalized conditioned density matrix
$\Sigma_P\equiv P\Sigma_{\mathrm{U}}P$ is
\begin{align}
	\overline{\exp[\Tr(S\Sigma_P)]}
	= \det(I_{Z_L}-PSP)^{-1},
	\label{eq:MGF_P}
\end{align}
which differs from \cref{eq:MGF_U}. Equivalently,
\begin{align}
	\Sigma_P \sim \mathcal{W}_{Z_L}^{\mathbb{C}}(1,P).
	\label{eq:prob_P}
\end{align}

The physical state after conditioning is described by the normalized density matrix
$\rho_{\scriptscriptstyle P}\coloneq \Sigma_P/\mathcal{O}_{P,I}$. From
\cref{eq:MGF_P,eq:prob_P}, it satisfies
\begin{align}
	\overline{\rho_{\scriptscriptstyle P}}
	&= \frac{1}{r_{\scriptscriptstyle P}}\,P,\label{eq:density_UP_1}
\\
	\overline{\rho_{\scriptscriptstyle P}\otimes\rho_{\scriptscriptstyle P}}
	&= \frac{I_{Z_L^2}+\hat S_{Z_L^2}}{r_{\scriptscriptstyle P}(r_{\scriptscriptstyle P}+1)}\,P\otimes P ,\label{eq:density_UP_2}
\end{align}
where $I_{Z_L^2}$ is the identity operator and $\hat S_{Z_L^2}$ is the swap operator on two copies of the system.
Consequently, for any $X\in M_{Z_L}(\mathbb{C})$,
\begin{equation}
\begin{aligned}
	\overline{\expval{X}_P}
	&= \frac{1}{r_{\scriptscriptstyle P}}\Tr(PXP),\\
	\mathrm{Var}(\expval{X}_P)
	&= \frac{1}{r_{\scriptscriptstyle P}+1}
	\Big[
	\frac{1}{r_{\scriptscriptstyle P}}\Tr(PXPXP)
	- \Big(\frac{1}{r_{\scriptscriptstyle P}}\Tr(PXP)\Big)^2
	\Big].
\end{aligned}
\label{eq:exp_UP}
\end{equation}

For a typical operator $X$ of nearly full rank, $\Tr(PXP)\sim r_{\scriptscriptstyle P}$, so the
ensemble fluctuations are suppressed by $\sqrt{1/r_{\scriptscriptstyle P}}$. As we will show in
\cref{sec:dS_entropy}, $r_{\scriptscriptstyle P}$ is precisely the mean value of the loop
partition function in the conditioned background. Thus, fluctuations are again controlled by the
background partition function, in agreement with the general discussion of \cref{sec:dS}.

One might worry about the regime $r_{\scriptscriptstyle P}\sim O(1)$, where fluctuations become
large. This is analogous to the low-rank projection regime encountered in
\cref{sec:subsystems}. Physically, such a regime does not arise for Schwarzschild--de Sitter
geometries, whose partition functions are bounded below by that of the Nariai solution. Even in
this toy model, achieving $r_{\scriptscriptstyle P}\sim O(1)$ would require conditioning on almost
the entire universe, which no realistic observer could perform. This regime is therefore
unphysical.

\subsubsection{De Sitter entropy as coarse-grained entropy}
\label{sec:dS_entropy}

By \emph{de Sitter entropy} $S_{\text{dS}}$ (or $S_{\text{dS},P}$ when conditioning is included), we mean the logarithm of the Euclidean partition function associated with a given classical background, with or without conditioning. In higher-dimensional gravitational theories that admit de Sitter solutions, this quantity is given by the horizon area. Although it has long been conjectured to represent an entropy, its precise microscopic interpretation is not fully understood.

In the present toy model, de Sitter entropy admits a simple and explicit interpretation.
As shown in \cref{sec:unify}, when $Z_L$ is an integer it equals the dimension of the largest
quantum-mechanical system that can be embedded in this de Sitter space.
Equivalently, $Z_L$ is the number of orthogonal states in the Hilbert space
$\mathcal{H}_{\mathrm{U}}$.
Since the Hartle--Hawking state $\ket{\psi_{\HH}}\in\mathcal{H}_{\mathrm{U}}$ is Haar-random,
its coarse-grained entropy is
\begin{align}
	S_{\mathrm{cg}} = \log Z_L .
\end{align}
On the other hand, in the absence of conditioning, $Z_L$ is also the mean value of the loop
partition function (see \cref{eq:prob_spin_identity,eq:prob_general_identity,eq:norm_square}). Thus, in this model, the analog of de Sitter entropy is precisely
\begin{align}
	S_{\text{dS}}=\log Z_L = S_{\mathrm{cg}}.
\end{align}

Now consider a more general background obtained by conditioning on a projection $P$.
The patch operator corresponding to the identity is
\begin{align}
	\mathcal{O}_{P,I} = \|P\ket{\psi_{\HH}}\|^2 .
\end{align}
Using \cref{eq:trace_conditioning,eq:rule_PI}, its distribution is
\begin{align}
	\mathcal{O}_{P,I} \sim \text{Gamma}(r_{\scriptscriptstyle P},1),
	\label{eq:loop_partition_new}
\end{align}
where $r_{\scriptscriptstyle P}=\operatorname{rank}(P)$. Note that $\mathcal{O}_{P,I}$ plays the role of the loop partition function of the conditioned
background, with mean value $r_{\scriptscriptstyle P}$.
For large $r_{\scriptscriptstyle P}$, the associated effective de Sitter entropy is therefore
\begin{align}
	S_{\text{dS}, P} = \log r_{\scriptscriptstyle P}.
\end{align}
On the other hand, from \cref{eq:condition_basis}, $P\ket{\psi_{\HH}}$ is a Haar-random state in the subspace
$P\mathcal{H}_{\mathrm{U}}$, which has dimension $r_{\scriptscriptstyle P}$.
The coarse-grained entropy of this conditioned state is therefore also 
\begin{align}
	S_{\text{cg},P}=\log r_{\scriptscriptstyle P} = S_{\text{dS},P}.
\end{align}
We conclude that, in this toy model, de Sitter entropy is precisely the coarse-grained entropy
of the underlying quantum state.

Finally, note that
\begin{align}
	\log\!\big(\overline{\mathcal{O}_{P,I}}\big)
	= \log\!\big(\overline{\mathcal{O}_{I}}\big)
	- \log\!\left(
	\frac{\operatorname{rank}(I_{Z_L})}{\operatorname{rank}(P)}
	\right)
	< \log\!\big(\overline{\mathcal{O}_{I}}\big).
	\label{eq:area_change}
\end{align}
In higher-dimensional de Sitter gravity, introducing features such as defects or black holes reduces the horizon area and hence the entropy. Equation~\eqref{eq:area_change} provides a direct analog of this effect in the toy model: conditioning on a random state reduces its coarse-grained entropy and, correspondingly, lowers the effective de Sitter entropy.

\subsubsection{Hartle--Hawking wave function of the universe}
\label{sec:Hartle_Hawking}

Historically, the Hartle--Hawking wave function evaluated on a surface $\mathcal{C}$ is defined via a gravitational path integral, obtained by summing over all manifolds that end on $\mathcal{C}$ \cite{Hartle:1983ai}. A natural question in our construction is therefore: what is the wave function of the universe under no-boundary conditions? A complication specific to our toy model is that a spatial slice of $1+0$-dimensional de Sitter space consists of two disconnected points. If one is interested only in an observer’s experience, it suffices to consider single-patch operators (see the discussion above \cref{eq:one_patch_only}). However, to discuss the wave function of the universe itself, one needs to consider products of two patch operators. This feature does not arise in higher-dimensional models and is discussed here solely to facilitate comparison with the Hartle--Hawking wave function. Except in this subsection, we will always restrict attention to one-patch operators.

A spatial slice of this de Sitter space contains two points, denoted $L$ and $R$. We define
\begin{align}
	\ket{\phi_{L},\phi_{R}}\equiv \ket{\phi_L}\ket{\phi_R}, \qquad 
	\ket{\phi_L},\ket{\phi_R}\in \mathcal{H}_{\mathrm{U}} .
\end{align}
The wave function of the universe is then
\begin{align}
	\Psi_{\HH}[\phi_L,\phi_R]\equiv \bra{\phi_L,\phi_R}\ket{\HH},
\end{align}
where we use standard bra--ket notation for $\ket{\HH}$ since it denotes a genuine quantum state in the bulk Hilbert space.
We first note that
\begin{align}
	\Psi_{\HH}[\tilde\phi_L,\tilde\phi_R]^*\;\Psi_{\HH}[\phi_L,\phi_R]
	&= \mathcal{O}_{|\tilde\phi_L\rangle\mkern-1.6mu\bra{\phi_L}}\,
	   \mathcal{O}_{\qty(|\tilde\phi_R\rangle\mkern-1.6mu\bra{\phi_R})^T}
	= \mathcal{O}_{|\tilde\phi_L\rangle\mkern-1.6mu\bra{\phi_L}}\,
	   \mathcal{O}_{\ket{\phi_R^*}\mkern-1.6mu\bra{\tilde\phi_R^*}} .
\end{align}

Using \cref{eq:HH_exp_0}, we obtain
\begin{equation}
\begin{aligned}
	\Psi_{\HH}[\tilde\phi_L,\tilde\phi_R]^*\;\Psi_{\HH}[\phi_L,\phi_R]
	=\ &\mathcal{O}_{|\tilde\phi_L\rangle\mkern-1.6mu\langle\phi_L|}\,
	     \mathcal{O}_{\ket{\phi_R^*}\mkern-1.6mu\bra{\tilde\phi_R^*}} \\
	=\ &\bra{\psi_{\HH}}\ket{\tilde\phi_L}\bra{\phi_L}\ket{\psi_{\HH}}\,
	     \bra{\psi_{\HH}}\ket{\phi_R^*}\bra{\tilde\phi_R^*}\ket{\psi_{\HH}} \\
	=\ &\bra{\psi_{\HH},\psi_{\HH}^*}\ket{\tilde\phi_L,\tilde\phi_R}\,
	     \bra{\phi_L,\phi_R}\ket{\psi_{\HH},\psi_{\HH}^*} .
\end{aligned}
\label{eq:wave_function_square}
\end{equation}

From \cref{eq:wave_function_square}, we identify the wave function of the universe as
\begin{equation}
\begin{aligned}
	\Psi_{\HH}[\phi_L,\phi_R]
	=\ &\bra{\phi_L,\phi_R}\ket{\psi_{\HH},\psi_{\HH}^*} \\
	=\ &\bra{\phi_L}\ket{\psi_{\HH}}\mkern-4mu\bra{\psi_{\HH}}\ket{\phi_R^*} .
\end{aligned}
\label{eq:wave_function_HH}
\end{equation}

This expression admits a simple intuitive interpretation, illustrated schematically as
\begin{equation}
	\adjincludegraphics[width=1.1in,valign=m]{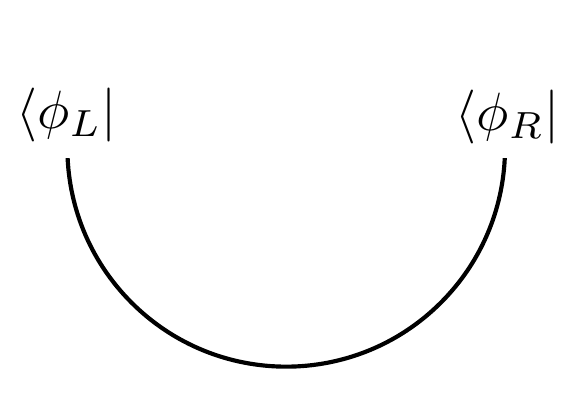}
	\quad\Longleftrightarrow\quad
	\adjincludegraphics[height=1.0in,valign=m]{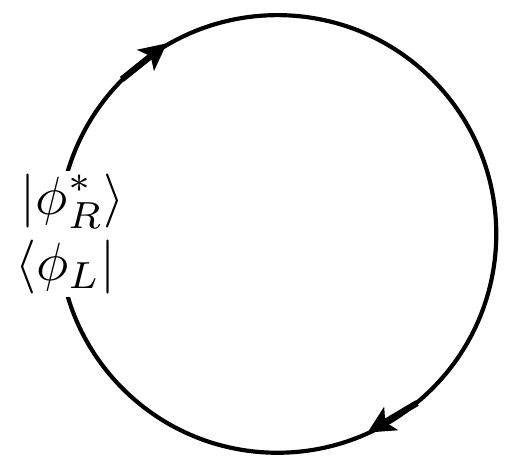}
	\quad\Longleftrightarrow\ 
	\adjincludegraphics[height=1.0in,valign=m]{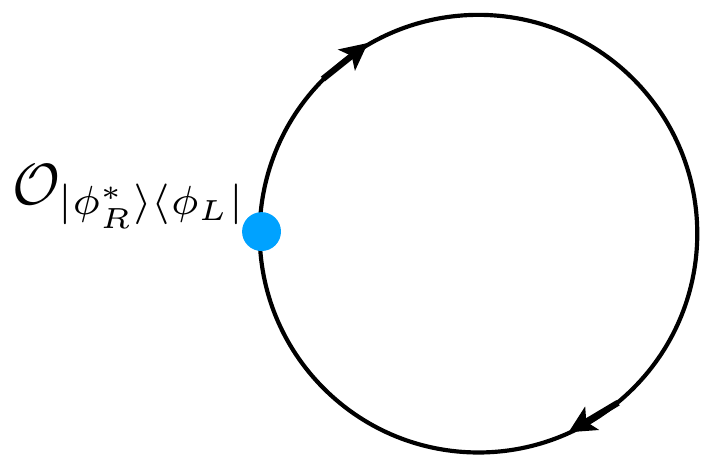} .
\label{eq:HH_picture}
\end{equation}
In the final diagram, the patch operator $\mathcal{O}_{|\phi_R^*\rangle\langle\phi_L|}$ precisely reproduces the wave function~\eqref{eq:wave_function_HH}.

The averaged value of the wave function is
\begin{align}
	\overline{\Psi_{\HH}[\phi_L,\phi_R]}
	= \phi_L^\dagger \,\phi_R^*
	= \bra{\phi_L}\ket{\phi_R^*} ,
\label{eq:wave_function_exp_HH}
\end{align}
and its variance is
\begin{align}
	\overline{\qty|\Psi_{\HH}[\phi_L,\phi_R]-\overline{\Psi_{\HH}[\phi_L,\phi_R]}|^2}
	= \|\phi_L\|^2 \|\phi_R\|^2 .
\label{eq:wave_function_HH_fluc}
\end{align}

Equation~\eqref{eq:wave_function_HH} gives the Hartle–Hawking wave function of the universe in this model. As illustrated by the figures in~\eqref{eq:HH_picture}, following the original Hartle--Hawking prescription~\cite{Hartle:1983ai} yields the averaged quantity~\eqref{eq:wave_function_exp_HH}. From \cref{eq:wave_function_HH_fluc}, one can see that in this simple toy model the fluctuations around the average are of order one.\footnote{This occurs as an artifact of this simple toy model. We do not expect it to occur in more realistic models, where a single classical background corresponds to a large number of microstates.}

The normalized wave function is given by
\begin{align}
	\widehat\Psi_{\HH}[\phi_L,\phi_R] = \frac{\mathcal{O}_{|\phi_R^*\rangle\mkern-1.6mu\langle\phi_L|}}{\mathcal{O}_I} = \frac{\bra{\phi_L}\ket{\psi_{\HH}}\mkern-4mu\langle\psi_{\HH}|\phi_R^*\rangle}{\bra{\psi_{\HH}}\ket{\psi_{\HH}}}, \label{eq:normalization_HH}
\end{align}
and it's easy to check that the normalization condition
\begin{align}
	\int d\phi_L \;d\phi_R \;\widehat\Psi_{\HH}[\phi_L,\phi_R]^*\;\widehat\Psi_{\HH}[\phi_L,\phi_R]=1 
\end{align}
holds exactly in each $\alpha$-sector.

Note that the normalization factor $\mathcal{O}_I$ does not depend on $\phi_{L/R}$. Moreover, from \eqref{eq:norm_square} its distribution is $\mathcal{O}_I\sim\text{Gamma}(Z_L,1)$, which becomes sharply peaked at large $Z_L$, with mean $\overline{\mathcal{O}_I}=Z_L$ and relative fluctuations of order $1/\sqrt{Z_L}$. Therefore, with high precision at large $Z_L$, the unnormalized wave function $\Psi_{\HH}$ differs from the normalized wave function $\widehat\Psi_{\HH}$ only by an overall $\phi_{L/R}\,$-independent factor, which depends on the $\alpha$-sector but is concentrated around the typical value $Z_L$.

Finally, note that the averaged wave function in \cref{eq:wave_function_exp_HH} can be written as
\begin{align}
	\overline{\Psi_{\HH}[\phi_L,\phi_R]}
	\equiv \langle\phi_L,\phi_R\overline{\ket{\HH}}
	= \bra{\phi_L}\ket{\phi_R^*} .
\label{eq:wave_function_exp_HH_2}
\end{align}
This expression suggests an averaged state of the form
\begin{align}
	\overline{\ket{\HH}}
	= \sum_{i=1}^{Z_L} \ket{i}_L \ket{i}_R ,
\end{align}
which is maximally entangled between the left and right points. On the other hand, our analysis and \cref{eq:wave_function_HH} show that within each $\alpha$ sector the underlying state is actually a product state,
\begin{align}
	\ket{\HH}_\alpha
	= \ket{(\psi_{\HH})_\alpha}_L \ket{(\psi_{\HH})_\alpha^*}_R .
\end{align}
The maximally entangled structure arises only after ensemble averaging and is absent in any fixed $\alpha$-sector. This indicates that ensemble averaging plays an essential role in this model: the resulting bulk physics is not merely a good approximation to that of any particular $\alpha$, but instead exhibits qualitative features that emerge only after averaging. We return to this point in \cref{sec:dis}.

\subsection{$\mathcal{CRT}$-invariant description}
\label{sec:CRT}

In \cref{sec:subsystems,sec:QM_U} we studied the emergence of quantum mechanics in this de Sitter space and recovered standard quantum mechanics in a complex Hilbert space $\mathcal{H}_{\mathrm{U}}$ of dimension $Z_L$. Throughout that analysis, we implicitly gauge fixed $\mathcal{CRT}$ by choosing a preferred direction of ambient time, as discussed in \cref{sec:one_bit,sec:set_up}. In this section, we present a $\mathcal{CRT}$-invariant formulation of the toy de Sitter model and show how the same physics can be represented in a real quantum code.

\subsubsection{A real Hilbert space description}
\label{sec:real_Hilbert_space}

As emphasized in \cite{Harlow:2023hjb}, the underlying $\mathcal{CRT}$-invariant Hilbert space of a closed universe must be real. It is well known that a complex Hilbert space of dimension $Z_L$ is equivalent to a real Hilbert space $\mathcal{H}_{\mathbb{R}}$ of dimension $2Z_L$ equipped with a compatible complex structure $J$. In this subsection we construct such a real description explicitly for the toy de Sitter model.

Let $\ket{\pm\hat y}_0\in\mathbb{C}^2$ denote the two states representing the two directions of ambient time, as introduced in \cref{sec:one_bit}. We enlarge the Hilbert space
$\mathcal{H}_{\mathrm{U}}\cong\mathbb{C}^{Z_L}$ to
$\mathbb{C}^2\otimes\mathcal{H}_{\mathrm{U}}\cong\mathbb{C}^{2Z_L}$ and define an embedding
\begin{equation}
\begin{aligned}
	\mathcal{E}:\ \mathcal{H}_{\mathrm{U}}
	&\longrightarrow \mathbb{C}^2\otimes\mathcal{H}_{\mathrm{U}},\\
	\ket{\psi}
	&\longmapsto
	\frac{1}{\sqrt{2}}
	\big(
	\ket{-\hat y}_0\ket{\psi}
	+\ket{+\hat y}_0\ket{\psi^*}
	\big).
\end{aligned}
\label{eq:map_to_real}
\end{equation}
We take $\mathcal{CRT}$ to exchange $\ket{+\hat y}_0$ and $\ket{-\hat y}_0$ and to act on
$\mathcal{H}_{\mathrm{U}}$ by complex conjugation.\footnote{With the phase convention
$\ket{-\hat y}_0=\frac{1}{\sqrt{2}}(1,-i)^T$ and
$\ket{+\hat y}_0=\frac{1}{\sqrt{2}}(1,i)^T$ in the $z$ basis, the two states are themselves
related by complex conjugation.}
With this choice, $\mathcal{E}\ket{\psi}$ is manifestly $\mathcal{CRT}$ invariant.

We define the real Hilbert space
\begin{align}
	\mathcal{H}_{\mathbb{R}}\coloneq \mathcal{E}\mathcal{H}_{\mathrm{U}},
\end{align}
with inner product induced from $\mathbb{C}^2\otimes\mathcal{H}_{\mathrm{U}}$.
The map $\mathcal{E}$ is real-linear but not complex-linear.
Let
\begin{align}
	J \coloneq -i(\sigma_y)_0\otimes I_{\mathcal{H}_{\mathrm{U}}}.
\end{align}
Then $J$ defines a complex structure on $\mathcal{H}_{\mathbb{R}}$, and we have
\begin{equation}
\begin{aligned}
	\mathcal{E}\big((a+ib)\ket{\psi}\big)
	&= a\,\mathcal{E}\ket{\psi}+b\,J\mathcal{E}\ket{\psi},
	\qquad a,b\in\mathbb{R},\\
	\braket{\phi}{\psi}_{\mathcal{H}_{\mathrm{U}}}
	&=
	\braket{\mathcal{E}\phi}{\mathcal{E}\psi}_{\mathcal{H}_{\mathbb{R}}}
	- i\braket{\mathcal{E}\phi}{J\mathcal{E}\psi}_{\mathcal{H}_{\mathbb{R}}}.
\end{aligned}
\label{eq:state_real}
\end{equation}

Operators are treated similarly. For
$X=X_1+iX_2\in M_{Z_L}(\mathbb{C})$ with
$X_1,X_2\in M_{Z_L}(\mathbb{R})$, define
\begin{align}
	\mathcal{F}(X)
	&\coloneq (I_2)_0\otimes X_1
	- i(\sigma_y)_0\otimes X_2 .
	\label{eq:operator_to_rel}
\end{align}
The map $\mathcal{F}$ preserves the operator algebra,
\begin{equation}
\begin{aligned}
	\mathcal{F}(I_{\mathcal{H}_{\mathrm{U}}})
	&= I_{\mathcal{H}_{\mathbb{R}}},\\
	\mathcal{F}((a+ib)X)
	&= a\,\mathcal{F}(X)+b\,J\mathcal{F}(X),\\
	\mathcal{F}(X)\mathcal{F}(Y)
	&= \mathcal{F}(XY),
\end{aligned}
\label{eq:operator_real}
\end{equation}
and is compatible with the embedding of states,
\begin{equation}
\begin{aligned}
	\mathcal{F}(X)\mathcal{E}\ket{\psi}
	&= \mathcal{E}(X\ket{\psi}),\\
	\bra{\phi}X\ket{\psi}_{\mathcal{H}_{\mathrm{U}}}
	&=
	\bra{\mathcal{E}\phi}\mathcal{F}(X)\ket{\mathcal{E}\psi}_{\mathcal{H}_{\mathbb{R}}}
	- i\bra{\mathcal{E}\phi}\mathcal{F}(X) J\ket{\mathcal{E}\psi}_{\mathcal{H}_{\mathbb{R}}}.
\end{aligned}
\label{eq:operator_state_real}
\end{equation}

\cref{eq:state_real,eq:operator_real,eq:operator_state_real}
show that quantum mechanics on the complex Hilbert space
$\mathcal{H}_{\mathrm{U}}$ is completely encoded in the real Hilbert space
$\mathcal{H}_{\mathbb{R}}$ equipped with the complex structure $J$.
The formulation in terms of $(\mathcal{H}_{\mathbb{R}},J)$ is manifestly
$\mathcal{CRT}$-invariant.

\subsubsection{Real quantum code}
\label{sec:code_real}

To make the real, $\mathcal{CRT}$-invariant description explicit in our example, note that the
embedding map~\eqref{eq:map_to_real} applied to the Hartle--Hawking state can be rewritten as
\begin{align}
	\mathcal{E}\ket{\psi_{\HH}}
	&=
	\frac{1}{\sqrt{2}}
	\Big(
	\ket{+\hat z}_0\,\frac{\ket{\psi_{\HH}}+\ket{\psi_{\HH}^*}}{\sqrt{2}}
	+\ket{-\hat z}_0\,\frac{\ket{\psi_{\HH}}-\ket{\psi_{\HH}^*}}{\sqrt{2}i}
	\Big).
\end{align}
An unnormalized Haar-random state $\ket{\psi_{\HH}}\in\mathbb{C}^{Z_L}$ can be specified by
$Z_L$ independent complex Gaussian variables $(\psi_{\HH})_i$, which can be written as
\begin{equation}
\begin{aligned}
	(\psi_{\HH})_i &= \frac{1}{\sqrt{2}}(u_i+iv_i),\\
	u_i,\,v_i &\overset{\text{i.i.d.}}{\sim}\mathcal{N}(0,1),
	\qquad i=1,\ldots,Z_L .
\end{aligned}
\label{eq:distribution_real}
\end{equation}
Treating $u\equiv(u_1,\ldots,u_{Z_L})^T$ and $v\equiv(v_1,\ldots,v_{Z_L})^T$ as real vectors, the embedded state
takes the simple form
\begin{align}
	\mathcal{E}\ket{\psi_{\HH}}
	= \frac{1}{\sqrt{2}}
	\big(
	\ket{+\hat z}_0\ket{u}
	+\ket{-\hat z}_0\ket{v}
	\big).
	\label{eq:state_real_2}
\end{align}
Thus, up to an overall normalization, $\mathcal{E}\ket{\psi_{\HH}}$ is a Gaussian random vector
in the real Hilbert space $\mathcal{H}_{\mathbb{R}}$.

Such a state can be explicitly represented using a real-valued quantum code,
\begin{equation}
	\mathcal{E}\ket{\psi_{\HH}}
	= \adjincludegraphics[width=0.48in,valign=m]{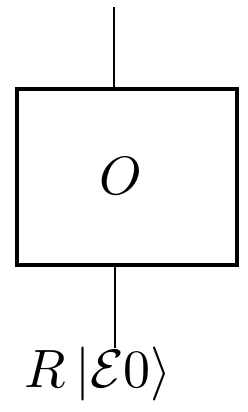}\ ,
	\label{eq:code_state_real}
\end{equation}
where $O\sim\text{Haar}\big(SO(2Z_L)\big)$ and
$R\sim\frac{1}{\sqrt{2}}\chi_{2Z_L}$. Here $\ket{0}$ is a fixed normalized state in $\mathcal{H}_{\mathrm{U}}$, and $\ket{\mathcal{E}0}$ denotes its image under $\mathcal{E}$.

Also note that for a Hermitian operator $X$, the corresponding operator $\mathcal{F}(X)$ defined in \cref{eq:operator_to_rel} is real and symmetric.

With these identifications, the quantum code in \cref{eq:code} now becomes
\begin{equation}
	\mathcal{O}_{P,X}
	= \adjincludegraphics[width=0.62in,valign=m]{code_1.pdf}
	= \adjincludegraphics[height=2.2in,valign=m]{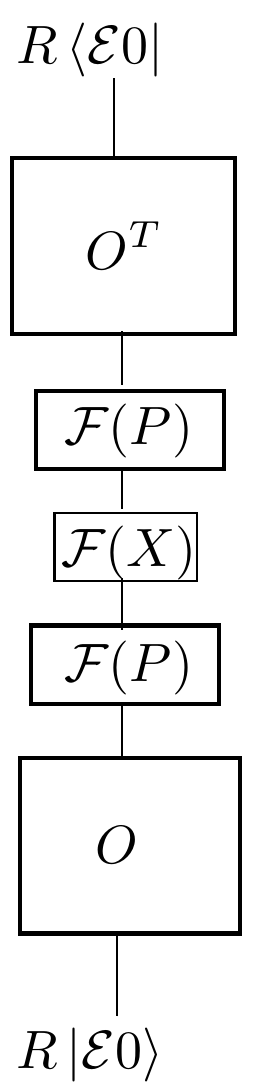},
	\qquad
	\expval{X}_P
	= \adjincludegraphics[width=0.76in,valign=m]{exp_X_2.pdf}
	= \adjincludegraphics[height=2.2in,valign=m]{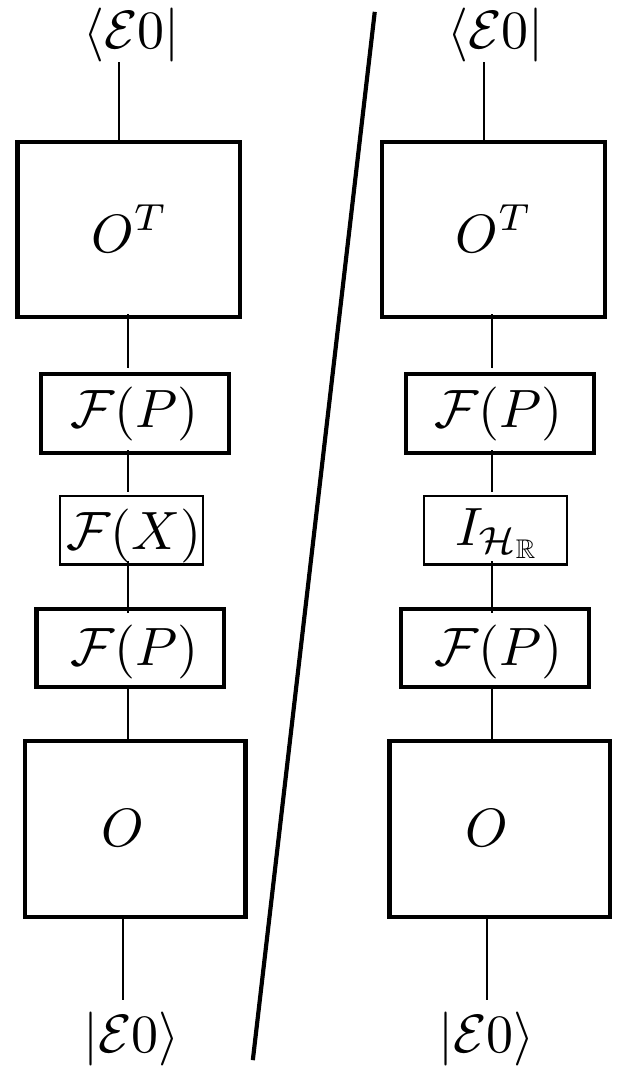}.
	\label{eq:code_real}
\end{equation}
The codes in \eqref{eq:code_state_real} and \eqref{eq:code_real} are manifestly real and
$\mathcal{CRT}$-invariant.

\subsection{``It from Bit'': Part II --- Baby-universe Hilbert space and de Sitter quantum mechanics}
\label{sec:baby_more}

With a concrete example in hand, we now return to the relationship and distinction between the quantum mechanics experienced by observers \emph{inside} de Sitter space and the  baby-universes Hilbert space, briefly discussed in \cref{sec:baby_1}.

\subsubsection{``Bit'' sector: Baby-universe Hilbert space as classical statistics}
Physically, baby-universe Hilbert space is constructed through gravitational path integral \cite{Marolf:2020xie}. In gravitational path integral, one can diagonalize a complete set of commuting boundary conditions and their simultaneous eigenstates are called $\alpha$-sectors. In other words, in each $\alpha$-sector all such boundary conditions take fixed values. The baby-universe Hilbert space $\mathcal{H}_{\mathrm{BU}}$ is a vector space spanned by these $\alpha$'s. In what follows, we first present the mathematical construction and then explain its physical meaning.

\paragraph{Hilbert space.}
Let $\Omega$ be the space of $\alpha$-parameters with probability distribution $p_{\alpha}$, and consider the Hilbert space of square-integrable functions on $\Omega$ with respect to measure $p_{\alpha}d\alpha$:
\begin{align}
	L^2(\Omega,p_{\alpha}d\alpha) \coloneq \Big\{|\mathcal{V}]\equiv \{\mathcal{V}_{\alpha}|\alpha\in\Omega\}:\!\int d\alpha p_{\alpha}\,|\mathcal{V}_{\alpha}|^2<\infty\Big\}\label{eq:baby_Hilbert_space}
\end{align}
with inner product between two vectors given by\footnote{We avoid calling elements of the baby-universe Hilbert space ``states,'' since we
do not perform quantum mechanics on this space and vector norms have physical meaning.} 
\begin{align}
	[\mathcal{V}_1|\mathcal{V}_2] = \int d\alpha p_{\alpha}\,\qty(\mathcal{V}_1)_{\alpha}^*\mkern1mu\qty(\mathcal{V}_2)_{\alpha}
	\label{eq:inner_baby}
\end{align}
As will be clear shortly, $L^2(\Omega, p_{\alpha}d\alpha)$ is the baby-universe Hilbert space $\mathcal{H}_{\mathrm{BU}}$ in our example. In a mathematically non-rigorous sense this Hilbert space is spanned by $\{|\alpha]:\alpha \in \Omega\}$.\footnote{This is like saying the position eigenstates $\ket{x}$ span the Hilbert space for a particle moving on a line, which is not mathematically accurate.} By definition in \eqref{eq:baby_Hilbert_space}, any vector $|\mathcal{V}]$ in this Hilbert space is specified by a set of numbers~$\{\mathcal{V}_{\alpha}:\alpha\in\Omega\}$. 

As a few examples, the vector $|\HH]$ corresponding to Hartle--Hawking no-boundary condition is specified by
\begin{align}
	\HH_{\alpha} = 1,\ \ \ \forall\alpha,
	\label{eq:HH_def}
\end{align}
reflecting the fact that the no-boundary insertion acts trivially in the path integral.

Inserting additional boundary condition $\mathcal{V}$ in the path integral produces the vector $\mathcal{V}|\HH]$ with components\footnote{$\mathcal{V}$ can be an asymptotic boundary as in AdS closed universe case in \cref{sec:AdS}, or a patch operator $\mathcal{O}_X$ as in our toy model.} 
\begin{align}
	(\mathcal{V}|\HH])_{\alpha} = \mathcal{V}_{\alpha}\HH_{\alpha} = \mathcal{V}_{\alpha}.
\end{align}

The $\alpha$-vectors are delta-function normalizable:
\begin{align}
	[\alpha'|\alpha] = \frac{1}{p_{\alpha}}\delta_{\alpha\alpha'}\label{eq:alpha_norm}.
\end{align}

\paragraph{Operator algebra.}
Because different $\alpha$-sectors do not mix, operators act by multiplication, thus an operator $\mathcal{O}$ is also specified by a set of numbers $\{\mathcal{O}_\alpha\}$ and acts as
$(\mathcal{O}|\mathcal{V}])_\alpha=\mathcal{O}_\alpha\mathcal{V}_\alpha$. The identity operator in this algebra is $|\HH][\HH|$.

To ensure $\mathcal{O}_1\cdots \mathcal{O}_k|\mathcal{V}]\in L^2(\Omega,p_\alpha d\alpha)$ for all
$|\mathcal{V}]$, it suffices that
\begin{align}
	\mathcal{O}\in \bigcap_{p<\infty} L^p(\Omega,p_\alpha d\alpha),\qquad
	L^p(\Omega,p_\alpha d\alpha)\coloneq
	\Big\{\{\mathcal{O}_\alpha\}:\int d\alpha\,p_\alpha\,|\mathcal{O}_\alpha|^p<\infty\Big\}.
	\label{eq:property_operator}
\end{align}
We will see shortly patch operators in the toy model satisfy this condition. Note that $\bigcap_{p<\infty}L^p(\Omega, p_{\alpha} d\alpha)$ is far from the full space of operators on $L^2(\Omega, p_{\alpha}d\alpha)$. It is a commutative subalgebra of multiplication operators. Again, this is the mathematical statement of one-state property, or no $\alpha$-mixing.

\paragraph{Path-integral interpretation.}
With vectors and operators in this Hilbert space defined above, now we have
\begin{align}
	[\mathcal{V}_1|\mathcal{O}_1\cdots \mathcal{O}_k|\mathcal{V}_2] = \int d\alpha p_{\alpha}\,\qty(\mathcal{V}_1)_{\alpha}^*\mkern1mu\qty(\mathcal{O}_1)_{\alpha}\cdots \qty(\mathcal{O}_k)_{\alpha}\mkern1mu\qty(\mathcal{V}_2)_{\alpha}\label{eq:exp_BUH}
\end{align}
When $\mathcal{V}_1$, $\mathcal{V}_2$, and the $\mathcal{O}_i$'s are boundary conditions of the gravitational path integral, the right-hand side is exactly the path-integral result,
\begin{align}
	\overline{\mathcal{V}_1^*\mkern1mu\mathcal{V}_2\mkern1mu\mathcal{O}_1\cdots \mathcal{O}_k}
	=
	\int d\alpha\,p_\alpha\,\qty(\mathcal{V}_1)_\alpha^*\mkern1mu\qty(\mathcal{V}_2)_\alpha\mkern1mu
	(\mathcal{O}_1)_\alpha\cdots (\mathcal{O}_k)_\alpha .
\end{align}
Hence
\begin{align}
	[\mathcal{V}_1|\mathcal{O}_1\cdots \mathcal{O}_k|\mathcal{V}_2]
	=
	\overline{\mathcal{V}_1^*\mkern1mu\mathcal{O}_1\cdots \mathcal{O}_k\mkern1mu\mathcal{V}_2},
	\label{eq:baby_GPI}
\end{align}
and in particular
\begin{align}
	[\HH|\mathcal{O}_1\cdots \mathcal{O}_k|\HH]
	=
	\overline{\mathcal{O}_1\cdots \mathcal{O}_k}.
	\label{eq:baby_GPI_HH}
\end{align}

Thus $\mathcal{H}_{\mathrm{BU}} = L^2(\Omega, p_{\alpha} d\alpha)$, equipped with the commutative algebra $\mathcal{A}_{\mathrm{BU}}$ of multiplication operators, reproduces all gravitational path-integral correlators. The pair $(\mathcal{H}_{\mathrm{BU}}, \mathcal{A}_{\mathrm{BU}})$ is therefore nothing more than a mathematical repackaging of these correlators. This is precisely the content of the one-state statement: there is no $\alpha$-mixing and no quantum mechanics on this Hilbert space. The baby-universe Hilbert space encodes only classical statistics of boundary conditions, including the statistics of patch operators.

\subsubsection{From statistics of patch operators (``Bit'') to quantum mechanics inside the de Sitter universe (``It'')}
\label{sec:HBU_to_HU}

As emphasized in \cref{sec:baby_1}, patch operators provide the bridge between the classical
statistics of baby universes and the quantum mechanics inside a single de Sitter universe.
This connection can be made fully explicit in our toy model.

In this model, the space of $\alpha$-parameters is
\begin{align}
	\Omega_{\mathrm{U}}
	= \mathbb{C}^{Z_L}/U(1)
	= \{\psi_{\HH}\in\mathbb{C}^{Z_L}:\psi_{\HH}\sim e^{i\theta}\psi_{\HH}\},
\end{align}
as given in \cref{eq:alpha_parameter_U}. The probability measure $p_\alpha\,d\alpha$ is induced by a
Gaussian measure on each component of $\psi_{\HH}$, as discussed in \cref{sec:unify}.

Boundary conditions in the gravitational path integral are specified by patch operators
$\mathcal{O}_X$ with $X\in M_{Z_L}(\mathbb{C})$.
In each $\alpha$-sector, the value of $\mathcal{O}_X$ is fixed:
\begin{align}
	(\mathcal{O}_X)_\alpha
	= \bra{(\psi_{\HH})_\alpha} X \ket{(\psi_{\HH})_\alpha}.
	\label{eq:value_patch_alpha}
\end{align}
Since the right-hand side of \cref{eq:value_patch_alpha} is polynomial in the Gaussian variables
$(\psi_{\HH})_i$ and the measure $p_\alpha\,d\alpha$ is exponentially decaying, all patch operators
satisfy the integrability condition in \cref{eq:property_operator}.
Therefore, in this toy model,
\begin{align}
	\mathcal{H}_{\mathrm{BU}} &= L^2(\Omega_{\mathrm{U}},\,p_\alpha\,d\alpha),\\
	\mathcal{A}_{\mathrm{BU}} &= \mathbb{C}\!\left[\{\mathcal{O}_X\}_{X\in M_{Z_L}(\mathbb{C})}\right],
\end{align}
where $\mathcal{A}_{\mathrm{BU}}$ is the commutative algebra generated by the patch operators
(all of which commute).\footnote{One can show that $|\HH]$, the vector corresponding to Hartle--Hawking no-boundary condition, is cyclic with respect to this algebra.}

As shown in \cref{sec:subsystems,sec:QM_U}, the statistics of the patch operators
$\{\mathcal{O}_X\}$ allow us to reconstruct the quantum mechanics \emph{inside} de Sitter space
under the Hartle--Hawking no-boundary condition.\footnote{This is quantum mechanics on $\mathcal{H}_{\mathrm{U}}$, not on $\mathcal{H}_{\mathrm{BU}}$.}
We now recast that construction in the language of the baby-universe Hilbert space.

Restricting \cref{eq:map_toy,eq:map_alpha_toy} to the case of $P=I$, we obtain
\begin{align}
	\mathcal{M}_{\HH}:\ \mathcal{A}_{\mathrm{U}}
	&\longrightarrow \mathcal{A}_{\mathrm{BU}},
	&
	X &\longmapsto \mathcal{O}_X,
	\label{eq:map_HH}\\
	(\mathcal{M}_{\HH})_\alpha:\ \mathcal{A}_{\mathrm{U}}
	&\longrightarrow \mathbb{C},
	&
	X &\longmapsto
	\bra{(\psi_{\HH})_\alpha}X\ket{(\psi_{\HH})_\alpha}.
	\label{eq:map_HH_alpha}
\end{align}
For each $\alpha$, the linear functional $(\mathcal{M}_{\HH})_\alpha$ specifies a quantum state
$\ket{(\psi_{\HH})_\alpha}$ in $\mathcal{H}_{\mathrm{U}}$.
The map $\mathcal{M}_{\HH}$ therefore gives a classical ensemble of such states, with
$\psi_{\HH}$ distributed according to the Gaussian measure described in \cref{sec:unify}.\footnote{
States are unnormalized and defined up to an overall phase.}
Expectation values are given by
\begin{align}
	\expval{X}_{\HH}
	= \frac{\mathcal{O}_X}{\mathcal{O}_I}
	= \frac{\bra{\psi_{\HH}}X\ket{\psi_{\HH}}}
	       {\bra{\psi_{\HH}}\ket{\psi_{\HH}}},
\end{align}
which fully characterizes the quantum mechanics inside de Sitter space under the Hartle--Hawking no-boundary condition.

We now consider boundary conditions beyond the no-boundary case.
In $\mathcal{H}_{\mathrm{BU}}$, choose a vector $|\mathcal{V}] = \mathcal{V}|\HH]$. Say, $\mathcal{V}$ can be a patch operator corresponding to some $Y_{\mathcal{V}}\in M_{Z_L}(\mathbb{C})$:
\begin{align}
	|\mathcal{V}] = \mathcal{V}|\HH] = \mathcal{O}_{Y_{\mathcal{V}}}|\HH],
\end{align}
Physically, this corresponds to inserting a patch operator $\mathcal{O}_{Y_\mathcal{V}}$ in the
gravitational path integral.
The claim is that such a vector $|\mathcal{V}]\in\mathcal{H}_{\mathrm{BU}}$ induces a new quantum state
$\ket{\psi_{\scriptscriptstyle \mathcal{V}}}\in \mathcal{H}_{\mathrm{U}}$ inside the de Sitter universe.

Analogously to the Hartle--Hawking case, define
\begin{align}
	\mathcal{M}_{\mathcal{V}}:\ \mathcal{A}_{\mathrm{U}}
	&\longrightarrow \mathcal{A}_{\mathrm{BU}},
	&
	X &\longmapsto \mathcal{V}^*\,\mathcal{O}_X\,\mathcal{V},
	\label{eq:map_lambda}\\
	(\mathcal{M}_{\mathcal{V}})_\alpha:\ \mathcal{A}_{\mathrm{U}}
	&\longrightarrow \mathbb{C},
	&
	X &\longmapsto
	|\mathcal{V}_\alpha|^2
	\bra{(\psi_{\HH})_\alpha}X\ket{(\psi_{\HH})_\alpha}.
	\label{eq:map_lambda_alpha}
\end{align}
These expressions have a direct path-integral interpretation: they coincide with gravitational
path integrals evaluated with additional boundary insertions
$\mathcal{O}_{Y_{\mathcal{V}}^*}$ and $\mathcal{O}_{Y_{\mathcal{V}}}$. In other words, the right hand side of \cref{eq:map_lambda_alpha} gives statistics of $\{\mathcal{O}_{Y_{\mathcal{V}}^*}\mathcal{O}_{X}\mathcal{O}_{Y_{\mathcal{V}}}:X\in \mathcal{A}_{\mathrm{U}}\}$.

For $|\mathcal{V}]\neq|\HH]$, the functional $(\mathcal{M}_{\mathcal{V}})_\alpha$ defines a different quantum state
\begin{align}
	\ket{(\psi_{\scriptscriptstyle\mathcal{V}})_\alpha}
	\equiv \mathcal{V}_\alpha \ket{(\psi_{\HH})_\alpha}
	\in \mathcal{H}_{\mathrm{U}},
\end{align}
and $\mathcal{M}_\mathcal{V}$ defines a classical ensemble of such states.

Expectation values are
\begin{align}
	\expval{X}_{\mathcal{V}}
	= \frac{\mathcal{V}^*\mathcal{O}_X\mathcal{V}}
	       {\mathcal{V}^*\mathcal{O}_I\mathcal{V}}
	= \frac{\bra{\psi_{\scriptscriptstyle\mathcal{V}}}X\ket{\psi_{\scriptscriptstyle\mathcal{V}}}}
	       {\bra{\psi_{\scriptscriptstyle\mathcal{V}}}\ket{\psi_{\scriptscriptstyle\mathcal{V}}}},
	\label{eq:exp_lambda}
\end{align}
which fully determines the quantum mechanics inside this de Sitter universe with the corresponding
boundary conditions.

The wave function of the universe under this modified boundary condition can be obtained analogously to the construction in \cref{sec:Hartle_Hawking}:
\begin{equation}
\begin{aligned}
	\Psi_{\mathcal{V}}[\phi_L,\phi_R]
	=\ & \bra{\phi_L,\phi_R}\ket{\psi_{\scriptscriptstyle\mathcal{V}},\psi_{\scriptscriptstyle\mathcal{V}}^*} \\
	=\ & \bra{\phi_L}\ket{\psi_{\scriptscriptstyle\mathcal{V}}}\mkern-4mu\bra{\psi_{\scriptscriptstyle\mathcal{V}}}\ket{\phi_R^*} .
\end{aligned}
\end{equation}

In summary, a vector $|\mathcal{V}]\in\mathcal{H}_{\mathrm{BU}}$ gives a state
$\ket{\psi_{\scriptscriptstyle\mathcal{V}}}$ in $\mathcal{H}_{\mathrm{U}}$, and hence a quantum state inside de Sitter space.
This highlights the sharp distinction between the two Hilbert spaces:
$\mathcal{H}_{\mathrm{BU}}$ supports only classical statistics (the one-state property),
while $\mathcal{H}_{\mathrm{U}}$ carries standard quantum mechanics.
Vectors in $\mathcal{H}_{\mathrm{BU}}$ do not represent quantum states themselves; rather,
they generate states in $\mathcal{H}_{\mathrm{U}}$ that encode the quantum mechanics
experienced by an observer inside de Sitter space. \cref{tab:BU_vs_U} is a comparison of some of their properties.

\begin{table}[!htbp]
\centering
\renewcommand{\arraystretch}{1.3}
\begin{tabular}{c c c}
\toprule
 & $\mathcal{H}_{\mathrm{BU}}$ & $\mathcal{H}_{\mathrm{U}}$ \\
\midrule
Hilbert space
 & $L^2(\Omega_{\mathrm{U}},\, p_\alpha\, d\alpha)$
 & $\mathbb{C}^{Z_L}$ \\
Dimension
 & $\infty$
 & $Z_L$ \\
Elements
 & vectors $|\mathcal{V}]$
 & quantum states $\ket{\psi_{\scriptscriptstyle\mathcal{V}}}$ \\
Operator algebra
 & $\mathbb{C}[\{\mathcal{O}_X\}_{X\in M_{Z_L}(\mathbb{C})}]$
 & $M_{Z_L}(\mathbb{C})$ \\
Identity operator
 & $|\HH][\HH|$
 & $\mathds{1}_{Z_L}$ \\
\bottomrule
\end{tabular}
\caption{Comparison between the baby-universe Hilbert space and de Sitter Hilbert space.}
\label{tab:BU_vs_U}
\end{table}

\section{Discussion}
\label{sec:dis}

\paragraph{Meaning of ensemble averaging}

Ensemble averaging is a standard tool in the study of complex physical systems. The traditional logic is that when a given system is too complicated to analyze directly, one instead studies a large ensemble of similar systems and focuses on their averaged behavior. If one is interested in a physical quantity $A$, the ensemble average $\overline{A}$ is computed, and as long as $A$ is self-averaging, $\overline{A}$ provides an accurate description of a typical system. A classic example is Wigner’s use of random-matrix models to describe the statistics of energy levels in heavy nuclei \cite{Wigner1955,Wigner1967}. More recently, the SYK model has been widely used to study aspects of black hole physics \cite{Kitaev2015KITP,Maldacena:2016hyu}. In such cases, the key point is that the observable $A$ is well-defined for each individual member of the ensemble; the ensemble average serves purely as a calculational convenience.

This logic does not apply in our setup. As a consequence of the one-state property of closed universes, an individual $\alpha$-sector contains no quantum mechanical structure and is nothing more than a set of numbers. A semi-classical description emerges only when one considers a large collection of $\alpha$-sectors and studies their statistical properties. In our toy model, the bulk theory is associated with the full $\alpha$-parameter space $\Omega_{\mathrm{U}}$ equipped with the probability measure $p_{\alpha}\, d\alpha$. If one restricts to a single value of $\alpha$, consisting of $2Z_L-1$ real numbers, the bulk picture ceases to exist. Unlike random-matrix models for nuclei or the SYK model for black holes, the ensemble here is not a calculational device acting on well-defined quantum systems; rather, physics in de Sitter space itself emerges only at the level of the ensemble.

\paragraph{Factorization in AdS/CFT}

The focus of this paper is the Hartle--Hawking state and de Sitter spacetime. In the context of AdS/CFT, the notion of a single closed-universe state is closely related to the factorization puzzle \cite{Maldacena:2004rf}. It is natural to ask whether any useful lessons can be drawn from our analysis.

A potential objection to our approach here is the absence of $\alpha$-parameters in higher-dimensional holographic theories such as supersymmetric Yang--Mills theory. However, in the model studied here, we do not require each $\alpha$-sector to define a holographic quantum mechanical theory on its own. Each $\alpha$-sector contains no quantum mechanical degrees of freedom and is characterized by $2Z_L-1$ real numbers. Taken together, the collection of $\alpha$-sectors instead forms a classical probability theory, which can be naturally packaged in the Hilbert-space language of baby universes.

In AdS/CFT, when open universes with asymptotic spatial boundaries exist, the holographic principle requires a genuine boundary quantum mechanical theory. However, this does not imply that ensemble averaging must involve many distinct holographic theories. One need not consider multiple boundary theories, each admitting a bulk dual with wormholes. Instead, apparent violations of factorization may emerge from classical statistical properties within a single holographic theory. Identifying such statistical structures is clearly challenging, but this viewpoint suggests an alternative to ensemble interpretations involving many different theories.

In the absence of many different theories, one may instead consider varying boundary conditions of bulk fields and ask whether wormholes can emerge from the classical statistics of these quantities. The rank of the gauge group $N$, discussed in \cite{Liu:2025cml,Kudler-Flam:2025cki,Liu:2025ikq}, provides one such example. However, if a structure analogous to that found here applies to AdS wormholes, a single continuous parameter is far from sufficient. Roughly speaking, the number of independent continuous parameters required should scale with the Hilbert space dimension associated with the wormhole. A single parameter such as $N$, even when treated as continuous, is therefore unlikely to support a semi-classical wormhole geometry.

\paragraph{AdS closed universes and evaporating black holes}

A natural question is whether the notion of patch operators can be applied to AdS closed universes or post-Page-time evaporating black holes. We consider two distinct scenarios.

In the first scenario, the evaporating black hole retains a large amount of bulk entanglement with its Hawking radiation \cite{Penington:2019npb, Almheiri:2019hni, Penington:2019kki, Almheiri:2019qdq}, or the AdS closed universe is highly entangled with a boundary CFT \cite{Antonini:2023hdh}. In this case, the black hole interior lies in the entanglement wedge of the radiation, while the AdS closed universe lies in the entanglement wedge of the boundary CFT. Entanglement wedge reconstruction can then be used to probe the physics of the black hole interior or the AdS closed universe \cite{Dong:2016eik, Almheiri:2019hni, Akers:2022qdl}.

In the second scenario, one performs a set of simple measurements on the radiation or the boundary CFT, removing the large amount of bulk entanglement. This case is more confusing. Nevertheless, we expect some of the lessons developed here to continue to apply. One of our main conclusions is that, in the absence of additional constraints, distinct patch operators all commute and have only classical correlations with one another; in particular, there is no quantum-mechanical structure in their mutual relations. From this perspective, the information specifying the quantum-mechanical structure must be encoded somewhere else, for example, in the definition of the patch operators themselves. We therefore expect that, for evaporating black holes and AdS closed universes, or more generally, for bulk spacetimes with known holographic duals, a more careful definition of patch operators may still reproduce semiclassical spacetime.

\paragraph{Future directions}

There are many directions for future work. The one-dimensional toy model studied in this paper is extremely simple and contains too little structure to be physically realistic. A natural next step is to construct more elaborate models that incorporate additional features, in particular to better understand the emergence of time in a cosmological setting.

Another important direction is to connect this framework more directly with cosmology. How might the ideas discussed here be related to our universe? Can they be applied in the context of slow-roll inflation \cite{Guth:1980zm,Linde:1981mu,Albrecht:1982wi}? Can they shed light on the Boltzmann brain problem \cite{Dyson:2002pf,Page:2006dt}, or help clarify the phenomenological challenges associated with the Hartle--Hawking wave function \cite{Maldacena:2024uhs}?

\section*{Acknowledgements}
I thank Daniel Harlow for initial collaboration and extensive discussions. I thank Ahmed Almheiri, Hong Liu, Juan Maldacena, Leonard Susskind, and Misha Usatyuk for helpful discussions. I also received very helpful comments from an anonymous referee. This work was supported by the John Templeton Foundation Award 41001491-013 (subagreement under prime agreement No. ID \# 63670), the DOE ``Quantised Award,'' and the ``Algebras and Complexity in Quantum Gravity'' award \#DE-SC0025937.

\appendix




\section{Wormhole effects for a bulk observer}
\label{app:wormhole_AdS}

In this appendix we review the consequence of wormhole corrections in the literature \cite{Coleman:1988cy,Giddings:1988cx,Polchinski:1994zs,Marolf:2020rpm,Marolf:2021ghr} and discuss their potential implications for observers in AdS closed universe. 

As discussed in \cite{Coleman:1988cy,Giddings:1988cx}, one effect of spacetime wormholes is the appearance of an ensemble of $\alpha$-sectors, in which couplings (or other parameters) take different values in different sectors. Crucially, these $\alpha$-sectors form a classical ensemble: distinct sectors do not mix. Once a bulk observer performs an experiment and infers that the parameters lie within some range, all subsequent observations will be consistent with that range.

One example involves black hole evaporation. Suppose a bulk observer prepares a black hole by collapsing matter in a pure state $\rho_i = \ket{\phi_i}\bra{\phi_i}$ and then lets it evaporate. Since the parameters of the theory depend on $\alpha$, a single experiment yields Hawking radiation in the mixed state
\begin{align}
\rho_f^{(1)} \;=\; \int d\alpha\, p_{\alpha}\, U_{\alpha}\rho_i U_{\alpha}^{\dagger}\, .
\end{align}
However, this does not mean that the observer would operationally witness information loss. With only one copy of the radiation, she cannot determine whether the state is pure or mixed unless she already knows which pure state to test against. Instead, she can repeat the experiment and collect the radiation from a second black hole prepared in identical initial state. The joint radiation state is then
\begin{align}
	\rho^{(2)}_f = \int d\alpha\, p_{\alpha}\, \bigl(U_{\alpha}\rho_i U_{\alpha}^{\dagger}\bigr)\otimes \bigl(U_{\alpha} \rho_i U_{\alpha}^{\dagger}\bigr)\, .
	\label{eq:BH_two}
\end{align}
Note that $\rho_f^{(2)}$ is not equal to $\rho_f^{(1)}\otimes \rho_f^{(1)}$. If the observer performs a swap test and measures the expectation value of the swap operator $S$, she finds $\Tr(S\rho^{(2)}_f)=1$ \cite{Marolf:2020rpm,Marolf:2021ghr}.

Alternatively, she could try to determine the state of the radiation by measuring it directly. A measurement on the first black hole would partially collapse the $\alpha$-distribution, i.e., the effective range of the $\alpha$-integration becomes narrower. A measurement on the second black hole would further collapse the distribution, and, in principle, after sufficiently many measurements she would effectively fix the relevant parameters and obtain radiation in a pure state $U_{\alpha}\rho_i U_{\alpha}^{\dagger}$ in all subsequent experiments. However, for this to happen the observer would need to determine $\mathcal{O}(e^{S_{\text{BH}}})$ parameters to high precision, which is not physically plausible.

Since spacetime wormholes introduce uncertainties in the parameters accessible to bulk observers, when wormhole effects are large (as in AdS closed universes) the resulting parameter distribution need not be sharply peaked. This alone does not imply inconsistencies among repeated experiments. However, the values one inferred after many experiments may deviate substantially from the predictions of effective field theory.

These issues in AdS closed universes can also be understood from a complementary perspective. To compute the expectation value of a patch operator $\mathcal{O}$ in this context, one needs both the patch operator and asymptotic Euclidean boundarie $\mathcal{Z}$, as in \cref{fig:AdS_state}(a) and \cref{eq:exp_bd_AdS}. However, the connection between $\mathcal{O}$ and $\mathcal{Z}$ is not guaranteed: the patch operator $\mathcal{O}$ can instead connect to other asymptotic boundaries, thereby probing a different bulk. In other words, attempts to impose a fixed boundary condition are not fully successful, and large fluctuations are unavoidable. As a result, for a bulk observer living in such a universe, the question of the ``initial state'' of her universe is not operationally meaningful; the best she can do is to infer the parameters of her universe through repeated experiments.

\section{Statistics of patch operators}

\subsection{Statistics of patch operators of a two-state classical system}
\label{app:classical}
In this appendix we study joint statistics of patch operators $\mathcal{O}_{\scriptscriptstyle (I_2)_0}$ and $\mathcal{O}_{\scriptscriptstyle (\sigma_y)_0}$ of a two-state classical system.

\subsubsection{Warm-up: Statistics of single patch operator}
\label{app:warmup_0}

As a warm-up, we first study the statistics of patch operators $\mathcal{O}_{\scriptscriptstyle (I_2)_0}$ and $\mathcal{O}_{\scriptscriptstyle (\sigma_y)_0}$ separately. 

Let's compute $\overline{\mathcal{O}_{\scriptscriptstyle (I_2)_0}^m}$ first. Say, there are $b_k$ loops each with $k$ patch operator $\mathcal{O}_{\scriptscriptstyle (I_2)_0}$ inserted. As we exclude vacuum bubbles, $k\geq 1$. $b_k\geq 0$.
\begin{align}
	 b_1+ 2b_2\cdots +k b_k+\cdots  =m
	\label{eq:condition}
\end{align}
Given $m$ and $\{b_k\}$ satisfying \cref{eq:condition}, let $N_1(m;\{b_k\})$ be the number of configurations, we have
\begin{align}
	\overline{\mathcal{O}_{\scriptscriptstyle (I_2)_0}^m} = \ &\sum_{b_1+\cdots +kb_k+\cdots  = m}Z_L^{b_1+\cdots +b_k}N_1(m;\{b_k\})
\end{align}
From some simple combinatorics, we have
\begin{align}
	N_1(m;\{b_k\}) =\ & \qty[{m \choose b_1}{b_1\choose 1}{b_1-1\choose 1}\cdots {1\choose 1}\frac{1}{b_1!}]\qty[{m-b_1 \choose 2b_2}{2b_2\choose 2}{2(b_2-1)\choose 2}\cdots {2\choose 2}\frac{1}{b_2!}\qty(\frac{2!}{2})^{b_2}]\cdots \nonumber\\
	&\ \ \ \ \ \ \qty[{m-b_1-2b_2-\cdots -(k-1)b_{k-1}\choose kb_k}{kb_k\choose k}{k(b_k-1)\choose k}..{k\choose k}\frac{1}{b_k!}\qty(\frac{k!}{k})^{b_k}]\cdots \nonumber\\
	=\ &\frac{m!}{b_1! b_2!\cdots b_k!\cdots }\frac{1}{1^{b_1} 2^{b_2}\cdots k^{b_k}\cdots }
	\label{eq:counting_1}
\end{align}

The moment generating function of $\mathcal{O}_{\scriptscriptstyle (I_2)_0}$ is given by
\begin{align}
	&\overline{\exp(s\mathcal{O}_{\scriptscriptstyle (I_2)_0})}=\sum_{m = 0}^{\infty}\frac{1}{m!}s^m\overline{\mathcal{O}_{\scriptscriptstyle (I_2)_0}^m}\nonumber\\
	=\ &\sum_{m = 0}^{\infty}\frac{1}{m!}s^m\sum_{b_1+\cdots +kb_k+\cdots  = m}Z_L^{b_1+\cdots +b_k}N_1(m;\{b_k\})\nonumber\\
	=\ &\sum_{\{b_k\geq0,k\geq 1\}} s^{b_1+\cdots +kb_k+\cdots }Z_L^{b_1+\cdots +b_k}\frac{1}{b_1! b_2!\cdots b_k!\cdots }\frac{1}{1^{b_1} 2^{b_2}\cdots k^{b_k}\cdots }\nonumber\\
	=\ &\prod_{k = 1}^{\infty}\qty(\sum_{b_k = 0}^{\infty}\frac{1}{b_k!}\qty(\frac{Z_L s^k}{k})^{b_k})\nonumber\\
		=\ &\frac{1}{\qty(1- s)^{Z_L}}
		\label{eq:MGF_I}
\end{align}
\cref{eq:MGF_I} is the moment generating function of a Gamma distribution. The probability density of $\mathcal{O}_{\scriptscriptstyle (I_2)_0}$ is given by
\begin{align}
	p_{\scriptscriptstyle\mathcal{O}_{\scriptscriptstyle (I_2)_0}}(x) = \frac{1}{\Gamma(Z_L)}x^{Z_L-1}e^{-x}\Theta(x)
\end{align}

The statistics of $\mathcal{O}_{\scriptscriptstyle (\sigma_y)_0}$ can be similarly computed. The only difference is that each circle needs to have even number of $\mathcal{O}_{\scriptscriptstyle (\sigma_y)_0}$ inserted. Consider $\overline{\mathcal{O}_{\scriptscriptstyle (\sigma_y)_0}^{2n}}$. Say, there are $b_k$ loops each with $2k$ $\mathcal{O}_{\scriptscriptstyle (\sigma_y)_0}$ inserted. 
\begin{align}
	2b_1+ 4b_2\cdots +2k b_k+\cdots  =2n
\end{align}
Given $n$ and $\{b_k\}$, the number of configurations is given by
\begin{align}
	&N_2(n;\{b_k\})\\
	 =\ & \qty[{2n \choose 2b_1}{2b_1\choose 2}{2(b_1-1)\choose 2}\cdots {2\choose 2}\frac{1}{b_1!}\qty(\frac{2!}{2})^{b_1}]\cdots \nonumber\\
	&\ \ \ \qty[{2n-2b_1-4b_2-\cdots -(2k-2)b_{k-1}\choose 2kb_k}{2kb_k\choose 2k}{2k(b_k-1)\choose 2k}..{2k\choose 2k}\frac{1}{b_k!}\qty(\frac{(2k)!}{2k})^{b_k}]\cdots \nonumber\\
	=\ &\frac{(2n)!}{b_1! b_2!\cdots b_k!\cdots }\frac{1}{2^{b_1} 4^{b_2}\cdots (2k)^{b_k}\cdots }
\end{align}
The moment generating function of $\mathcal{O}_{\scriptscriptstyle (\sigma_y)_0}$ is given by
\begin{align}
&\overline{\exp(t\mathcal{O}_{\scriptscriptstyle (\sigma_y)_0})} = \sum_n\frac{t^{2n}}{(2n)!}\overline{\mathcal{O}_{\scriptscriptstyle (\sigma_y)_0}^{2n}}\nonumber\\ 
=\ &\sum_n \frac{t^{2n}}{(2n)!}\sum_{b_1+2b_2+\cdots +kb_k+\cdots  = n}N_2(n;\{b_k\})Z_L^{b_1+\cdots +b_k}\nonumber\\
	=\ &\frac{1}{(1- t^2)^{Z_L/2}}
	\label{eq:MGF_F}
\end{align}

The probability density of $\mathcal{O}_{\scriptscriptstyle (\sigma_y)_0}$ is given by
\begin{align}
	p_{\mathcal{O}_{\scriptscriptstyle (\sigma_y)_0}}(y) = \frac{1}{\sqrt{\pi}\Gamma\qty(\frac{Z_L}{2})}\qty(\frac{|y|}{2})^{\frac{Z_L}{2}-\frac{1}{2}}K_{\frac{Z_L}{2}-\frac{1}{2}}\qty(|y|)
\end{align}

\subsubsection{Joint statistics of two patch operators}

Now we come back to the evaluation of joint statistics of both patch operators $\mathcal{O}_{\scriptscriptstyle (I_2)_0}$ and $\mathcal{O}_{\scriptscriptstyle (\sigma_y)_0}$. The combinatorics is similar to that of Appendix~\ref{app:warmup_0} but slightly more complicated. Consider $\overline{\mathcal{O}_{\scriptscriptstyle (\sigma_y)_0}^{2n}\mathcal{O}_{\scriptscriptstyle (I_2)_0}^m}$. Say, there are $b_{k,l}$ loops with $2k$ $\mathcal{O}_{\scriptscriptstyle (\sigma_y)_0}$ insertions and $l$ $\mathcal{O}_{\scriptscriptstyle (I_2)_0}$ insertions. 
\begin{align}
	\sum_{k,l}2 k b_{k,l} =2 n,\ \ \ \sum_{k,l} lb_{k,l} = m
\end{align}

Given $n$, $m$, and $\{b_{k,l}\}$, the number of choices is given by 
\begin{align}
	&N_3(n,m;\{b_{k,l}\})\nonumber\\
	 =\ &\Bigg\{\qty[{2n \choose 2b_{1,0}}{2b_{1,0}\choose 2}{2(b_{1,0}-1)\choose 2}\cdots {2\choose 2}\frac{1}{b_{1,0}!}\qty(\frac{2!}{2})^{b_{1,0}}]\qty[{m \choose b_{0,1}}{b_{0,1}\choose 1}{b_{0,1}-1\choose 1}\cdots {1\choose 1}\frac{1}{b_{0,1}!}]\Bigg\}\nonumber\\
 \ &\Bigg\{\qty[{2n-2b_{1,0}\choose 4b_{2,0}}{4b_{2,0}\choose 4}{4(b_{2,0}-1)\choose 4}\cdots {4\choose 4}\frac{1}{b_{2,0}!}\qty(\frac{(4+0)!}{4+0})^{b_{2,0}}]\nonumber\\
 \  &\qty[{2n-2b_{1,0}-4b_{2,0}\choose 2b_{1,1}}{m-b_{0,1}\choose b_{1,1}}{2b_{1,1}\choose 2}{b_{1,1}\choose 1}\cdots {2\choose 2}{1\choose 1}\frac{1}{b_{1,1}!}\qty(\frac{(2+1)!}{2+1})^{b_{1,1}}]\nonumber\\
\   &\qty[{m-b_{0,1}-b_{1,1}\choose 2b_{0,2}}{2b_{0,2}\choose 2}{2(b_{0,2}-1)\choose 2}\cdots {2\choose 2}\frac{1}{b_{0,2}!}\qty(\frac{2!}{2})^{b_{0,2}}]\Bigg\}\cdots \nonumber\\
   =\ &\frac{(2n)!m!}{b_{1,0}!b_{0,1}!\cdots b_{k,l}!\cdots }\qty(\frac{2!}{2!\cdot 2})^{b_{1,0}}\qty(\frac{4!}{4!\cdot 4})^{b_{2,0}}\qty(\frac{(2+1)!}{2! 1!(2+1)})^{b_{1,1}}\qty(\frac{2!}{2!\cdot 2})^{b_{0,2}}\cdots \qty(\frac{(2k+l)!}{(2k)!l!(2k+l)})^{b_{k,l}}\cdots 
\end{align}
The moment generating function is given by
\begin{align}
	&\overline{\exp(t\mathcal{O}_{\scriptscriptstyle (\sigma_y)_0}+s\mathcal{O}_{\scriptscriptstyle (I_2)_0})}=\sum_{n,m}\frac{t^{2n}s^m}{(2n)!m!}\overline{\mathcal{O}_{\scriptscriptstyle (\sigma_y)_0}^{2n}\mathcal{O}_{\scriptscriptstyle (I_2)_0}^m}\nonumber\\
	=\ &\sum_{n,m}\frac{t^{2n}s^m}{(2n)!m!}\sum_{\substack{\sum_{k,l} k b_{k,l} = n\\ \sum_{k,l} lb_{k,l} = m,\ b_{k,l}\geq 0}
}N_3(n,m;\{b_{k,l}\})Z_L^{\sum_{k,l}b_{k,l}}\nonumber\\
				=\ &\prod_{k,l}\exp(\frac{(2k+l)!t^{2k}s^l}{(2k)!l!(2k+l)}Z_L)\nonumber\\
				=\ &\frac{1}{\qty[(1-s)^2- t^2]^{\frac{Z_L}{2}}}
				\label{eq:MGF_join}
\end{align}
As a consistency check, we can see that \cref{eq:MGF_join} reduces to \cref{eq:MGF_I} or \cref{eq:MGF_F} once we set $t = 0$ or $s = 0$. The joint probability distribution of $\mathcal{O}_{\scriptscriptstyle (I_2)_0}$ and $\mathcal{O}_{\scriptscriptstyle (\sigma_y)_0}$ is given by
\begin{align}
p_{\scriptscriptstyle\mathcal{O}_{\scriptscriptstyle (I_2)_0}, \mathcal{O}_{\scriptscriptstyle (\sigma_y)_0}}(x, y) = \frac{\Theta(x-|y|)}{2^{Z_L-1}\Gamma\qty(\frac{Z_L}{2})^2}(x^2-y^2)^{\frac{Z_L}{2}-1}e^{-x}	\label{eq:prob_join_0}
\end{align}

As a classical two-state system, the physically interesting quantities are statistics of
\begin{align}
	(p_+)_0\equiv\frac{\mathcal{O}_{\scriptscriptstyle (P_{+})_0}}{\mathcal{O}_{\scriptscriptstyle (I_2)_0}},\ \ \ \ (p_-)_0\equiv\frac{\mathcal{O}_{\scriptscriptstyle (P_{-})_0}}{\mathcal{O}_{\scriptscriptstyle (I_2)_0}}
	\label{eq:clock}
\end{align}
which gives the probability of the classical system in state $\ket{+\hat y}_0$ or state $\ket{-\hat y}_0$.

From \cref{eq:prob_join_0} we get
\begin{align}
    (p_+)_0\sim \text{Beta}\qty(\frac{Z_L}{2},\frac{Z_L}{2}),\ \ (p_-)_0 = 1-(p_+)_0,
    \label{eq:prob_classical_1}
\end{align}
or explicitly,
\begin{align}
	p_{\scriptscriptstyle (p_+)_0,(p_-)_0}(u,v) = \frac{\Gamma(Z_L)}{\Gamma\qty(\frac{Z_L}{2})^2}u^{\frac{Z_L}{2}-1}(1-u)^{\frac{Z_L}{2}-1}\mathbf{1}_{[0,1]}(u)\delta(1-u-v)
	\label{eq:prob_classical_2}
\end{align}
We see that the distribution is symmetric between $(p_+)_0$ and $(p_-)_0$, i.e., there is no bias toward either direction. At large $Z_L$ the probability is highly peaked at $\frac{1}{2}$.

\subsection{Statistics of patch operators for a spin}
\label{app:one_spin}
In this section we study the joint statistics of four patch operators $\qty(\mathcal{O}_{\sigma_0},\mathcal{O}_{\sigma_1},\mathcal{O}_{\sigma_2},\mathcal{O}_{\sigma_3})$.

We want to compute the moment generating function
\begin{equation}
\begin{aligned}
	&\overline{\exp(t_0\mathcal{O}_{\sigma_0} +t_1\mathcal{O}_{\sigma_1}+t_2\mathcal{O}_{\sigma_2}+t_3\mathcal{O}_{\sigma_3})}\\
	=\ &\sum_{n_0,n_1,n_2,n_3}\frac{1}{n_0!n_1!n_2!n_3!}\ t_0^{n_0}t_1^{n_1}t_2^{n_2}t_3^{n_3}\ \overline{\mathcal{O}_{\sigma_0}^{\,n_0}\mathcal{O}_{\sigma_1}^{\,n_1}\mathcal{O}_{\sigma_2}^{\,n_3}\mathcal{O}_{\sigma_3}^{\,n_3}}
\end{aligned}
\label{eq:MGF_spin_def}
\end{equation}

Let $n_0+n_1+n_2+n_3 = m$. Consider the set of ordered $m$-tuples 
\begin{align}
	\mathcal{P}^m = \{\vec a\equiv (a_1,\cdots a_m)| a_i\in \mathcal{P}\}
\end{align}
 where $\mathcal{P}$ is the Pauli group containing four elements $\{\sigma_0\equiv I, \sigma_1,\sigma_2,\sigma_3\}$.
 
 Note that the permutation group $S_m$ can act on $\mathcal{P}^m$ as follows:
 \begin{align*}
 	\pi\cdot (a_1,\cdots ,a_m) = \qty(a_{\pi^{-1}(1)},\cdots ,a_{\pi^{-1}(m)}),\ \ \ \pi\in S_n
 \end{align*}
 Under the action of $S_m$, $\mathcal{P}^m$ is divided into different orbits. Each orbit is labelled by $\mathbf{ n}(\vec a)\equiv (n_0,n_1, n_2, n_3)$ with $n_0+n_1+n_2+n_3 = m$. $n_{\mu}(\vec a)$ is the number of $\sigma_{\mu}$'s contained in the element $\vec a\in\mathcal{P}^m$. $\vec a$ and $\vec b$ are in the same orbit if and only if $\mathbf{n}(\vec a) = \mathbf{n}(\vec b)$. The length of each orbit is given by
 \begin{align}
 	L(\mathbf{n}) = \frac{m!}{n_0!n_1!n_2!n_3!}
 \end{align}

 Now define the following function $\mathcal{F}_m$ on $\mathcal{P}^m$:
 \begin{equation}
 \begin{aligned}
 	\mathcal{F}_m: \ &\mathcal{P}^m\longrightarrow \mathcal{C}\\
 	&\vec a\ \longmapsto \overline{\mathcal{O}_{a_1}\cdots \mathcal{O}_{a_m}}
 \end{aligned}
 \end{equation}
As all patch operators commute, it is easy to see that $\mathcal{F}_m$ is invariant under the action of $S_m$ and $\mathcal{F}_m(\vec a)$ only depends on $\mathbf{n}(\vec a)$: 
 \begin{align}
 	\mathcal{F}_m(\vec a) = \mathcal{F}_m(\pi\cdot \vec a)\equiv \mathcal{F}_m(\mathbf{n}(\vec a)),\ \ \ \forall \vec a \in \mathcal{P}^m
 \end{align}
 Now look at $\sum_{\vec a\in \mathcal{P}^m}\mathcal{F}_m(\vec a)$. We first do the sum over the elements on the same orbit labelled by $\mathbf{n}$, then sum over different orbits:
\begin{align}
 	\sum_{\vec a\in \mathcal{P}^m}\mathcal{F}_m(\vec a) =\sum_{\mathbf{n}:n_0+n_1+n_2+n_3 = m}\mathcal{F}_m(\mathbf{n})L(\mathbf{n})=\sum_{\mathbf{n}:n_0+n_1+n_2+n_3 = m}\frac{m!}{n_0!n_1!n_2!n_3!}\mathcal{F}_m(\mathbf{n}).
 \end{align}
 
We see that up to polynomial factors of $t_{\mu}$'s, the quantity $\displaystyle\sum_m\frac{1}{m!}\sum_{\vec a\in \mathcal{P}^m}\mathcal{F}_m(\vec a)$ is exactly the moment generating function in \cref{eq:MGF_spin_def}.

Now let's study the structure of $\displaystyle \mathcal{F}_m(\vec a)=\overline{\mathcal{O}_{a_1}\cdots \mathcal{O}_{a_m}}$. The boundary condition of this quantity consists of $m$ distinct patch operators $\mathcal{O}_{a_1},\cdots,\mathcal{O}_{a_m}$. In evaluating the path integral, we distribute these $m$ patch operators to oriented circles with cyclic symmetry on each circle. In fact, there is one-to-one correspondence between such configurations and elements in permutation group $S_m$. Given boundary condition labeled by $\vec a$, the sum over configurations becomes the sum over $\tau\in S_m$:
\begin{align}
	\mathcal{F}_m(\vec a) = \sum_{\tau\in S_m}F_m(\vec a, \tau)
	\label{eq:cycle_0}
\end{align}
where $F_m(\vec a,\tau)$ denotes the contribution from the configuration corresponding to $\tau\in S_m$.

To see this, we take an element $\tau\in S_m$ and write it in cycle decomposition as $(132)(45)\cdots $. Each cycle in the composition corresponds to one circle in the path integral configuration. The numbers in each cycle indicates how patch operators are arranged on the corresponding circle.

To be more explicit, say, $\tau\in S_m$ contains $b_k$ cycles of length $k$: $\sum_k kb_k = m$, $b_k\geq 0$ and $k\geq 1$. We write the cycle decomposition of $\tau$ as $\tau = \tau^1_{1}..\tau^{1}_{b_1}\cdots \tau^{k}_{1}\cdots \tau^k_{b_k}\cdots $ where the superscript labels length of the cycle. We will also use the following notation. Consider one particular cycle $\tau^k$ of length $k$ in the decomposition: $\tau^k = (c_1\cdots c_k)$ where $c_j\in\{1,\cdots ,m\}$.
\begin{align}
	\tr(\vec a_{\tau^k})\coloneq \tr(a_{c_1}\cdots a_{c_k})
\end{align}

With this notation and using \cref{eq:rule_PI}, the path integral configuration corresponding to \(\tau=\tau^1_{1}\cdots \tau^{1}_{b_1}\cdots \tau^{k}_{1}\cdots \tau^k_{b_k}\cdots \in S_m\) gives contribution
\begin{align}
	F_m(\vec a,\tau)= Z_L^{b_1+\cdots +b_k}\tr(\vec a_{\tau^1_1})\cdots \tr(\vec a_{\tau^1_{b_1}})\cdots \tr(\vec a_{\tau^k_{\sigma_1}})\cdots \tr(\vec a_{\tau^k_{b_k}})\cdots 
	\label{eq:cycle}
\end{align}

Combining everything together, the moment generating function without the polynomial factors is given by
\begin{align}
	&\sum_m\frac{1}{m!}\sum_{\vec a\in \mathcal{P}^m}\mathcal{F}_m(\vec a)\nonumber\\
	=\ &\sum_m\frac{1}{m!}\sum_{\vec a\in \mathcal{P}^m}\sum_{\tau\in S_m} Z_L^{b_1+\cdots +b_k}\tr(\vec a_{\tau^1_1})\cdots \tr(\vec a_{\tau^1_{b_1}})\cdots \tr(\vec a_{\tau^k_{\sigma_1}})\cdots \tr(\vec a_{\tau^k_{b_k}})\cdots \nonumber\\
	=\ &\sum_m\frac{1}{m!}\sum_{\sum_k kb_k = m}N_1(m;\{b_k\})Z_L^{b_1+\cdots +b_k+..}\nonumber\\
	&\qquad\qquad\qquad\qquad\cdot\qty[\sum_{a_1,\cdots ,a_m}\tr(\vec a_{\tau^1_1})\cdots \tr(\vec a_{\tau^1_{b_1}})\cdots \tr(\vec a_{\tau^k_{\sigma_1}})\cdots \tr(\vec a_{\tau^k_{b_k}})\cdots ]\nonumber\\
		=\ &\sum_{\{b_k\}}\frac{1}{b_1!\cdots b_k!\cdots }\cdots \frac{1}{1!\cdots k!\cdots }\qty(Z_L\sum_a\tr(a))^{b_1}\cdots \qty(Z_L\sum_{a_1,\cdots ,a_k}\tr(a_1\cdots a_k))^{b_k}\cdots \nonumber\\
		 =\ &\exp(Z_L\sum_{k = 1}^{\infty}\frac{1}{k}\sum_{a_i\in \mathcal{P}}\tr(a_1\cdots a_k))
		 \label{eq:MGF_spin_0}
\end{align}
Going from the first to second line we combined \cref{eq:cycle_0} and \cref{eq:cycle}. Going to the third line we exchanged the order of sum over $\vec a$ and $\tau$, and used the fact that after $\vec a$ summations, what's inside the square bracket no longer depends on $\tau^k_j$'s. 
$N_1(m;\{b_k\})$ denotes the number of elements in $S_m$ whose cycle decompositions contain $b_k$ cycles of length $k$. The function $N_1$ was computed in \cref{eq:counting_1}.

To incorporate factors of $t_{\mu}$'s. We let $t_a\coloneq t_{\mu}$ for $a = \sigma_{\mu}$, $\mu = 0,1,2,3$. Then we just replace $a$ by $a\,t_a$ in \cref{eq:MGF_spin_0}. The moment generating function \cref{eq:MGF_spin_def} is given by
\begin{equation}
\begin{aligned}
	&\overline{\exp(t_0\mathcal{O}_{\sigma_0} +t_1\mathcal{O}_{\sigma_1}+t_2\mathcal{O}_{\sigma_2}+t_3\mathcal{O}_{\sigma_3})}=\exp(Z_L\sum_{k = 1}^{\infty}\frac{1}{k}\sum_{a_i\in \mathcal{P}}\tr(a_1\cdots a_k)t_{a_1}\cdots t_{a_k})
\end{aligned}
\label{eq:MGF_spin}
\end{equation}

With $2$ by $2$ matrices we can evaluate \cref{eq:MGF_spin} explicitly. From
\begin{align}
	\sum_{a_i\in \mathcal{P}}\tr(a_1\cdots a_k)t_{a_1}\cdots t_{a_k}=\ &\tr\qty[(t_0\sigma_0+t_1\sigma_1+t_2\sigma_2+t_3\sigma_3)^k]\nonumber\\
		=\ &\frac{1}{2}\qty[\qty(t_0-\sqrt{t_1+t_2^2+t_3^2})^k+\qty(t_0+\sqrt{t_1+t_2^2+t_3^2})^k],
\end{align}
we obtain the moment generating function
 \begin{equation}
\begin{aligned}
	&\overline{\exp(t_0\mathcal{O}_{\sigma_0} +t_1\mathcal{O}_{\sigma_1}+t_2\mathcal{O}_{\sigma_2}+t_3\mathcal{O}_{\sigma_3})}=\frac{1}{\qty[(1- t_0)^2-(t_1^2+t_2^2+t_3^2)]^{\frac{Z_L}{2}}}.
\end{aligned}
\label{eq:MGF_spin_answer}
\end{equation}

The joint probability distribution is given by
\begin{align}
	&p_{\mathcal{O}_{\sigma_0},\mathcal{O}_{\sigma_1},\mathcal{O}_{\sigma_2},\mathcal{O}_{\sigma_3}}(x_0,x_1,x_2,x_3)\nonumber\\
	 =\ & \frac{\Gamma\qty(\frac{Z_L}{2}+\frac{1}{2})}{\pi^{\frac{3}{2}}\Gamma\qty(\frac{Z_L}{2}-1)\Gamma(Z_L)}e^{-x_0}\qty(x_0^2-x_1^2-x_2^2-x_3^2)^{{\frac{Z_L}{2}-2}}\Theta(x_0)\Theta(x_0^2-x_1^2-x_2^2-x_3^2)
	 \label{eq:prob_spin_app}
\end{align}

\subsection{Statistics of patch operators for a system of general dimension $d$}
\label{app:general_d}
Recall $\mathcal{T} \coloneq \{T_0 = I_d, T_1, T_2,\cdots  T_{d^2-1}\}$ is a set of Hermitian generators of $M_d(\mathbb{C})$ satisfying $\Tr(T_{\mu}T_{\nu}) =d \delta_{\mu\nu}$. We compute the moment generating function $\overline{\exp(\sum_{\mu}s_{\mu}\mathcal{O}_{T_{\mu}})} $. We can follow the same derivation as in Appendix~\ref{app:one_spin} and obtain the analog of \cref{eq:MGF_spin}:
\begin{align}
	\overline{\exp(\sum_{\mu}s_{\mu}\mathcal{O}_{T_{\mu}})} = \ &\exp(Z_L\sum_{k = 1}^{\infty}\frac{1}{k}\sum_{a_i\in \mathcal{T}}\tr(a_1\cdots a_k)s_{a_1}\cdots s_{a_k})  \label{eq:MGF_general_0}
\end{align}
Define matrix $S_d\coloneq \sum_{\mu}s_{\mu}T_{\mu}$ and recall $\Sigma_{d} = \frac{1}{d}\sum_{\mu}\mathcal{O}_{T_{\mu}}T_{\mu}$. The left hand side of \cref{eq:MGF_general_0} becomes $\overline{\exp\qty[\Tr(S_d\Sigma_{d})]}$ and the right hand said can be written as
\begin{align}
		\exp(Z_L\sum_{k = 1}^{\infty}\frac{1}{k}\tr(S_d^k))
	=\ &\det(I_d-S_d)^{-\frac{Z_L}{d}}
\end{align}
where the factor of $\frac{1}{d}$ comes from the difference between normalized trace and canonical trace. 

We obtain the moment generating function for matrix $\Sigma_{d}$.
\begin{align}
	\overline{\exp\qty[\Tr(S_d\Sigma_{d})]} = \det(I_d-S_d)^{-\frac{Z_L}{d}}
\end{align}
This is the moment generating function for complex Wishart distribution:
\begin{align}
	\Sigma_{d}\sim \mathcal{W}_d^{\mathbb{C}}\qty(\frac{Z_L}{d},I_d).
\end{align}

\bibliographystyle{JHEP}
\bibliography{reference}

\end{document}